\documentclass[openright,10pt,twoside]{book}

\usepackage{CV}
\usepackage{eurosym} 

\usepackage{tabularx}
\usepackage{textcomp}
\usepackage[square,authoryear]{natbib}

\usepackage{footmisc}

\usepackage[b5paper,twoside,marginpar=0cm,includemp=false,dvips=true,pdftex=false,bottom=3cm]{geometry}
\usepackage[dvips]{graphicx}

\DeclareGraphicsExtensions{.eps}

\usepackage{color}
\usepackage{calc}
\definecolor{titles}{rgb}{0,0,0}
\definecolor{headers}{rgb}{0,0,0}
\definecolor{listbackground}{rgb}{0.75,0.75,1}
\definecolor{inlinelistbackground}{rgb}{0.75,1,0.75}
\definecolor{bostonuniversityred}{rgb}{0.8, 0.0, 0.0}
\definecolor{aurometalsaurus}{rgb}{0.33, 0.33, 0.33}
\definecolor{Black}{rgb}{0.0, 0.0, 0.0}

\usepackage{soul}

\usepackage{titlesec}
\titleformat{\section}[block]{\normalfont\sffamily}{\color{titles}\thesection}{.5em}{\color{titles}\titlerule\newline\Large\bfseries\color{titles}}
\titleformat{\subsection}[block]{\normalfont\sffamily}{\color{titles}\thesubsection}{.5em}{\large\bfseries\color{titles}}
\titleformat{\subsubsection}[block]{\normalfont\sffamily}{}{.5em}{\normalsize\bfseries\color{titles}}
\titleformat{\chapter}[block]{\normalfont\sffamily}{\color{titles}\chaptername{} \thechapter}{.5em}{\color{titles}\titlerule\newline\Huge\bfseries\color{titles}}
\titleformat{\part}[display]{\normalfont\sffamily\huge\bfseries}{\color{titles}\centering\partname{} \thepart\newline\titlerule}{0.5cm}{\centering\color{titles}\Huge\bfseries\color{titles}}

\usepackage{fancyhdr}
\fancypagestyle{groovy}{
\fancyhf{}
\fancyhead[RO]{\bfseries \textcolor{headers}{\rightmark}}
\fancyhead[LE]{\bfseries \textcolor{headers}{\leftmark}}
\fancyfoot[RO,LE]{\textcolor{headers}{\thepage}}

}

\fancypagestyle{plain}{
\fancyhf{}
\fancyfoot[RO,LE]{\textcolor{headers}{\thepage}}

}

\makeatletter
\def\cleardoublepage{\clearpage\if@twoside \ifodd\c@page\else
    \hbox{}
    \vspace*{\fill}
    \vspace{\fill}
    \thispagestyle{empty}
    \newpage
    \if@twocolumn\hbox{}\newpage\fi\fi\fi}
\makeatother

\makeatletter
\def\@endpart{\vfil\newpage
              \if@twoside
                \null
    \hbox{}
    \vspace*{\fill}
    \begin{center}
    \end{center}
    \vspace{\fill}
                \thispagestyle{empty}%
                \newpage
              \fi
              \if@tempswa
                \twocolumn
              \fi}
\makeatother

\makeatletter
\def\cleartorecto{\clearpage\if@twoside \ifodd\c@page\else
  \hbox{}\thispagestyle{cleared}%
  \newpage\if@twocolumn\hbox{}\newpage\fi\fi\fi}
\makeatother

\makeatletter
\def\cleartoverso{\clearpage\if@twoside
  \ifodd\c@page\hbox{}\thispagestyle{cleared}%
  \newpage\if@twocolumn\hbox{}\newpage\fi\fi\fi}
\makeatother

\let\stdappendix\appendix
\renewcommand{\appendix}{\stdappendix\renewcommand{\chaptername}{Appendix}}

\usepackage{nomencl}
\usepackage{acronym}

\makeglossary

\let\stdtoc\tableofcontents
\renewcommand{\tableofcontents}{\parskip=0.0cm
\stdtoc
\parskip=0.2cm
}

\usepackage[T1]{fontenc}

\usepackage{amssymb}
\usepackage{amsmath}
\usepackage{bbm}

\usepackage{pandora}

\usepackage{pxfonts}

\usepackage[TABTOPCAP]{subfigure}

\usepackage{here}

\usepackage[thinspace,thinqspace,amssymb]{SIunits}
\usepackage[pdftex]{dropping}

\renewcommand{\thechapter}{\arabic{chapter}}
\renewcommand{\thepart}{\Roman{part}}

\usepackage{algorithmic}

\newcommand{\mpstart}{%

 \setlength{\pdfpagewidth}{176mm}
 \setlength{\pdfpageheight}{250mm}

\DeclareGraphicsExtensions{.jpg,.eps,.pdf,.png}

}

\usepackage[b5paper,dvipdfm,plainpages=false,colorlinks=true,hyperindex=true,linkcolor=black,bookmarks=true,citecolor=black,bookmarksopen=true,urlcolor=black,bookmarksnumbered=true,pdfborder=0 0 0]{hyperref}

\usepackage{float}
\floatstyle{plain}

\newfloat{Listing}{tbh}{lis}[chapter]

\usepackage[centerlast,small,bf]{caption}
\usepackage{afterpage}

\usepackage{footnote}
\usepackage{subfigure}
\usepackage{float}
\usepackage{midpage}
\usepackage{colortbl}
\usepackage{graphicx}
\usepackage{graphics}
\usepackage{epsfig}
\usepackage{bookmark}
\usepackage[dvips]{graphicx}
\usepackage{amsmath, amssymb}
\usepackage[tight]{units}
\usepackage{wrapfig}
\usepackage[utf8x]{inputenc}
\usepackage{float}
\usepackage[toc]{appendix}
\usepackage{titlesec}
\titleformat{\chapter}[display]
  {\normalfont\Large\raggedleft}
  {\MakeUppercase{\chaptertitlename \ \ \ \fontsize{55pt}{55pt}\selectfont{\thechapter}}%
      \rlap{  \rule{5cm}{1.5cm}}}
  {10pt}{\Huge}
\titlespacing*{\chapter}{0pt}{30pt}{20pt}

\titleformat{\part}[display]
{\normalfont\filcenter}
{\Huge Part \thepart}
{1ex}
{\vspace{2ex}%
\Huge}

\titleformat{\section}[display]
{\normalfont\bfseries}
{\normalsize\thesection}
{0.1ex}
{\titlerule[1pt]
\vspace{0.1ex}
\Large}

\titleformat{\subsection}[hang]
{\normalfont\bfseries}
{\large\thesubsection}
{0.1ex}
{\hspace{0.1ex}
\large}

\usepackage{url}
\usepackage{lipsum}

\newcommand{\HRule}{\rule{\linewidth}{0.5mm}}

\makeatletter
\renewcommand\maketitle{
\begin{titlepage}%
    \addcontentsline{toc}{chapter}{Titlepage}%
  \let\footnotesize\small
  \let\footnoterule\relax
  \let \footnote \thanks

\begin{figure}
\begin{center}
  \includegraphics[width=0.59\textwidth]{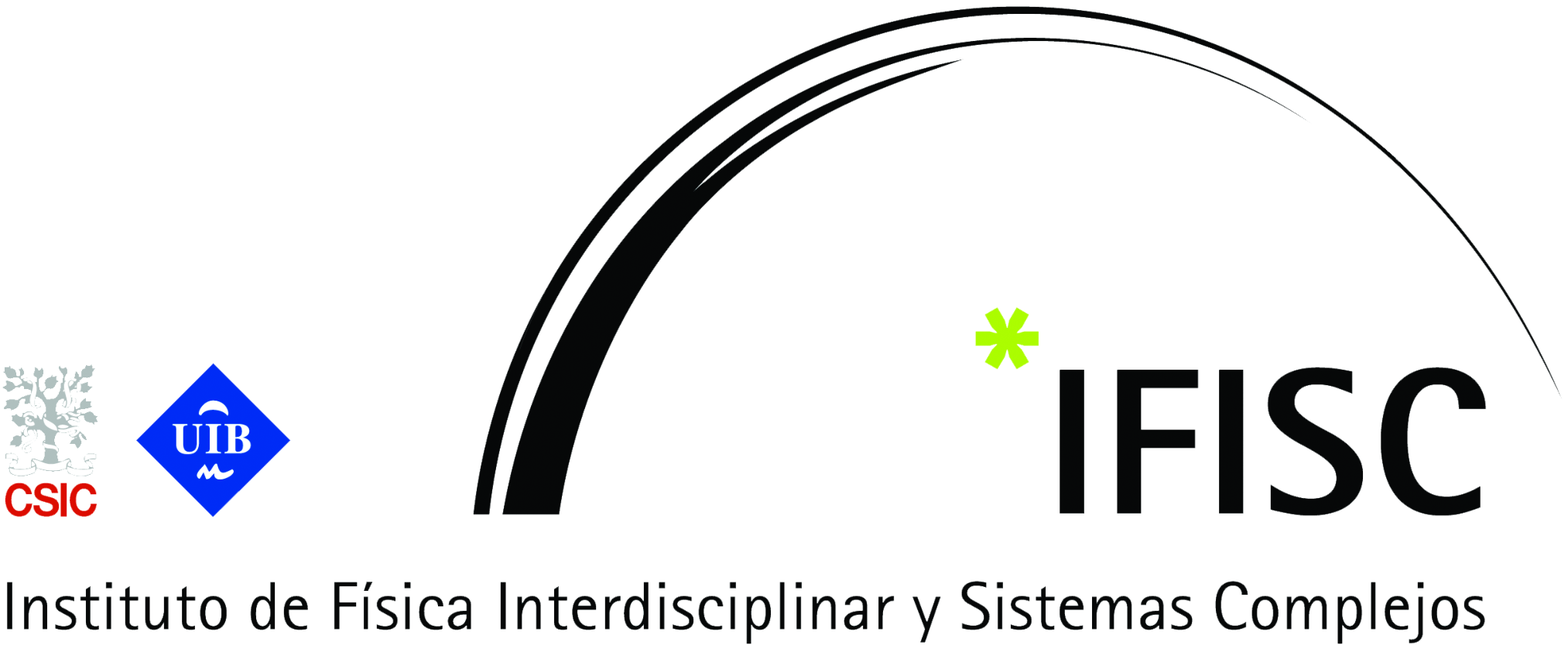}
\end{center}
\end{figure}

  \begin{center}
    \HRule \\[0.2cm]
     {\LARGE \textbf{Nonequilibrium Statistical Physics in Ecology: Vegetation Patterns, Animal Mobility and Temporal Fluctuations}}
    \HRule \\[0.2cm]
    \vskip 5em%
    {\LARGE PhD THESIS}
    \vskip 3.5em%
    {\Large
     \lineskip .75em%
      \begin{tabular}[t]{c}
        \@author
      \end{tabular}\par}%
      \vskip 14em%
    {\large Director:}\\
    \vskip 1.2em
    {\large Dr. Crist\'obal L\'opez}\\
    {\large 2014}
  \end{center}\par
  \end{titlepage}
  \setcounter{footnote}{0}%
  \global\let\thanks\relax
  \global\let\maketitle\relax
  \global\let\@thanks\@empty
  \global\let\@author\@empty
  \global\let\@date\@empty
  \global\let\@title\@empty
  \global\let\title\relax
  \global\let\author\relax
  \global\let\date\relax
  \global\let\and\relax
}
\makeatother



\makeatletter
   {%
    \end{minipage}
   \end{lrbox}%
\colorbox{listbackground}{\usebox{\@tempboxa}}}%
\makeatother

\makeatletter
   {%
    \end{minipage}
   \end{lrbox}%
\colorbox{inlinelistbackground}{\usebox{\@tempboxa}}}%
\makeatother

\def\citenum#1{{\def\@cite##1##2{##1}\cite{#1}}}

\newcommand{\acps}[1]{%
\expandafter\ifx\csname fn@#1\endcsname\relax%
 \textbf{!!!#1!!!}%
\else%
 #1s%
\fi}

\newcommand{\acpf}[1]%
  {\csname fn@#1\endcsname{}s\nolinebreak[3] (\acps{#1})}%

\newcommand{\acpl}[1]{\csname fn@#1\endcsname{}s}%

\renewcommand{\acp}[1]{%
\expandafter\ifx\csname ac@#1\endcsname\relax%
 \acpf{#1}%
 \expandafter\gdef\csname ac@#1\endcsname{x}%
\else%
 \acps{#1}%
\fi}

\newcommand{\cpp}[1]{\lstinline!#1!}

\newcommand{\bx}{{\bf  x}}
\newcommand{\br}{{\bf  r}}
\newcommand{\bk}{{\bf  k}}

\newcommand{\bs}{{\bf  s}}

\let\tmp\oddsidemargin
\let\oddsidemargin\evensidemargin
\let\evensidemargin\tmp
\reversemarginpar

\begin{document}
\pagenumbering{roman}
\mpstart

\frontmatter
\title{\textbf{Nonequilibrium Statistical Physics in Ecology: Vegetation Patterns, Animal Mobility and Temporal Fluctuations}}

\author{Ricardo Martínez García}

\maketitle

 \pagestyle{empty}
 \noindent
 \begin{footnotesize}
 \parbox{12cm}{
\textsc{\textbf{Nonequilibrium Statistical Physics in Ecology: Vegetation Patterns, Animal Mobility and Temporal Fluctuations}}\\
\\
Ricardo Martínez García\\
Tesis realizada en el Instituto de Física Interdisciplinar y Sistemas Complejos, IFISC (CSIC-UIB).\\
Presentada en el Departamento de Física de la Universitat de les Illes Balears.} \\
\vspace{\stretch{1}}
\\
\parbox{12cm}{\textbf{For an updated version of this thesis please contact:\\
ricardo@ifisc.uib-csic.es or rmtzgarcia@gmail.com}\\
\\
PhD Thesis\\}
\\
\begin{flushright}
\parbox{12cm}{
Director: Dr. Cristóbal López\\
\\
Palma de Mallorca, 2 de Mayo de 2014.}
\end{flushright}
The cover shows two aerial photographies of vegetation patterns taken from \cite{Hardenberg}, and a Mongolian gazelle offspring taken from www.lhnet.org.
\vspace{0.45cm}

 \end{footnotesize}
%
\newpage
 \vspace*{0.1cm}
\HRule \\[0.7cm]
{Cristóbal López Sánchez, Profesor Titular de Universidad}

  \vskip 1.5em
  {CERTIFICA:}
  
\vskip 1.5em
  {que esta tesis doctoral ha sido realizada por el doctorando Sr. Ricardo Martínez García bajo su dirección
  en el Instituto de Física Interdisciplinar y Sistemas Complejos y, para que conste, firma la presente}

  \vskip 3.5em
  \hskip 15em
  Director:  

  \vskip 7em
  \hskip 22em
  Dr. Cristóbal López Sánchez

  \vskip 4em
  \hskip 15em
Doctorando:

  \vskip 7em
  \hskip 22em
  Ricardo Martínez García
  
  \vskip 3em

    \hskip 23em
  Palma de Mallorca,
  \vskip 0.7em
      \hskip 25em
  2 de Mayo de 2014.
  
\vspace{0.6cm}

\HRule \\[0.7cm]

\newpage
\thispagestyle{empty}
  \vspace*{7cm}
  \begin{flushright}
   
  \textit{\textbf{A mis padres y hermano.}}
  
  \vspace{1.5cm}
  
  \color{aurometalsaurus}
  \textit{Look up to the sky.\\
   You will never find rainbows\\
   if you are looking down.}\\
   
  \vspace{0.7cm}
  
  \textit{Charles Chaplin.}
    \color{black}
  \end{flushright}
\color{black}
 \cleardoublepage


\pagestyle{headings}

 \chapter*{Agradecimientos}
\begin{flushleft}
  \rule{\textwidth}{0.3mm}
\end{flushleft}

En primer lugar, quisiera agradecer a Cristóbal el haberme dado la oportunidad de hacer esta tesis doctoral. Gracias por todo lo aprendido, pero sobre todo por tu buen humor 
y continuo apoyo. Gracias también por darme la libertad necesaria para equivocarme y tener mis propias ideas,
o al menos intentarlo. I also want to thank Justin Calabrese who I have considered my second PhD supervisor since my stay in Front 
Royal. Thanks for your friendliness and for teaching me so many things. No quisiera olvidarme de Emilio Hernández-García y Federico Vázquez, siempre
habéis tenido un rato para sentaros conmigo. Gracias especialmente a tí, Fede, por tantas horas juntos delante del ordenador 
buscando power laws. Finally, thank you Miguel Ángel Mu\~noz, Chris Fleming, Thomas Mueller and Kirk Olson for spending part of your time on me.

Empezando por el principio, quiero dedicar unas palabras a Juan Francisco, por dedicarme tanto tiempo en el colegio y por despertar el interés por la física
en mí. También a Daniel Alonso, por ayudarme a recuperarlo, y a Santiago Brouard, por iniciarme en el mundo de la investigación en La Laguna.

En el IFISC he encontrado el ambiente ideal para disfrutar de esta tesis. Gracias a todos por hacerlo posible. En particular a 
Marta, Inma, Rosa Campomar y Rosa Rodríguez por hacer de la burocracia algo más sencillo. A Edu, Rubén, Antònia y David, porque sin su ayuda seguramente
esta trabajo no habría ido para delante. Espero no haber dado mucho la lata, y si lo he hecho, os debo un desayuno. Siempre estaré agradecido a 
todos mis compañeros en la profundidad de la S07. Al Dr. Luis F. Lafuerza, por su acogida durante mi primer verano
y por estar siempre dispuesto a echarme una mano. A Luca: Calviá y Milán nunca estuvieron tan cerca.
A Pablo y Víctor, por nuestras conversaciones sobre arte precolombino y vuestro sentido del humor, y a 
Enrico y Simone, por las clases gratis de italiano. Muchas gracias también a todos los demás: Przemyslaw (espero haberlo escrito bien),
Miguel Ángel, Pedro, Adrián, Julián, Neus, Juan, Xavi, Marie, Ismael, Leo, Alejandro, Toni Pérez...

Part of this Thesis has been done in other institutions. I have really enjoyed these experiences, but they would not have been the same without all the 
nice people that I could meet. Thanks to Nat, Ben, Fan, Meng, Tuya, Leah, Caroline, Bettina, Christian, Jan...
for the time in Front Royal, y gracias a Pablo por alguna escapada por Dresden.

A nivel institucional, gracias al CSIC y a la Univeristat de les Illes Balears, pues sin sus fondos nunca podría haber completado
este doctorado, y a los proyectos FISICOS e INTENSE@COSYP.

Estos años en Mallorca han sido mucho más que años de trabajo. A través del fútbol sala he podido conocer a muchísima gente. Gracias a todos, a los del Bahia's
y Ciutat de Mallorca, pero especialmente
a mis compañeros de La Salle Pont d'Inca. Habéis sido una familia desde el primer día. Especialmente, me gustaría
acordarme de Alberto, por abrirme las puertas y por tener siempre un rato para liarse. A los demás, ¿qué pasa ...?
Nombraros a todos ocuparía mucho espacio, pero lo voy a hacer: Bily, Biel, Raúl,
Xevi, Héctor, Colo, Edu, Toni, Roberto, Berni, Baia (gran futuro delante de las cámaras),
Toni Mir, Pitu, Majoni, Rafa, Perillas y Bernat. Muchísimas gracias también
a todos los niños que he tenido la inmensa fortuna de entrenar. Sobre todo a los más pequeños, de quienes tenemos muchísimo que aprender. 

He tenido la gran suerte de vivir con grandes compañeros. Miguel, uno de los mejores amigos que se pueden tener y probablemente la persona con mayor
visión de futuro.
Mario, importador a Mallorca de la última tecnología riojana: el motorabo. ¡Dale qué suene! Siempre. Juntos hemos compartido muchos de los mejores momentos. Gracias.
Gracias Àngel por enseñarme la isla y por nuestros momentos de pesca. 
No me olvido de vosotros, Luis, Gloria, Kike, Laura y Adrián porque por vuestra culpa llegar a Mallorca fue fácil e irse será difícil.

Estas últimas líneas van para la gente que más quiero. A mis amigos de Tenerife, los de siempre. A Paula, por ser como eres, por tu sonrisa,
y por acompañarme. A Guillermo, Bárbara y Miki, por preocuparos por mi. Pero sobre todo dedico este trabajo a mis padres, por ser el mejor ejemplo y por
creer y confiar en mi. Ya que no os lo recuerdo muy a menudo, aprovecho estas palabras para deciros lo mucho que os quiero.
A mi hermano, el mejor compañero y un ejemplo como científico. Gracias por mis primeros trabajos de campo a los 2 años, poca
gente tiene la suerte de saber qué es un coleóptero antes de ir a la guardería\footnote{Orden de insectos masticadores que poseen un caparazón duro y dos alas,
también duras, que cubren a su vez dos alas membranosas}. Por último, a mis abuelos,
a los que ya no están, Alejandro, Rosa e Isidro y a Cuquita, por poder disfrutarte cada día.

 \addcontentsline{toc}{chapter}{Agradecimientos}
 
\chapter*{Resumen}
\begin{flushleft}
  \rule{\textwidth}{0.3mm}
\end{flushleft}

Esta  tesis doctoral se centra en la aplicación de técnicas propias
de la física estadística del no equilibrio al estudio de problemas con trasfondo ecológico.

En la primera parte se presenta una breve introducción con el fin de contextualizar el uso de modelos cuantitativos 
en el estudio de problemas ecológicos. Para ello, se revisan los fundamentos teóricos y las herramientas matemáticas utilizadas en los trabajos
que ocupan los capítulos siguientes. En primer lugar, se explican las distintas maneras de describir matemáticamente 
 este tipo de sistemas, estableciendo relaciones entre ellas y explicando
las ventajas e incovenientes que presenta cada una. En esta sección también se introducen la terminología
y la notación que se emplearán más adelante.

En la segunda parte se comienzan a presentar resultados originales. Se estudia la formación de patrones
de vegetación en sistemas en los que el agua es un factor que limita la aparición de nuevas plantas.
Esta parte se divide en dos capítulos. 
\begin{itemize}
\item El primero se centra en el caso particular de sabanas
mésicas, con una precipitación media anual intermedia, y en las que
los árboles coexisten con otros tipos de vegetación más baja (arbustos y hierbas). Se presenta un modelo en el que se incluyen los
efectos de la competición por recursos y la presencia de incendios. En este último caso, la protección que los árboles
adultos proporcionan a los jóvenes contra el fuego supone una interacción de facilitación a muy corto alcance entre
la vegetación. El principal
resultado de este estudio concluye que, incluso en el límite en el que los mecanismos facilitativos tienen un alcance muy corto
(local), aparecen patrones en el sistema. Finalmente, incluyendo la naturaleza estocástica de la dinámica de nacimiento y muerte de los árboles se
recuperan estructuras con formas más parecidas a las observadas en sabanas reales.

\item El segundo capítulo de esta parte estudia la formación de patrones en sistemas áridos, cuyas formas son mucho más regulares 
que en las sabanas mésicas. Además, las precipitaciones 
también son más escasas. El origen de estas estructuras se 
atribuye tradicionalmente a la presencia de diferentes interacciones entre las plantas que actúan en distintas 
escalas espaciales. En particular, muchos de los trabajos previos defienden que se deben a 
la combinación de mecanismos que facilitan
el crecimiento de vegetación  a corto alcance (facilitación) con otros, de mayor alcance, que lo inhiben (competición).
En este capítulo se presentan modelos
en los que únicamente se incluyen interacciones competitivas, a pesar de lo cual se recupera la secuencia típica de patrones obtenida en modelos previos.
Se introduce el concepto de \textit{zonas de exclusión} como mecanismo biológico responsable de la formación de patrones.
\end{itemize}

En la tercera parte de la tesis se presentan modelos para el estudio del movimiento y comportamiento colectivo de animales. En concreto, se investiga
la influencia que tiene la comunicación entre individuos en los procesos de búsqueda que estos llevan a cabo, con especial énfasis en la búsqueda de recursos.
Consta de dos capítulos.
\begin{itemize}
 \item En primer lugar, se analiza desde un punto de vista teórico la influencia de la comunicación en los tiempos de búsqueda.
En general, comunicaciones a escalas intermedias resultan en tiempos de búsqueda menores, mientras que alcances más cortos o más largos proporcionan 
una cantidad de información insuficiente o excesiva al resto de la población.
Esto impide a los individuos decidir correctamente en qué dirección moverse, lo cual da lugar a tiempos de búsqueda mayores. 
El capítulo se completa estudiando la influencia que
tiene el tipo de movimiento de los individuos (browniano o L\'evy) en los resultados del modelo.
 
\item Esta parte finaliza presentando una aplicación del modelo desarrollado en el capítulo anterior
al caso de las gacelas que habitan las estepas centroasiáticas (\textit{Procapra gutturosa}).
En los últimos años, se ha observado un gran decrecimiento en la población de esta especie. Esto se debe
a la caza masiva de estos animales y a una pérdida y fragmentación de su hábitat provocada por la acción del hombre.
Conocer sus hábitos migratorios y comportamiento resulta, por tanto, fundamental para desarrollar estrategias de conservación
eficientes. En particular, en este capítulo se estudia la búsqueda de pastos por parte de estas gacelas, utilizando mapas reales de vegetación
y medidas GPS del posicionamiento de un grupo de individuos. Se presta especial atención al efecto de la comunicación
vocal entre animales, midiendo la eficiencia de la búsqueda en términos de su duración y de la formación de grupos en las zonas más
ricas en recursos. Las gacelas encuentran buenos pastos de una manera óptima cuando se comunican emitiendo sonidos cuyas frecuencias
coinciden con las obtenidas en medidas reales hechas en grabaciones de estos animales. Este resultado sugiere la posibilidad de que a lo largo de su evolución la gacela
\textit{Procapra gutturosa} haya optimizado su tracto vocal para facilitar la comunicación en la estepa.
\end{itemize}

En la cuarta parte, que consta de un único capítulo, se analiza el efecto que tiene un medio externo cuyas propiedades cambian estocásticamente en el tiempo sobre
diferentes propiedades de un sistema compuesto por muchas partículas que interaccionan entre sí. Se estudian los tiempos de paso 
cuando el parámetro de control del problema fluctúa en torno a un valor medio.
Se encuentra una región finita del diagrama de fases en la cual los tiempos escalan como una ley de potencia con el tamaño del sistema. Este resultado
es contrario al caso {\it puro}, en el que el parámetro de control es constante y esto únicamente ocurre en el punto critico. 
Con estos resultados se extiende el concepto de 
\textit{Fases Temporales de Griffiths} a un mayor número de sistemas.

La tesis termina con las conclusiones del trabajo y señalando posibles líneas de investigación que 
toman como punto de partida los resultados obtenidos.
 \addcontentsline{toc}{chapter}{Resumen}
 
\chapter*{Abstract}
\begin{flushleft}
  \rule{\textwidth}{0.3mm}
\end{flushleft}

This thesis focuses on the applications of mathematical tools and concepts brought from nonequilibrium statistical
physics to the modeling of ecological problems.

The first part provides a short introduction where the theoretical concepts and mathematical tools that are going to be 
used in subsequent chapters are presented. Firstly, the different levels of description usually employed
in the models are explained. Secondly, the mathematical relationships among them 
are presented. Finally, the notation and terminology
that will be used later on are explained.

The second part is devoted to studying vegetation pattern
formation in regions where precipitations are not frequent
and resources for plant growth are scarce. This part comprises two chapters.

\begin{itemize}
 \item The first one studies the case of mesic savannas. These systems are characterized by
 receiving an intermediate amount of water and by a long term coexistence of layer made of grass and shrubs interspersed with
 irregular clusters of trees. A minimalistic model considering only long range competition among plants
 and the effect of possible fires is presented. In this later case, adult trees protect the growth of juvenile 
 individuals against the fires by surrounding them and creating an antifire shell. This introduces a local facilitation
 effect for the establishment of new trees. Despite the range of facilitative interactions is taken to its infinitesimally short limit,
 the spectrum of patterns obtained in models with competitive and facilitative nonlocal interactions is recovered. 
 Finally, considering the stochasticity in the birth and death dynamics 
 of trees, the shapes of the structures reproduce the irregularity observed in aerial photographs of mesic savannas.
 
 \item The second chapter investigates the formation of patterns in arid regions,
 that are typically more regular than in mesic savannas.
 Previous studies attribute the origin of these structures to the existence of
 competitive and facilitative interactions among plants acting simultaneously but at different spatial scales.
 More precisely, to the combination of a short-range facilitation and a long-range competition (\textit{scale-depedent feedback}). The findings of this chapter
 are based on the study of a theoretical model that assumes only long-range competitive interactions and shows the existence of vegetation
 patterns even under these conditions. This result suggests that the role of facilitative interactions could be superfluous in the development of 
 these spatial structures. The biological concept of \textit{exclusion areas} is proposed as an alternative to 
 conventional \textit{scale-dependent feedback}.
\end{itemize}

The third part of the thesis develops a series of mathematical models describing the collective movement and behavior of some animal species.
Its primary objective is to investigate the effect that communication among foragers has on searching
times and the formation of groups. It consists of two chapters:

\begin{itemize}
 \item In the first one, the model is established and its properties studied from a theoretical point of view.
 The main novelty of this work is the inclusion of communication among searchers  to share information
 about the location of the targets. Communication and amount of shared information are directly connected through the range 
 of the signals emitted by successful searchers. In this context, searching processes are optimized in terms of duration 
 when the individuals share intermediate amounts of information, corresponding to mid-range communication.
 Both a lack and an excess of information may worsen the search. The first implies an almost uninformed search, while the latter 
 causes a loss in the directionality of the movement since individuals are overwhelmed with information coming from many targets.  
 Finally, the influence of the type of movement on the search efficiency is investigated, comparing the Brownian and L\'evy cases. 
 Some analytical approximations and a continuum description of the model are also presented.
 
 \item This part ends with an application of the previous model to the foraging behavior of Mongolian gazelles (\textit{Procapra gutturosa}).
 The population of this species has
 decreased in the last century because of massive hunting and a progressive habitat degradation and fragmentation caused by human disturbances in the 
 Eastern steppe of Mongolia. Studying their mobility patterns and social behavior improves the development of conservation strategies.
 This chapter suggests possible searching strategies used by these animals to increase their forage encounters rate. The study 
 is supported by the use of real vegetation maps 
 based on satellite imagery and GPS data tracking the position of a group of gazelles. The main focus is on the effect that nonlocal vocal communication 
 among individuals has on foraging times and group formation in the areas with better resources. According to the results of the model, the searching 
 time is minimized 
 when the communication takes place at a frequency that agrees with measurements made in gazelle's acoustic signals. This suggests
 that, through its evolution, \textit{Procapra gutturosa} may have optimized its vocal tract in order to facilitate the 
 communication in the steppe.
\end{itemize}

The fourth part covers the effect of stochastic temporal disorder, mimicking climate and environmental variability,
on systems formed by many interacting particles. These models may serve as an example of ecosystems. The temporal disorder
is implemented making the control parameter fluctuating around a mean value close to the critical point. The effect of this
external variability is quantified using passage times. 
The results show a change in the behavior of this magnitude compared
with the pure case, that is, in the absence of external fluctuations. 
Within a finite region of the phase diagram, close to the critical point,
the passage times scale as a power law with continuously varying exponent.
In the pure model this behavior is only observed at the critical point.
After these results, the concept of 
\textit{Temporal Griffiths Phases}, introduced in the spreading of epidemics, is extended to a vast range of models.

The thesis ends with a summary and devising future research lines.
 \addcontentsline{toc}{chapter}{Abstract}
 
\tableofcontents
\cleardoublepage

\pagenumbering{arabic}

\mainmatter

\pagestyle{groovy}



\part[\textit{\textsc{Introduction}}]{\textbf{\textit{\textsc{Introduction \\}}}
         }   

 \label{part:intro}
\chapter*{}

\Huge
  \hskip 0.5em S\normalsize tatistical physics focuses on the study of those systems 
that comprise a large number of simple components. Regardless of the particular nature of these
fundamental entities, it describes the interactions among them and the global properties that appear at a macroscopic scale.
These emergent phenomena are the hallmark of complex systems. Such systems are used to model processes in several disciplines,
most of the times, far from the physical sciences. That's why, during the last few years, statistical physics
has become a powerful cross disciplinary tool, supplying the theoretical framework and the mathematical
techniques that allow the study of many different problems in biology, economics or sociology. It provides a scenario that makes possible to
encapsulate the huge number of microscopic degrees of freedom of a complex system
into just a few collective variables.

On the other hand, ecology is concerned with the study of the relationships between organisms and their 
environment. In terms of this thesis, it is a paradigmatic example of complexity science. Ecological systems are formed
by a huge number of heterogeneous constituents that interact and evolve stochastically in time. In addition, they 
are subject to changes and fluctuations in the surroundings that apart them from equilibrium\footnote{Here equilibrium
refers to the thermodynamic equilibrium. It is a state of balance characterized by the absence of fluxes and currents in the system¡.}.

Because of this complex nature, ecology was originally an empirical science with purely descriptive purposes.
Ancient Greek philosophers such as Hippocrates and Aristotles laid the foundations of ecology in their studies on natural history. 
However, over the years, the need for a mathematical formalism to tighten all the observations increased, until ecology adopted a
more analytical approach in the late 19th century. The first models attracted the attention of many physicist and mathematicians that started developing
new techniques and tools. Nowadays, theoretical ecology is a well established discipline that deals with several topics related not
only with environmental conservation but also with evolutionary biology, ethology and genetics.
It constitutes, together with recent technological
advances, a potent instrument to better understand the natural environment.

Ecological systems show characteristic variability on a range of spatial, temporal and organizational scales \citep{levin1992problem}.
However, when we observe them, we do it in a limited range. Theoretical studies aim to comprehend how 
information is transferred from one level to other. They permit understanding natural phenomena in terms of the processes
that govern them, and consequently develop management strategies. Without this knowledge, each stress must be evaluated 
separately in every system, and it would not be possible to extrapolate the knowledge obtained from one situation to another.
But, what is the role of statistical physics in this task? On the one hand, most ecological systems can exhibit multistability,
abrupt transitions, patterns or self-organization when a control parameter is varied. These concepts are
characteristic of nonlinear systems, that have been traditionally studied by statistical physicists. Particularly interesting 
are those cases in which the dynamics at one level of organization can be understood as a consequence of the collective behavior
of multiple similar identities. This reminds the definition of the systems that are the 
focus of statistical physics, which serves for developing simple models that retain and condense
the essential information, omitting unnecessary details. 

There is a large list of recent developments that may serve as examples of this relationship \citep{fort2013statistical}:
collective animal movement \citep{cavagna2010scale}, demographic stochasticity in multiple species systems \citep{mckane2005predator,butler}, 
evolutionary theory \citep{chia2011statistical}, population genetics \citep{de2011contribution}, species distribution \citep{harte2008maximum,volkov2003neutral},
complex ecological networks \citep{montoya2006ecological,bastolla2009architecture}, animal foraging \citep{mendez2014random,libroforaging},
or species invasion \citep{seebens2013risk}. In this thesis,
I will abroad different problems within the framework
of statistical physics, in particular vegetation pattern formation, animal behavior and ecosystem's robustness. It is important to remark
the diverse nature of each of these systems. Plants are inert, and so the development of patterns is a consequence of the interaction with the
environment and the birth-death dynamics. On the other hand, animals usually show large migratory displacements and 
tend to form groups of individuals by coming together. Gathering these problems, the objective of
this dissertation is to emphasize the connection between statistical physics and
environmental sciences and its role in the establishment of ecological models.
 
The powerful of statistical physics as a cross disciplinary tool allows to tackle different questions 
depending on the particularities of each system.
Here we wonder how external variability affects robustness and evolution of ecosystems and the mean lifetime
of the species. We are also interested in disentangling the different facilitative and competitive interaction among plants in
vegetation systems to unveil its role in the formation of patterns. Are both needed to maintain these regular structures? How efficient are
inhomogeneous distributions of vegetation to avoid desertification in water-limited systems? Finally, we will try to shed light on the relationship
between communication and foraging efficiency. This is one of the less investigated topics in the study of searching strategies. How can different
communication mechanisms affect searching processes? Is the mean searching time a good metrics to quantify search efficiency? Does it exist an 
optimal communication range that accelerates the search? How does sharing information affect the collective use of a heterogeneous landscape?
Answering these and other issues will be the goal of this thesis.

The results of each chapter can be found in the following publications:
\begin{itemize}

 \item Chapter 2:
 \begin{itemize}
  \item R. Mart\'inez-Garc\'ia, J.M. Calabrese, and C. L\'opez, (2013), \textit{Spatial patterns in mesic savannas: the local facilitation limit and the role of demographic stochasticity, Journal of Theoretical Biology}, {\bf 333}, 156-165.
 \end{itemize}
 
 \item Chapter 3:
  \begin{itemize}
  \item R. Mart\'inez-Garc\'ia, J.M. Calabrese, E. Hern\'andez-Garc\'ia and C. L\'opez, (2013), \emph{Vegetation pattern formation in semiarid systems without facilitative mechanisms}, Geophysical Research Letters, {\bf 40}, 6143-6147.
  \item R.Martínez-García, J.M. Calabrese, E. Hernández-García and C. López, (2014), \emph{Minimal mechanisms for vegetation patterns in semiarid regions}, Reviewed and resubmitted to Philosophical Transactions of the Royal Society A.
  \end{itemize}
  
 \item Chapter 4:
 
  \begin{itemize}
   \item R. Martínez-García, J.M. Calabrese, T. Muller, K.A. Olson, and C. López, (2013), \emph{Optimizing the Search for Resources by Sharing Information: Mongolian Gazelles as a Case Study}, Physical Review Letters, {\bf 110}, 248106.
   \item R. Martínez-García, J.M. Calabrese, and C. López, (2014), \emph{Optimal search in interacting populations: Gaussian jumps versus L\'evy flights}, Physical Review E, {\bf 89}, 032718,
  \end{itemize}

 \item Chapter 5:
 
  \begin{itemize}
   \item R. Martínez-García, J.M. Calabrese, T. Muller, K.A. Olson, and C. López, (2013), \emph{Optimizing the Search for Resources by Sharing Information: Mongolian Gazelles as a Case Study}, Physical Review Letters, {\bf 110}, 248106. 
  \end{itemize}

 \item Chapter 6:
 
  \begin{itemize}
   \item R. Martínez-García, F. Vázquez, C. López, and M.A. Muñoz, (2012) \emph{Temporal disorder in up-down symmetric systems}, Physical Review E, {\bf 85}, 051125.
 \end{itemize}
 
\end{itemize}
\chapter{Methods and tools}\label{methods}

\section{From Individual Based to Population Level Models}
\subsection{The Master equation}\label{sec:masterq}
The master equation provides a complete description of a stochastic dynamics. It encapsulates,
in the evolution of the probability of finding the system in a particular state,
all the processes that occur with given transition rates.
Let us consider an arbitrary system 
with $N$ possible states jumping from to other with exponentially distributed waiting times. In addition,
let us consider that the state of the system at a given time only depends on the previous state, which is called the
Markovian assumption. The probability of finding it in
a particular one, $c$, at a time $t+\Delta t$ is
\begin{equation} \label{master7}
 P_{c}(t+\Delta t)=\left(1-\sum_{c'}\omega_{c \rightarrow c'}\Delta t\right)P_{c}(t)+\sum_{c'}\omega_{c' \rightarrow c}\Delta tP_{c'}(t),
\end{equation}
where $c'$ in the first term denotes the set of states that can be reached from $c$ while
in the second one it refers to the states from which $c$ can be reached.
The first term in Eq.~(\ref{master7}) is the probability of having the system in the state $c$ at time $t$ 
and still remaining there at time $t+\Delta t$ (no transitions occur in the time interval $\Delta t$). The second one
gives the probability of finding the system at any state $c'$ at time $t$ and then jumping to $c$ in a time interval $\Delta t$.

In the limit of infinitely short time steps, $\Delta t \rightarrow dt$, Eq.~(\ref{master7}) becomes 
an evolution equation for the probability of finding the system at each state $c$. This is the master equation:
\begin{equation} \label{master2}
\frac{\partial P_{c}(t)}{\partial t}=\sum_{c'}\omega_{c' \rightarrow c}P_{c'}(t)-\sum_{c'}\omega_{c \rightarrow c'}P_{c}(t).
\end{equation}

Gain and loss terms in Eq.~(\ref{master2}) balance each other, so the probability distribution remains normalized. In addition, the 
coefficients $\omega_{c \rightarrow c'}$ are rates rather than probabilities, so they have units of omverse of time and may be greater 
than one.

Master equations are often hard to solve because they involve a set of several, many times infinite, coupled first order 
ordinary differential equations. The most common techniques to obtain analytical solutions are based on the use of integral transformations such as the 
generating function, the Fourier or the Laplace transform \citep{redner}. However, only in few simple cases the general time dependent solution $P_{c}(t)$ 
can be found, and most of the times numerical simulations of the underlying stochastic dynamics are done \citep{gillespie1977exact}.

\begin{figure}
\begin{center} 
\includegraphics[width=\textwidth]{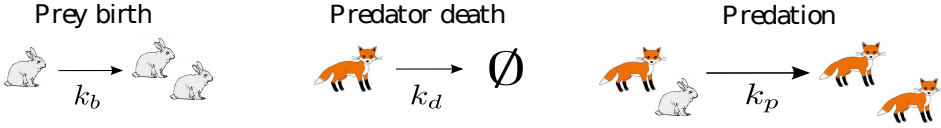}
\caption{Events that may take place in a Lotka-Volterra Individual Based Model with their corresponding rates. Rabitts play 
the role of preys and foxes of predators.}
\label{esquema_ibm}
\end{center}
\end{figure}

To illustrate all the derivarions shown in this chapter, we will use a Lotka-Volterra model 
as a paradigmatic case of a stochastic dynamics that can be modelled at different levels. 
As it is shown in Fig.~\ref{esquema_ibm}, several events can take place with given rates: a birth of a prey
with rate $k_{b}$, a death of a predator with rate $k_{d}$ and a predation and birth of a predator with rate $k_{p}$.

The elementary processes occuring
in the time interval $(t,t+dt)$ that contribute to $P(n,p;t+dt)$ are the following:

 1) The population was $(n,p)$ at time $t$ and nothing happened.
 
 2) The population was $(n-1,p)$ at time $t$ and a rabbit reproduced.
 
 3) The population was $(n,p+1)$ at time $t$ and a fox died.
 
 4) The population was $(n+1,p-1)$ at time $t$ and a fox ate a rabbit and reproduced.

These contributions imply a probability of having $n$ prey and $p$ predators at time $t+dt$ given by
\begin{eqnarray}
 P(n,p,t+dt)&=&P(n,p;t)(1-k_{b}ndt)(1-k_{d}pdt)(1-k_{p}npdt), \ \ \ \ \ \ \ \ \ \ {\rm Event \ 1} \nonumber \\
 &+&P(n-1,p;t)k_{b}(n-1)dt,	\ \ \ \ \ \ \ \ \ \ \ \ \ \ \ \ \ \ \ \ \ \ \ \ \ \ \ \ \ \ \ \ \ \ \ \ \ \ \ \ \  {\rm Event \ 2} \nonumber \\
 &+&P(n,p+1;t)k_{d}(p+1)dt, \ \ \ \ \ \ \ \ \ \ \ \ \ \ \ \ \ \ \ \ \ \ \ \  \ \ \ \ \ \ \ \ \ \ \ \ \ \ \ \ \  {\rm Event \ 3} \nonumber \\
 &+&P(n+1,p-1;t)k_{p}(n+1)(p-1)dt,	\ \ \ \ \ \ \ \ \ \ \ \ \ \ \ \ \ \ \ \ \ \ \  {\rm Event \ 4} \nonumber \\
\end{eqnarray}
that in the limit $dt\rightarrow 0$, and retaining linear terms in $dt$, gives
\begin{eqnarray}\label{masterLV}
 \frac{\partial P(n,p;t)}{\partial t}&=&-(k_{b}n+k_{d}p+k_{p}np)P(n,p;t)+k_{d}(p+1)P(n,p+1;t) \nonumber \\
 &+&k_{b}(n-1)P(n-1,p;t)+k_{p}(n+1)(p-1)P(n+1,p-1;t). \nonumber \\
\end{eqnarray}

The master equation contains all the information about the stochastic dynamics, so it is possible to know
the probability of finding the system in a particular state as a function of time. However, due to the difficulties
that one usually finds to obtain its complete solution, many numerical techniques and analytical approximations
have been developed to deal with it. This is the case of the Gillespie algorithm and the mean-field approximation, that will be explained next.

\subsubsection{The Gillespie algorithm}\label{sec:gillespie}
The Gillespie algorithm \citep{gillespie1977exact} is a Monte Carlo method used to simulate Poissonian \footnote{Exponentially distributed
waiting times between events} stochastic processes where transitions 
from one state to another take place with different rates.
The main objective of the algorithm is to calculate the time until the next transition takes place and the state where the system will move to. 
In principle, one should obtain the time at which every transition occurs, then select the one that happens first, and execute it. 
The advantage of Gillespie method is that it avoids simulating all the transitions and,
instead, only the one that takes place first has to be reproduced.

The algorithm can be explained in four steps:
\begin{enumerate}
 \item Considering that the system is initially in one of the possible $M$ states,$i$, we obtain the total escape rate from it
 \begin{equation}
  \Omega_{i}=\sum_{j\neq i}\omega_{i\rightarrow j}, \ \ \ \ \ \ i=1,\ldots,M
 \end{equation}
 where $j$ is the set of accesible states from $i$ and $\omega_{i\rightarrow j}$ are the individual transition rates from $i$ 
 to each of the states labelled by $j$.
 
 \item The time until the next jump, $dt$, is computed. It is drawn from an exponential distribution of mean $1/\Omega_{i}$. To
 this aim one generates a random number uniformly distributed, $u_{0}$, and computes $dt$ as
 \begin{equation}
  dt=\frac{-{\rm ln} u_{0}}{\Omega_{i}}.
 \end{equation}

 \item The final state has to be determined. Each of the possible transitions takes place with a probability $p_{i\rightarrow j}$ that is 
 proportional to the corresponding rate $\omega_{i\rightarrow j}$,
 \begin{equation}
  p_{i\rightarrow j}=\frac{\omega_{i\rightarrow j}}{\Omega_{i}}
 \end{equation}
 
 \item The time is updated $t \rightarrow t+dt$

\end{enumerate}

\begin{figure}
\begin{center} 
\includegraphics[width=.7\textwidth]{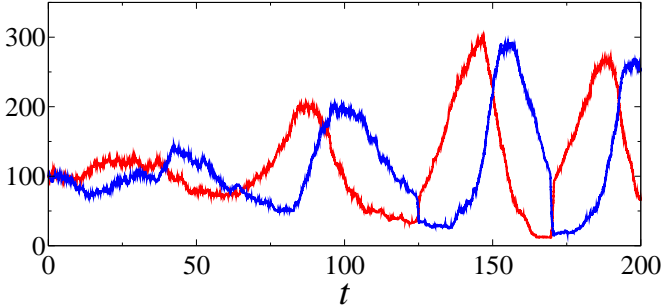}
\caption{Evolution of the population of preys (red line) and predators (blue line) from numerical simulations of the stochastic dynamics in Fig.~\ref{esquema_ibm}
using Gillespie algorithm. Initial condition $100$ preys (rabbits) and $100$ predators (foxes)}
\label{sto_lv}
\end{center}
\end{figure}

When simulated, a Gillespie realization represents
a random walk trajectory for the stochastic variables that exactly represents the distribution of the master equation.
It can be used, for instance, to reproduce the dynamics of the individual based Lotka-Volterra model of the Fig.~\ref{esquema_ibm}, where
birth, death or predation can be interpreted as a transition from a state with $n$ prey and $p$ predators to a new one with 
different population sizes depending on which event has occured. In Fig.~\ref{sto_lv} 
a simulation of the stochastic Lotka-Volterra dynamics using the Gillespie algorithm is shown.

\subsubsection{Mean-field approximation}\label{sec:metmean}
It is the simplest analytical approximation to deal with a master equation.
It allows the derivation of deterministic
differential equations for the mean values of the stochastic variables and establishes the simplest class of population-level models.
Referring to the Lotka-Volterra model as a guiding example, we will derive the equations for the evolution of the mean number of preys, $n$, and predators, $p$.
Given a multivariate probability density function with discrete variables, as it is $P(n,p;t)$, the expected values 
are defined as
\begin{eqnarray}
 \langle n(t)\rangle=\sum_{p,n=0}^{\infty}nP(n,p;t), \ \ \ \ \ \  \langle p(t)\rangle=\sum_{p,n=0}^{\infty}pP(n,p;t).
\end{eqnarray}
Multiplying the master equartion, Eq.~(\ref{masterLV}), by $n$ and $p$ respectively and making the summation over both 
variables, one gets the equations for the temporal evolution of the mean values coupled to the higher moments $ \langle n(t)p(t)>$
\begin{eqnarray}
 \frac{d}{dt} \langle n(t)\rangle=k_{b} \langle n(t)\rangle-k_{p} \langle n(t)p(t)\rangle \nonumber \\
 \frac{d}{dt} \langle p(t)\rangle=k_{p} \langle n(t)p(t)\rangle-k_{d} \langle p(t)\rangle.
\end{eqnarray}
It is possible to obtain the equation for the temporal evolution of $ \langle n(t)p(t)\rangle$, but it would be again coupled to higher 
moments, leading to an infinite system of coupled differential equations. The main assumption of the mean-field approximation
is to consider that both populations are independent, $ \langle n(t)p(t)\rangle= \langle n(t)\rangle\langle p(t)\rangle$, so 
it is possible to write a closed system of deterministic 
differential equations for the mean value of preys and predators
\begin{eqnarray}\label{lvmodel}
 \frac{dN}{dt}&=&N(k_{b}-k_{p}P), \nonumber \\
 \frac{dP}{dt}&=&P(k_{p}N-k_{d}),
\end{eqnarray}
where $N(t)\equiv \langle n(t)\rangle$ and $P(t)\equiv  \langle p(t)\rangle$.

For simplicity, the set of equations (\ref{lvmodel}) can be nondimensionalised by writing \citep{Murray2002}
\begin{equation}
u(\tau)=\frac{k_{p}N}{k_{d}}, \ \ \ \ \ \  v(\tau)=\frac{k_{p}P}{k_{b}}, \ \ \ \ \ \  \tau=k_{b}t, \ \ \ \ \ \  \alpha=\frac{k_{d}}{k_{b}}, 
\end{equation}
and it becomes,
\begin{eqnarray}\label{lvmodel2}
 \frac{du}{dt}=u(1-v), \nonumber \\
 \frac{dv}{dt}=\alpha v(u-1).
\end{eqnarray} 

\begin{figure}
\begin{center} 
\includegraphics[width=.6\textwidth]{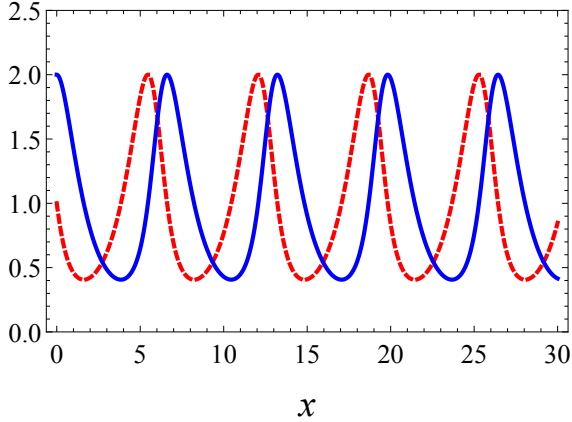}
\caption{Numerical solutions of the nondiemsional Lotka-Volterra equations (\ref{lvmodel2}) with an initial condition $u(0)=1$ and $v(0)=2$. $\alpha=1$.
The red-dashed line corresponds to the evolution of preys and the blue-full line to predators.}
\label{lvfigure}
\end{center}
\end{figure}

The nondimensional system (\ref{lvmodel2}) can be solved analytically, although this is not the general case for nonlinear systems.
Most of the times one has to use linear approximations and other techniques developed in the study of dynamical systems. Additionally,
it is always possible to numerically integrate the equations. This has been done for equations (\ref{lvmodel2}) and the results are shown in 
Fig.~\ref{lvfigure}.

The mean-field equations are a simplified version of the complete stochastic dynamics, but still contain most of the relevant information
of the system. For instance, the oscillations in the populations are preserved for the Lotka-Volterra model.
However, there are many other approximations that, although more complicated, are able to keep the inherent stochasticity of the
system. The Fokker-Planck and the Langevin equations are two of them.

\subsection{The Fokker-Planck equation}\label{sec:fpe}

The master equation describes the dynamics of a physical system as a sequence of jumps from one state to another.
We present in this section an approximation that considers the limit where these jumps are
very short and the evolution of the system can be seen as a diffusive process.  This leads to a simpler
description in terms of the Fokker-Planck equation. The accuracy of this method is better the smaller are the jumps, so the master equation becomes
a Fokker-Planck in the limit of infinitely small jumps.

There are many ways of deriving the Fokker-Planck equation. In this section, we focus on
the Kramers-Moyal expansion \citep{N.vanKampen2007,gardiner}. This is not a completely rigourous derivation from a mathematical point of view,
in fact many alternatives have been used in this thesis, but it is still one of the most 
common and intuitive approaches.

To begin with, we consider a system with several possible states. To ensure the accuracy of the expansion,
we assume that all the jumps between two states are small enough, so that the set of possible states
of the system can be considered as a continuous in the master equation,
\begin{equation}\label{mastercont}
 \frac{\partial P_{c}(t)}{\partial t}=\int\left[\omega_{c' \rightarrow c}P_{c'}(t)-\omega_{c \rightarrow c'}P_{c}(t)\right]dc'.
\end{equation}

Next, we write, in Eq.~(\ref{mastercont}), the transition rates as a function of the size of the jump, $r$, and of the starting point, $c$,
 \begin{equation}
\omega_{c\rightarrow c'}=\omega(c;r),
\end{equation}
with $r=|c'-c|$. Then, the master equation, Eq.~(\ref{mastercont}), becomes
\begin{equation}
  \frac{\partial P_{c}(t)}{\partial t}=\int\omega(c-r;r)P_{c-r}(t)dr-P_{c}(t)\int\omega(c;-r)dr.
\end{equation}

At this point, two assumptions have to be introduced to allow the expansion of the transition rates:
\begin{enumerate}
 \item Only small jumps occur. That is, $\omega(c';r)$ is a sharply peaked function of
$r$ but varies smoothly with $c'$. Mathematically, it means that
\begin{eqnarray}
 &\omega(c';r)\approx0\ \ \ \ &\mbox{for } |r|>\delta, \\
 &\omega(c'+\Delta c;r)\approx\omega(c';r) \ \ \ \ &\mbox{for } |\Delta c|<\delta. \label{varyc}
\end{eqnarray}

 \item The solution, $P_{c}(t)$, varies slowly with $c$ as it is expressed by Eq.~(\ref{varyc}). 
\end{enumerate}

Therefore, one can do a Taylor expansion up to 
second order in Eq.~(\ref{mastercont}) to deal with the shift from $c$ to $c-r$:
\begin{eqnarray}\label{expansion}
 \frac{\partial P_{c}(t)}{\partial t}&=&\int\omega(c;r)P_{c}(t)dr-\int r\frac{\partial}{\partial c}[\omega(c;r)P_{c}(t)]dr+ \nonumber \\
&+&\frac{1}{2}\int r^{2}\frac{\partial^{2}}{\partial c^{2}}[\omega(c;r)P_{c}(t)]dr-\int\omega(c;-r)P_{c}(t)dr.
\end{eqnarray}

The first and fourth term in the right-hand side of Eq.~(\ref{expansion}) cancel each other, and defining the jump moments
\begin{equation}
 \alpha_{\nu}(c)=\int_{-\infty}^{+\infty}r^{\nu}\omega(c;r)dr,
\end{equation}
the final result can be written as
\begin{equation}\label{fp}
 \frac{\partial P_{c}(t)}{\partial t}=-\frac{\partial}{\partial c}[\alpha_{1}(c)P_{c}(t)]+\frac{1}{2}\frac{\partial^{2}}{\partial c^{2}}[\alpha_{2}(c)P_{c}(t)].
\end{equation}

This is the Fokker-Planck equation. It is important to remark that we have not shown a completely rigurous derivation.
The election of the small parameter to perform the Taylor expansion has not been justified and
there are many processes in which this expansion fails. This is the case of systems with jump size $\pm 1$ or
some small integer, whereas typical sizes of the variable may be large, e.g., the number of molecules in a chemical reaction or 
the position of a random walker on a long lattice. In those cases expansions where the small parameter is explicitly taken are much
more appropiate (See Chapter \ref{chap:tempdis} for a rigorous derivation of the Fokker-Planck equation).
Nevertheless, this description provides a good first contact with the Fokker-Planck equation,
that allows the development of a large variety of population level spatial models. 

On the other hand, many ecological systems, such as groups of animals and vegetation landscapes that will be studied in this 
thesis, are formed by many particles. Let us now suppose that we have a suspension of a very large number of identical
individuals, and denote its local density by $\rho({\bf x}, t)$. If the suspension
is sufficiently diluted, to the extent that particles can be considered
independent, then $\rho({\bf x}, t)$ will obey the same Eq.~(\ref{fp}) \citep{peliti2011statistical}.
This family of models based on the density of individuals is the basis of the
studies on vegetation patterns shown in the Part \ref{part:pattern} of this thesis.

In either case, and independently of the way used to write it, the Fokker-Planck equation describes a large class of stochastic
dynamics in which the system has a continuous sample path. The state of the system
can be written as a stochastic and continuous function of time. From this picture, it seems obvious to seek a
description in some direct probabilistic way and in terms of stochastic differential
equations for the path of the system. This procedure is discused next.

\subsection{The Langevin equation}

In some cases it is useful to describe a system
in terms of a differential equation, that gives the stochastic evolution of its state as a trajectory in the phase space. This is the Langevin equation, that 
has the general form
\begin{equation}\label{langevin7}
 \frac{dc}{dt}=f(c,t)+g(c,t)\eta(t),
\end{equation}
where $c$ is a stochastic variable that gives the state of the system at every time. $f(c,t)$ and $g(c,t)$ are known functions
and $\eta(t)$ is a rapidly fluctuating term whose average over single realizations is equal to zero, $\langle\eta(t)\rangle=0$.
Any nonzero mean can be absorbed into the definition of $f(c,t)$. An idealization of a term like $\eta(t)$ must be that in which if
$t \neq t'$, $\eta(t)$ and $\eta(t')$ are statistically independent (white noise), so
\begin{equation}
 \langle\eta(t)\eta(t')\rangle=\Gamma \delta(t-t'),
\end{equation}
where $\Gamma$ gives the strength of the random function.

To be rigorous, the differential equation (\ref{langevin7}) is not properly defined, although the corresponding integral equation,
\begin{equation} \label{inc}
 c(t)-c(0)=\int_{0}^{t}f[c(s),s]ds+\int_{0}^{t}g[c(s),s]\eta(s)ds,
\end{equation}
can be consistently defined understanding the integral of the white noise as a Wiener process $W(t)$ \citep{N.vanKampen2007,gardiner}:
\begin{equation}
 dW(t)\equiv W(t+dt)-W(t)=\eta(t)dt.
\end{equation}
Hence
\begin{equation}\label{wiener}
 c(t)-c(0)=\int_{0}^{t}f[c(s),s]ds+\int_{0}^{t}g[c(s),s]dW(s),
\end{equation}
where the second integral can be seen  like a kind of Riemann integral with respect to a sample 
function $W(t)$. 

The definition of the Langevin equation (\ref{langevin7}), requires a careful interpretation due to this lack of mathematical rigor.
When the noise term appears multiplicatively, that is, $g(c,t)$ is not a constant, 
ambiguities appear in some mathematical expressions. Giving a sense to the undefined expressions constitutes
one of the main goals when integrating a Langevin equation. The most widely used interpretations are those of It\^{o} 
and Stratonovich (Appendix \ref{apitost}). The It\^{o} integral
is preferred by mathematicians \citep{N.vanKampen2007}, but it is not always the most natural
choice from a physical point of view. The Stratonovich integral is more suitable, for instance, when $\eta(t)$ is a real noise with finite 
correlation time where the vanishing correlation time limit wants to be taken. (In the Appendix \ref{apitost} we show a more detailed discussion).
The matter is not what is the right definition of the stochastic integral, but how stochastic processes can model real systems. That is,
in what situations either It\^{o} or Stratonovich choice is the most suitable.

Langevin equations are also valid to go beyond a mean-field description. In these cases a new term enters in the equation
to include diffusion, besides other spatial couplings and degrees of freedom. The variable $c(t)$ becomes 
a continuous field $\phi({\bf r},t)$ that depends on space and time. The Langevin equation becomes a stochastic partial differential equation
of the type
\begin{equation}
 \frac{\partial \phi({\bf r},t)}{\partial t}=f(\phi({\bf r},t),t)+\nabla^{2}\phi({\bf r},t)+g(\phi({\bf r},t),t)\eta({\bf r},t).
\end{equation}
This approach is quite useful for spatially extended systems or to study the formation of patterns.

\subsubsection{From the Fokker-Planck to Langevin equation and vice versa.}

To close this overview on the modeling of stochastic systems, we will show the relationship between
Fokker-Planck and Langevin equations. Starting from a Fokker-Planck equation for the probability distribution of the 
variable $c$
\begin{equation}\label{fpe12}
   \frac{\partial P(c,t)}{\partial t}=-\frac{\partial}{\partial c}\alpha_{1}(c)P(c,t)+\frac{1}{2}\frac{\partial^{2}}{\partial c^{2}}\alpha_{2}(c)P(c,t),
\end{equation}
it is easy to write down a Langevin equation of the type (\ref{langevin7}) \citep{gardiner,N.vanKampen2007}
\begin{equation}
 \frac{dc}{dt}=f(c,t)+g(c,t)\eta(t), 
\end{equation}
where $\eta(t)$ is a white, Gaussian and zero mean noise.

The coefficients of the equations are related according to
\begin{eqnarray}
f(c,t)&=&\alpha_{1}(c,t), \\
g(c,t)&=&\sqrt{\alpha_{2}(c,t)}.
\end{eqnarray}
provided that the It\^{o} interpretation is chosen.

The first term in Eq.~(\ref{fpe12}) is called {\it drift}, because it leads to the deterministic part of the Langevin
equation, and the second one, the {\it diffusion term}, since it determines the stochastic part of the Langevin equation.

In the Stratonovich scheme an additional drift appears,
\begin{equation}
 \frac{dc}{dt}=f(c,t)+\frac{1}{2}g(c,t)\frac{\partial g(c,t)}{\partial c}+g(c,t)\eta(t).
\end{equation}

On the other hand, if the starting point is a Langevin equation
\begin{equation}
 \frac{dc}{dt}=f(c,t)+g(c,t)\eta(t),
\end{equation}
to obtain the Fokker-Planck equation one has to specify if the It\^{o} or the Stratonovich calculus will be used.
In the Stratonovich interpretation the Fokker-Planck is
\begin{equation}
    \frac{\partial P(c,t)}{\partial t}=-\frac{\partial}{\partial c}f(c)P(c,t)+\frac{1}{2}\frac{\partial}{\partial c}g(c)\frac{\partial}{\partial c}g(c)P(c,t),
\end{equation}
while in the It\^{o} case it is
\begin{equation}
    \frac{\partial P(c,t)}{\partial t}=-\frac{\partial}{\partial c}f(c)P(c,t)+\frac{1}{2}\frac{\partial^{2}}{\partial c^{2}}[g(c)]^{2}P(c,t).
\end{equation}

The diffusion term vanishes typically with the number of components as $N^{-1/2}$, so it is negligible if the system is large enough.
Therefore, in the thermodynamic limit where $N$ and the volume $V$ tend to infinity
keeping $N/V$ finite, a deterministic mean-field approximation gives an accurate description.
Sometimes, this way is walked on the inverse sense. One may start with a deterministic equation and, using
heuristic arguments, add noise to obtain the Langevin equation. Then, following the steps that have been explained in this
section it is possible to get a Fokker-Planck equation.


\section{Linear stability analysis}

Linear stability analysis is the simplest analytical tool used to study the formation of patterns
in deterministic spatially extended systems. It assumes an ideal infinite system and uses Fourier analysis to 
investigate the stability of its homogeneous state. We will consider in this section the two dimensional case.
The starting point is the equation for the evolution of a field $\phi$ 
\begin{equation}\label{modeleqext}
 \frac{\partial \phi(x,y,t)}{\partial t}=f\left(\phi(x,y,t),\frac{\partial\phi}{\partial x},\frac{\partial\phi}{\partial y}
 ,\frac{\partial^{2}\phi}{\partial x^{2}},\frac{\partial^{2}\phi}{\partial y^{2}},\frac{\partial^{2}\phi}{\partial x\partial y};R\right)
\end{equation}
where $R$ is the control parameter. The linear stability analysis assumes that the system is 
at the homogeneous (spatially independent) stationary state $\phi({\bf x},t)=\phi_{0}$ and studies its stability against small perturbations
that will be denoted by $\psi({\bf x},t)$, with $|\psi|\ll1$. The technique is applied in the Appendix
\ref{deriv} to one particular case and the calculations explained in detail. In this section we will 
introduce and discuss the theoretical basis and the main results that can be obtained. 
Plugging the ansatz $\phi({\bf x},t)=\phi_{0}+\psi({\bf x},t)$ into the model Eq.~(\ref{modeleqext}) and retaining only 
linear terms in the perturbation, one obtains a linear equation for the evolution of the perturbation at short times that
can be solved using the Fourier transform. Then, the final task is to solve the transformed equation for the perturbation,
$\hat{\psi}({\bf k},t)$. Assuming that at short time scales the temporal dependence is $\hat{\psi}({\bf k},t)\propto\exp(\lambda({\bf k})t)$,
 where $\lambda$ is the growth rate,
then $\hat{\psi}({\bf k},t)=\lambda({\bf k})\hat{\psi}({\bf k},t)$. Finally an expression for $\lambda({\bf k})$ can be obtained.
It is called the dispersion relation and contains all the information about the evolution of the Fourier modes of 
$\hat{\psi}({\bf k},t)$. The modes ${\bf k}$ with a negative growth rate will be stable while those corresponding
to $\lambda\geq 0$ are unstable and lead to perturbations growing in time and, therefore, to spatial patterns in the system.
The dispersion relation also allows to obtain the characteristic wavelength of the pattern through the value of the most
unstable Fourier mode, ${\bf k}_{c}$, that most of the times corresponds with the one with the highest growth rate.

\begin{figure}
\begin{center} 
\includegraphics[width=\textwidth]{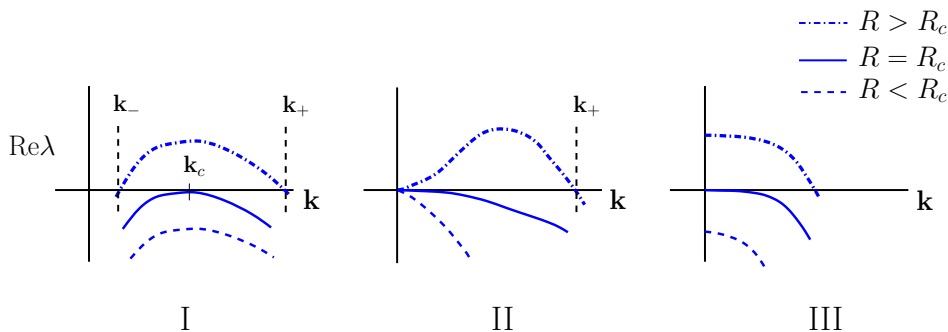}
\caption{Different types of linear instabilities depicted in the real part of the dispersion relation.}
\label{inst}
\end{center}
\end{figure}

Depending on the functional form of the dispersion relation, it is possible to establish a classification of the different types of linear
instabilities appearing in natural systems \citep{Halperin1977}. These classes are shown in Fig.~\ref{inst}, where the real part of $\lambda$
is sketched as a function of the wave number, ${\bf k}$. They are:
\begin{itemize}
 \item Type I. For $R<R_{c}$ the homogeneous state is stable and ${\rm Re}\lambda<0$, whereas for $R=R_{c}$ the instability
 sets in (${\rm Re}\lambda=0$) at a wave vector $k_{c}$. For $R>R_{c}$
 there is a band of wave vectors $k_{-}<k<k_{+}$ for which the uniform state is unstable. The patterns observed in these system will be dominated by a
 wavelength given by one of this unstable modes, typically by that with the highest growth rate, $k_{c}$. This case is represented in the left panel
 of Fig.~\ref{inst}.
 
 \item Type II. This is a different type of instability appearing when, for some reason (usually a conservation law), ${\rm Re}\lambda(k=0)=0$ independently
 of the value of the control parameter $R$. This corresponds with the central panel of Fig.~\ref{inst}. The critical wave vector, 
 the one that becomes unstable by the first time, is now $k_{c}=0$, and a band of unstable modes appears between $0$ and $k_{+}$ for $R>R_{c}$.
 The pattern occurs on a long length scale. This case is remarkable because the critical wave vector is different from that
 with the highest growth rate.
 
 \item Type III. In this case both the instability and the maximum growth rate occur at $k_{c}=0$. There is not an intrinsic length scale, 
 and patterns will occur over a length scale defined by the system size or the dynamics. This situation is depicted in the right panel of 
 Fig.~\ref{inst}.
\end{itemize}

Finally, there are two subtypes for each type of instability depending on the temporal instability:
stationary if ${\rm Im}\lambda=0$, and oscillatory if ${\rm Im}\lambda\neq0$. 

Linear stability analysis provides analytical results about the formation of patterns in spatially extended systems, such as the dominant wavelength and
the type of instability leading the structure. However, it is important to remark that the analysis assumes that the perturbations of
the uniform state are small. This assumption is good at short times and for an initial condition that has a small magnitude, but at long times 
the nonlinear terms left out in the linear approximation become important \citep{crosslibro}. One effect of nonlinearity is to quench the assumed exponential growth.
Further analysis, such as weakly nonlinear stability analysis \citep{cross}, must be used in these cases.

\section{First-passage times processes}

First-passage phenomena are of high relevance in stochastic processes that are triggered by a first-passage event \citep{redner}
and play a fundamental role quantifying and limiting
the success of different processes that can be mapped into random walks.
Ecology and biology offer some examples such as the lifetime of a population or the duration of a search
or a biochemical reaction.

In this section we will present some results on first-passage times in the simple case of a discrete symmetric random walk moving in 
a finite interval $[x_{-},x_{+}]$ \citep{redner}. The extension to higher dimensions is straightforward. Let us denote the mean time to exit the interval
starting at $x$ by $T(x)$. This quantity is equal to the exit time of a given trajectory times the probability of that path, averaged over all 
the trajectories,
\begin{equation}
 T(x)=\sum_{p}\mathcal{P}_{p}t_{p}(x),
\end{equation}
where $t_{p}$ is the exit time of the trajectory $p$ that starts at $x$ and $\mathcal{P}_{p}$ the probability of the path.
Because of the definition of a symmetric random walk on a discrete space with jump length $\delta x$ and time step $\delta t$,
the mean exit time also obeys 
\begin{equation}\label{recurr}
 T(x)=\frac{1}{2}\left\lbrace\left[T(x+\delta x)+\delta t\right]+\left[T(x-\delta x)+\delta t\right]\right\rbrace,
\end{equation}
with boundary conditions $T(x_{-})=T(x_{+})=0$ which correspond to a mean exit time equal to zero if the particle
starts at either border of the interval. $\delta x$ is the jumping length. This recursion relation expresses the mean exit time
starting at $x$ in terms of the outcome one step in the future, for which the initial walk can be seen as restarting in
$x\pm\delta x$ (each with probability $1/2$) but also with the time incremented by $\delta t$.

Doing a Taylor expansion to the lowest nonvanishing order in Eq.~(\ref{recurr}), and considering the limit of 
continuous time and space, it yields
\begin{equation}\label{time1d}
 D\frac{d^{2}T}{dx^{2}}=-1,
\end{equation}
where $D=\delta x^{2}/2\delta t$ is the difussion constant. In the case of a two dimensional domain Eq.~(\ref{time1d}) is 
\begin{equation}\label{timesfin}
 D\nabla^{2}T({\bf x})=-1.
\end{equation}

These results can be extended to the case of general jumping processes with a single-step jumping probability given by $p_{\mathbf{x}\rightarrow\mathbf{x}'}$.
The equivalent of Eq.~(\ref{recurr}) is
\begin{equation}
 T(\mathbf{x})=\sum_{\mathbf{x}'}p_{\mathbf{x}\rightarrow{\mathbf{x}'}}[T(\mathbf{x}')+\delta t],
\end{equation}
that provides an analog of Eq.~(\ref{timesfin}) that is
\begin{equation}
 D\nabla^{2}T({\bf x})+{\bf v}({\bf x})\cdot\nabla T({\bf x})=-1,
\end{equation}
where ${\bf v}({\bf x})$ is a local velocity that gives the mean displacement after a single step when starting from ${\bf x}$ in the hopping process.
This equation can be solved in each particular case. We have used it in this thesis as an starting point of many of the calculations in the Part \ref{part:tempfluc}. 
See Appendix~\ref{App-A} for a detailed calculation.


\part[\textit{\textsc{Vegetation Patterns}}]{\textbf{\textit{\textsc{Vegetation Patterns \\}}}\label{part:pattern}
         }   

   \label{part:patterns}
   \chapter{Mesic savannas} \label{chap:mesic}

In this chapter we propose a continuum description for the dynamics of tree density in mesic savannas \cite{MartinezGarciaJTB}
inspired on the individual based model introduced in \cite{calabrese}.
It considers only long-range competition among trees and the effect of fires resulting in a local facilitation mechanism.
Despite short-range facilitation is taken to the local-range limit, 
the standard full spectrum of spatial structures 
obtained in general vegetation models is recovered.
Long-range competition is thus the key ingredient for the development of patterns.
This result opens new questions on the role that facilitative interactions
play in the maintenance of vegetation patterns.
The long time coexistence between trees and grass, the effect of fires on 
the survival of trees as well as the maintenance of the patterns are also studied. 
The influence of demographic noise is analyzed. The stochastic system,
under parameter constraints typical of more humid landscapes, shows irregular
patterns characteristic of realistic situations.  
The coexistence of trees and grass still remains at reasonable noise
intensities.

\section{Introduction}

Savanna ecosystems are characterized by the long-term coexistence between a continuous grass
layer and scattered or clustered trees \citep{sarmiento1984}. Occurring in many regions of the
world, in areas with very different climatic and ecological conditions, the spatial structure,
persistence, and resilience of savannas have long intrigued ecologists \citep{scholes, sankaran2005, BorgognoRG,belsky1994influences}.
However, despite substantial research, the origin and nature of savannas have not yet been fully resolved and much remains to be learned. 

Savanna tree populations often exhibit pronounced, non-random spatial
structures \citep{skarpe1991spatial, barot1999demography, jeltsch1999detecting, caylor2003tree, scanlon2007positive}.
Much research has therefore focused on explaining how spatial patterning in savannas arises
\citep{jeltsch,jeltsch1999detecting,scanlon2007positive,skarpe1991spatial,calabrese,Vazquez-2010}. In most natural plant systems
both facilitative and competitive processes are simultaneously
present \citep{scholes,vetaas} and hard to disentangle \citep{Veblen,Barbier}. Some studies have pointed toward the existence of 
short-distance facilitation \citep{caylor2003tree, scanlon2007positive}, while others have demonstrated evidence of competition
\citep{skarpe1991spatial, jeltsch1999detecting, barot1999demography}, with conflicting reports sometimes arriving from the same regions.

Different classes of savannas, which can be characterized by how much rainfall they typically receive, should be affected 
by different sets of processes. For example, in semiarid savannas water is extremely limited (low mean annual precipitation)
and competition among trees is expected to be strong, but fire plays little role because there is typically not enough grass
biomass to serve as fuel. In contrast, humid savannas should be characterized by weaker competition among trees, but also by
frequent and intense fires. In-between these extremes, in mesic savannas, trees likely have to contend with intermediate levels
of both competition for water and fire \citep{calabrese,sankaran2005, sankaran2008, bond2003, bond2008, bucini}. 

Competition among trees is mediated by roots that typically extend well beyond the crown \citep{BorgognoRG,Barbier}. Additionally, fire
can lead to local facilitation due to a protection effect, whereby vulnerable juvenile trees placed near adults are
protected from fire by them \citep{holdo}. We are particularly interested in how the interplay between these mechanisms
governs the spatial arrangement of trees in mesic savannas, where both mechanisms may operate. 
On the other side, it has frequently been claimed
that pattern formation in arid systems can be explained by a combination of long-distance competition and short-distance facilitation 
\citep{klausmeier, LefeverLejeune, lefeverJTB2009, Lefever2012,TheAmNat2002,Hardenberg, D'Odoricob}. This combination
of mechanisms is also known to produce spatial structures in many other natural systems \citep{cross}. Although mesic
savannas do not display the same range of highly regular spatial patterns that arise in arid systems (e.g., tigerbush),
similar mechanisms might be at work. Specifically, the interaction between long-range competition and short-range facilitation
might still play a role in pattern formation in savanna tree populations, 
but only for a limited range of parameter values and possibly modified by demographic stochasticity.

Although the facilitation component has often been thought to be a key component in previous vegetation
models \citep{D'Odoricob, D'Odoricoa, TheAmNat2002, scanlon2007positive}, Rietkerk and Van de Koppel \citep{RietkerkTrends}, 
speculated, but did not show, that pattern formation could occur without short-range facilitation in the particular example of tidal freshwater marsh.
In the case of savannas, as stated before, the presence of adult trees favor the establishment of new trees in the area, protecting the juveniles against fires. Considering this effect, 
we take the facilitation component to its infinitesimally short spatial limit, and study its effect in the emergence of spatially periodic structures of trees.
To our knowledge, this explanation, and the interrelation between long-range competition and local facilitation,
has not been explored for a vegetation system. 

To this aim, we develop a minimalistic model of savannas that considers two of
the factors, as already mentioned,  
thought to be crucial to structure
mesic savannas: tree-tree competition and fire, with a primary focus on spatially nonlocal competition. 
Employing standard tools used in the study of pattern formation phenomena
in physics (stability analysis and the structure function) \citep{cross}, we explore
the conditions under which the model can produce non-homogeneous spatial distributions.
A key strength of our approach is that we are able to provide a complete and rigorous analysis of 
the patterns the model is capable of producing, and we identify which among these correspond to situations that are relevant for mesic savannas.
We further examine
the role of demographic stochasticity in modifying both spatial patterns and the conditions under which trees persist in the system
in the presence of fire, and discuss the implications of these results for the debate on whether the balance of processes affecting savanna
trees is positive, negative, or is variable among systems. 
This is the framework of our study: the role of long-range competition, 
local facilitation and demographic fluctuations in the spatial structures of mesic savannas.

\section{The deterministic description} \label{deterministic}

In this section we derive the deterministic equation for the local density of trees,
such that dynamics is of the logistic type and we only consider
tree-tree competition and fire. We study the formation of patterns via stability analysis and provide numerical simulations,
showing the emergence of spatial structures.

\subsection{The nonlocal savanna model}\label{savannamodel}

\cite{calabrese} introduced a simple discrete-particle lattice savanna model that considers the birth-death
dynamics of trees, and where tree-tree competition and fire
are the principal ingredients. These mechanisms act on the probability of
establishment of a tree once a seed lands at a particular point on the lattice. In the discrete model, 
seeds land in the neighborhood of a parent tree with a rate $b$, and establish as adult trees if they are able to survive both competition neighboring trees and fire.
 As these two phenomena are independent, the probability of establishment is $P_{E}=P_{C}P_{F}$, where
$P_{C}$ is the probability of surviving the competition, and $P_{F}$ is the probability of surviving a fire event.
From this dynamics, we write a 
deterministic differential equation describing
the time evolution of the global density of trees (mean field), $\rho (t)$, where the population
has logistic growth at rate $b$, and an exponential death term at rate $\alpha$.
It reads:
\begin{equation}
\frac{d\rho}{dt}=bP_{E}(\rho) \rho(t)\left( 1-\rho(t)\right) -\alpha\rho(t).
\label{eq:mf}
\end{equation}

Generalizing Eq.~(\ref{eq:mf}), we propose an evolution equation for the space-dependent (local) density of trees,
$\rho (\bx, t)$:
\begin{equation}
\frac{\partial \rho(\bx, t)}{\partial t}=bP_{E}\rho(\bx, t)(1-\rho(\bx, t))-\alpha\rho(\bx, t).
\label{eq:PDE}
\end{equation}
We allow the probability of overcoming competition to 
depend on tree crowding in a local neighborhood, decaying exponentially with the
density of surrounding trees as
\begin{equation}
 P_{C}=\exp\left(-\delta\int G(\bx-\br)\rho(\br,t)d\br\right),
\label{probc}
\end{equation}
where $\delta$ is a parameter that modulates the strength of the competition,
and $G(\bx)$ is a positive kernel function that introduces a finite range of influence.
This model is related to earlier one of pattern formation in arid systems \citep{LefeverLejeune}, and subsequent works \citep{lefeverJTB2009, Lefever2012},
but it differs from standard kernel-based models in that the kernel function accounts for the interaction neighborhood, and not
for the type of interaction with the distance. Note also that the nonlocal term enters nonlinearly in the equation.

Following \cite{calabrese}, $P_F$ is assumed to be a saturating function of grass biomass, $1-\rho(\bx,t)$, similar to the
implementation of fire of Jeltsch {\it et al.} in \citep{jeltsch}
\begin{equation}
 P_{F}=\frac{\sigma}{\sigma+1-\rho(\bx,t)},
\label{probfire}
\end{equation}
where $\sigma$ governs the resistance to fire, so $\sigma=0$ means no resistance to fires. 
Notice how our model is close to the one in \citep{calabrese} through the definitions of $P_{C}$ and $P_{F}$, although
we consider the probability of surviving a fire depending on the local density of trees, and in \citep{calabrese} it
depends on the global density.
The final deterministic differential equation that considers tree-tree competition
and fire for the spatial tree density is
\begin{equation}\label{sav1}
 \frac{\partial \rho(\bx, t)}{\partial t}= b_{eff}(\rho)
\rho(\bx, t)\left(1 -\rho(\bx, t)\right)-\alpha\rho(\bx, t),
\end{equation}
where
\begin{equation}
 b_{eff}(\rho)=\frac{b{\rm e}^{-\delta\int G(\bx-\br)\rho(\br,t)d\br}\sigma}{\sigma+1-\rho(\bx,t)}.
\end{equation}

Thus, we have a logistic-type equation with an effective growth rate that depends
nonlocally on the density itself, and which is a combination of long-range competition and local facilitation mechanisms (fire).
The probability of surviving a fire is higher when the local density of trees is higher, as can be seen from the definition
in Eq.~(\ref{probfire}).

\begin{figure}
\centering
\includegraphics[width=0.6\textwidth]{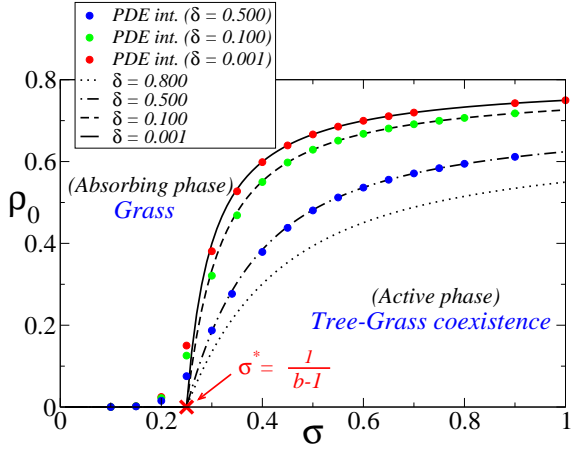}
\caption{Grass-coexistence phase transition. Stationary tree density, $\rho_{0}$, as a function
of the resistance to fires parameter, $\sigma$. The lines come from the mean field solution, Eq.~(\ref{steq}), and the dots
from the numerical integration of Eq.~(\ref{sav1}) over a square region of $1~ha$. We have chosen $\alpha=1$, and $b=5$.
In the case of the spatial model, $\rho_{0}$ involves an average of the density of trees over
the studied patch of savanna.}
\label{transition}
\end{figure}

In Fig.~\ref{transition} we show  numerical solutions for the mean field
Eq.~(\ref{eq:mf}) (lines) and the spatially explicit model (equation \ref{sav1}) (dots) in the stationary state
$(t\rightarrow\infty)$ using different values of the competition. We have used a top-hat function
as the competition kernel, $G(\bx)$ (See Sec.~\ref{LSA} for more details on the kernel choice).
We observe a very good agreement of both descriptions which becomes worse when we get closer to the critical point $\sigma^{*}$,
where the model presents a phase transition from a tree-grass coexistence to a grassland state.
This disagreement appears because while the mean field equation describes an infinite system, 
the Eq.~(\ref{sav1}) description forces us to choose a size for the system. 

The model reproduces the long-term coexistence between grass and trees that is characteristic of savannas.
To explore this coexistence, we study the long-time behavior of the system and 
analyze the homogeneous stationary solutions of Eq.~(\ref{sav1}), which has two fixed points.
The first one is the absorbing state representing the absence of trees,
$\rho_{0}=0$, and the other can be obtained, in the general case, by numerically solving  
\begin{equation}\label{steq}
 b_{eff}(\rho_{0})(1-\rho_{0})-\alpha=0.
\end{equation}
In the regime where $\rho_{0}$
is small (near the critical point), if competition intensity, $\delta$, is also small,
it is possible to obtain an analytical expression
for the critical value of the probability of surviving a fire, $\sigma^{*}$, 
\begin{equation}
\sigma^{*}=\frac{\alpha}{b-\alpha}.
\end{equation}
Outside of the limit where $\delta\ll 1$, we can solve Eq.~(\ref{steq}) numerically 
in $\rho_{0}$ to show that the critical value of the fire resistance parameter, $\sigma^{*}$, does not depend on 
competition. A steady state with trees is stable for higher fire survival probability 
(Fig.~\ref{transition}).

There is, then, a transition from a state where grass is the only form of vegetation to
another state where trees and grass coexist at $\sigma^{*}$. 
In what follows, we fix $\alpha=1$, so we choose our temporal scale
in such a way that time is measured in units of $\alpha$. This choice does not qualitatively affect our results.

\subsection{Linear stability analysis}
\label{LSA}

The spatial patterns can be studied by 
performing a linear stability analysis \citep{cross} of the stationary
homogeneous solutions of Eq.~(\ref{sav1}),
$\rho_{0}=\rho_0 (\sigma,\delta)$. 
The stability analysis is performed by considering
small harmonic perturbations around $\rho_0$,
 $\rho(\bx,t)=\rho_{0}+\epsilon{\rm e}^{\lambda t-i\bk\cdot \bx}$, $\epsilon\ll 1$.
After some calculations\footnote{A linear stability analysis in a similar equation modeling vegetation in arid systems is shown in detail in Appendix \ref{deriv}.},
one arrives at a perturbation growth rate given by
\begin{equation}\label{reldisper}
\lambda(k; \sigma, \delta)=b_{eff}\left[(\rho_{0})\frac{1+\sigma(1-2\rho_{0})}{\sigma-\rho_{0}+1}-
\frac{\rho_{0}\left[2-\rho_{0}+\delta {\hat G}(k)(\rho_{0}-1)(\rho_{0}-1-\sigma)\right]}{(\sigma-\rho_{0}+1)}\right]-1,
\end{equation}
where ${\hat G}(k)$, $k =|\bk |$, is the Fourier transform of the kernel,
\begin{equation}\label{foudef}
 {\hat G}(\bk)=\int G(\bx){\rm e}^{-i\bk\cdot \bx}d\bx.
\end{equation}

The critical values of the parameters of the transition to pattern,
$\delta_c$ and $\sigma_c$, and the fastest
growing wavenumber $k_c$, 
are obtained from the simultaneous solution of
\begin{eqnarray}
\lambda (k_c; \sigma_c, \delta_c )&=&0, \label{eq:first}\\
\left(\frac{\partial \lambda}{\partial k}\right)_{k_c; \sigma_c, \delta_c }&=&0. \label{eq:sec}
\end{eqnarray}
Note that $k_c$ represents the most unstable mode of the system, which means that it grows faster
than the others and eventually dominates the state of the system. Therefore,
it determines the length scale of the spatial pattern.
These two equations yield the values of the parameters $\delta$ and $\sigma$ at 
which the maximum of the curve $\lambda (k)$, right at $k_c$, starts becoming positive. This signals
the formation of patterns in the solutions of Eq.~(\ref{sav1}). As
Eq.~(\ref{eq:sec}) is explicitly written as
\begin{equation}\label{derlam}
 \lambda'(k_{c})=b_{eff}(\rho_{0})\delta\rho_{0}{\hat G}'(k_{c})(\rho_{0}-1),
\end{equation}
the most unstable wavenumber $k_c$ can be obtained by 
evaluating the zeros of the derivative of the Fourier transform of the kernel.

Eq.~(\ref{reldisper}) shows that competition, through the kernel function,
fully determines the formation of patterns in the system. The local facilitation 
appears in  $b_{eff}(\rho_{0})$ and it is not relevant in the formation of spatial structures.
If the Fourier transform
of $G$ never takes positive values, then $\lambda (k; \sigma, \delta)$ is always
negative and only the homogeneous solution is stable. However, when $\hat G$ can take
negative solutions then patterns may appear in the system. What does this mean in biological
terms? Imagine that we have a family of kernels described by a parameter $p$: 
$G(\bx )= \exp(-|(\bx )/R |^p)$ ($R$ gives the range of competition).
The kernels are more peaked around $\bx =0$ for $p<2$ and more box-like when $p>2$. It turns out that this family
of functions has non-negative Fourier transform for $0 \leq p <2$, so 
that no patterns appear in this case. A lengthy discussion of this property in the context
of competition of species can be found in \cite{pigo}.
Thus, the shape of the competition kernel dictates whether or not patterns will appear in the system. If pattern formation is possible, then the values of the fire and competition parameters
govern the type of solution (see Sec.~\ref{sec:NS}).

Our central result for nonlocal competition is that, contrary to conventional wisdom, it
can, in the limit of infinitesimally short (purely local) facilitation, promote the clustering of trees. 
Whether or not this occurs depends entirely on the shape 
of the competition kernel. For large $p$ we have a box-like shape, and in these
cases trees compete strongly with other trees, roughly within a distance 
$R$ from their position. The mechanism behind this counterintuitive result is that trees farther than $R$ away from a resident tree area are not able 
to {\it invade} the zone defined by the radius R around the established tree (their seeds do not establish there), so that
an exclusion zone develops around it. For smaller $p$ there is less competition
and the exclusion zones disappear. We will develop longer this concept in the next chapter.

For a more detailed analysis, one must choose an explicit form for the kernel function.
Our choice is determined by the original $P_C$ taken in \citep{calabrese}, so
that it decays exponentially with the number of trees in a neighborhood of radius $R$ around a
given tree. Thus, for $G$ we take the step function (limit $p\rightarrow\infty$)
\begin{eqnarray}\label{kerneldd}
G(|\br|)= \left\{ \begin{array}{lcc}
             1 &   if  & |\br| \leq R \\
             \\ 0 &  if & |\br| > R. \\
             \end{array}
   \right.
\end{eqnarray}
As noticed before, the idea behind the nonlocal competition is to capture
the effect of the long roots of a tree. The kernel function defines the area of influence
of the roots, and it can be modeled at first order with the constant function of Eq.~(\ref{kerneldd}).
Thus the parameter R, which fixes the {\it nonlocal} interaction scale, must be of the order of the length of the roots \citep{BorgognoRG}.
Since the roots are the responsible for the adsorption of
resources (water and soil nutrients), a strong long-range competition term
implies strong resource depletion.
For this kernel the Fourier transform is \citep{lopezhdez}
${\hat G}(k)=2\pi R^{2} J_{1}(kR)/kR$ 
 and its derivative is
${\hat G}'(k)=-2\pi R^{2} J_{2}(kR)/k$, 
where $ k\equiv |\bk|$, and $J_{i}$ is the $i^{th}$-order Bessel function.
Since ${\hat G}(k)$ can take positive and negative values, pattern solutions
may arise in the system, that will in turn depend on the values of
$\delta$ and $\sigma$.
The most unstable mode is numerically obtained as the first zero of $\lambda'(k)$, Eq.~(\ref{derlam}), which means the first zero
of the Bessel function $J_{2}(kR)$. This value only depends on $R$, being independent of
the resistance to fires and competition, and it is $k_c=5.136/R$.
Because a pattern of $n$ cells is characterized by a wavenumber 
$k_c=2\pi n /L$, where $L$ is the system size,
the typical distance between clusters,
$d_t=L/n$, using the definition of the critical wavenumber is given by $d_t \approx 1.22 R$. In other words, it is approximately
the range of interaction $R$. This result is also independent
of the other parameters of the system.  

Since we are interested in the effect of competition and fire on the distribution of savanna trees,
we will try to fix all the parameters but $\sigma$ and $\delta$. We will explore the effect of different values of these parameters
on the results.
First, we have chosen, as in \cite{calabrese}, the death rate $\alpha=1$, and solving
Eq.~(\ref{steq}) we will roughly estimate the birth rate, $b$. We will work in the limit of intermediate to high mean annual
precipitation, so water is non-limiting and thus we
can neglect the effects of competition ($\delta=0$). At this intermediate to high mean annual precipitation the empirically 
observed upper limit of savanna tree cover is approximately
$\rho_{0}=0.8$ \citep{sankaran2005, bucini}. To reach this upper limit in the tree cover, disturbances must also be absent, implying
no fire ($\sigma\rightarrow\infty$). In this limit, the mean field Eq.~(\ref{eq:mf}) is quantitatively accurate, as it is shown in Fig.~\ref{transition},
and the stationary mean field solution of the model depends only on the birth rate
\begin{equation}
\rho_{0} (\sigma \to \infty) =\frac{b-1}{b}.
\end{equation}
It can be solved for $b$ for a fixed $\rho_{0}=0.8$, and it yields $b=5$ \citep{calabrese}. In the
following we just consider the dependence of our results on $\delta$
and $\sigma$. In particular, $\rho_0=\rho_0 (\sigma, \delta)$.

\begin{figure}
\centering
\includegraphics[width=0.55\textwidth]{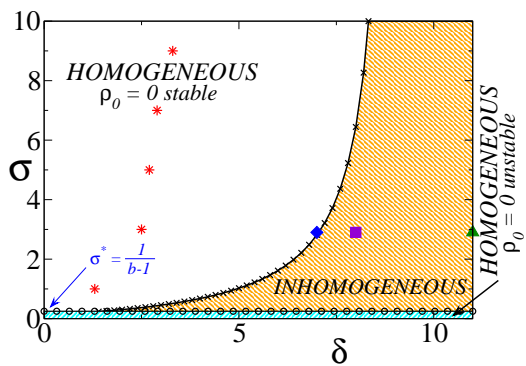}
\caption{Phase diagram of the mean field equation ~(\ref{sav1}) for $b=5.0$, $\alpha=1.0$, and a step kernel.
The absorbing-active transition is shown at $\sigma^{*}$ with circles (o). The homogeneous-pattern transition (Eq.~(\ref{criline})) is
indicated with crosses (x). The diamond, the square, and the up-triangle
show the value of the parameters $\sigma$ and $\delta$ 
taken in Figures~\ref{patterns}(a)-(c)
respectively. The stars point out the transition to
inhomogeneous solutions in the stochastic model as described in Sec.~\ref{stochastic}, with $\Gamma=0.2$.}
\label{phspace}
\end{figure}

The phase diagram of the model, computed
numerically, is shown in Fig.~\ref{phspace},
where we plot 
the spatial character of the steady solution (homogeneous or inhomogeneous) as a function of $\delta$ and $\sigma$.
Note that increasing competition enhances the inhomogeneous or pattern solution.
 This is because, as we are now in the case of a kernel giving rise to clusters, increasing
$\delta$ makes it more difficult to enter the exclusion zones in-between the clusters. 
For very strong competition (high, unrealistic, $\delta$), fire has no influence on the pattern. 

The critical line separating these two solutions (pattern and homogeneous) can be obtained analytically as a function of
the parameters $\delta$, $\sigma$, $\rho_0$ and  $\hat{G}(k_{c})$. Taking $b=5$ and $\alpha=1$, it is
\begin{eqnarray}\label{criline}
&\sigma_c=\frac{(\rho_{0}-1)[5(\rho_{0}-1)
(\delta \hat{G}(k_{c})\rho_{0}-1)-2{\rm e}^{\delta\pi R^{2}\rho_{0}}]+(\rho_{0}-1)\sqrt{5\left[5(\rho_{0}-1)^{2}
(\delta \hat{G}(k_{c})\rho_{0}-1)^{2}-4{\rm e}^{\delta\pi R^{2}\rho_{0}}\rho_{0}\right]}}{10\left[1-2\rho_{0}+
\delta \hat{G}(k_{c})\rho_{0}(1+\rho_{0})-{\rm e}^{\delta\pi R^{2}\rho_{0}}/5\right]}.& \nonumber \\
\end{eqnarray}

This complicated expression must be evaluated numerically together with the 
solution of Eq.~(\ref{steq}) for the stationary density of trees, which is also
a function of the competition and fire parameters. We show the results 
in Fig.~\ref{phspace}, where the curve, represented with the black crosses, 
fits perfectly with the numerical results from the linear stability analysis. 

With $b=5$, in the absence of fire ($\sigma\rightarrow\infty$), and
for weak competition, we can take the limits $\delta\rightarrow 0$ and $\sigma\rightarrow\infty$ of 
the dispersion relation Eq.~(\ref{reldisper}), leading to
\begin{equation}\label{limitceroinfinity}
\lambda(k;\delta\rightarrow 0, \sigma\rightarrow\infty)=4-10\rho_{0}.
\end{equation}
In Fig.~\ref{transition}, for large $\sigma$,  it can be seen that typically $\rho_{0}> 0.4$, so Eq.~(\ref{limitceroinfinity}) becomes negative.
This result means that in this limit, trees are uniformly distributed in the system as there is no competition, 
and space does not play a relevant role in the
establishment of new trees. Such situation could be interpreted as favorable to forest
leading to a fairly homogeneous density of trees.
This result agrees with the phase plane plotted in Fig.~\ref{phspace}. 
In biological terms, there are no exclusion zones in the system because there is no competition.

\subsection{Numerical simulations}
\label{sec:NS}

\begin{figure}
\centering
\includegraphics[width=0.65\textwidth]{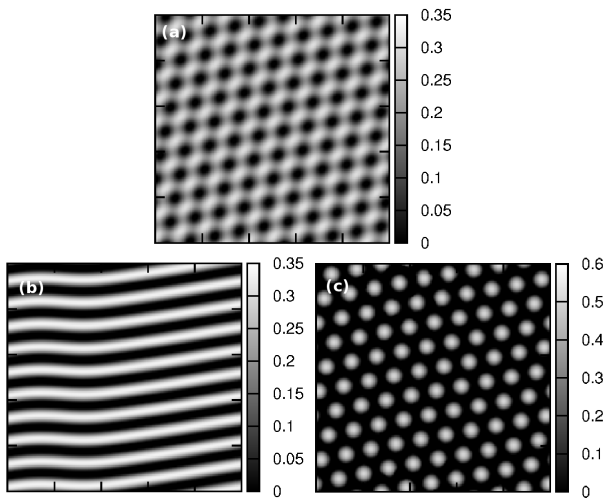}
\caption{(a) Grass spots ($\delta=7.0$), (b) striped grass vs. tree ($\delta=8.0$), and (c) tree spots ($\delta=11.0$) patterns in 
the deterministic model in a square patch of savanna of $1~ha$. $\sigma=2.9$, $R=8.0~m$, $b=5.0$ and $\alpha=1.0$ in all the plots.}
\label{patterns}
\end{figure}

The previous analysis provides information, depending on the competition and fire parameters,
about when the solution is spatially homogenous and when trees arrange in clusters.
However, the different shapes of the patterns have to 
be studied via numerical simulations \citep{dodoricolibro} of the whole equation
of the model. 
We have taken a finite square region of savanna with an area of $1$ ha., allowed competition to occur 
in a circular area of radius $R=8 \ m$, and employed periodic boundary conditions and a finite differences algorithm
to obtain the numerical solution. Similarly to what has been observed in studies of semiarid water limited systems
\citep{D'Odoricob,TheAmNat2002}, different structures, including gaps, stripes, and tree spots,
are obtained in the stationary state as we increase the strength of competition
for a fixed value of the fire parameter or, on the other hand, as we 
decrease the resistance to fires for a given competition intensity. 
In both equivalent cases, we observe this 
spectrum of patterns as far as we go to a more dry state of the system, where resources
(mainly water) are more limited (see Figs.~\ref{patterns}(a)-\ref{patterns}(c)) and competition is consequently stronger. 
This same sequence of appearance of patterns has been already observed in the presence of different short-range
facilitation mechanisms 
\citep{LejeuneTlidi, TheAmNat2002}. 
It indicates that, when  $\delta$ is increased (i.e. the probability of surviving competition is decreased), new trees cannot establish
in the exclusion areas so clustering is enhanced.

On the other hand, in the case of fire-prones savannas, previous works had only shown either tree spot \citep{LejeunePRE2002} or 
grass spots \citep{dodofire}. Therefore, at some values of the parameter space (see Fig. \ref{patterns}b), 
the patterns in our deterministic approach are not observed
in mesic savannas, and should correspond to semiarid systems. 
However, we will show in the following sections that under the parameter constraints of a mesic savanna, 
and considering the stochastic nature of the tree growth dynamics in the system
(i.e. demographic noise), our model shows realistic spatial structures.

\begin{figure}
\begin{center} 
\includegraphics[width=0.55\textwidth]{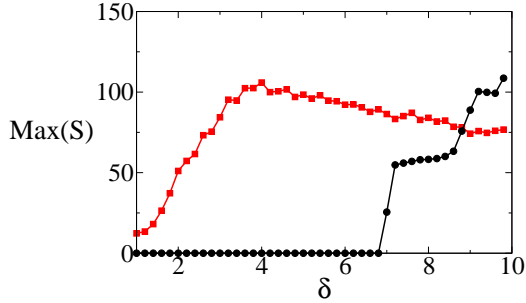}
\caption{Maximum of the structure function for different values of the competition
parameter $\delta$ at long times. The fire parameter is fixed at $\sigma=2.9$. Black
circles refers to the deterministic model and red squares to
the stochastic model, $\Gamma=0.20$.}
\label{maxstructure}
\end{center}
\end{figure}

A much more quantitative analysis of the periodicity in the patterns can be performed
via the structure function. This will be helpful to check the previous results and, especially,
for the analysis of the data of the stochastic model of the next section, for which we will not 
present analytical results.
The structure function is defined as the modulus of the spatial
Fourier transform of the density of trees in the stationary state,
\begin{equation}\label{strucfunc}
S(k)=\left\langle \left| \int d\bx{\rm e}^{i\bk \cdot \bx}\rho(\bx,t \to \infty ) \right| \right\rangle,
\end{equation}
where the average is a spherical average over the wavevectors with modulus $k$. 
The structure function is helpful to study spatial periodicities in the system, similar to the power spectrum of a temporal signal. 
Its maximum
identifies dominant periodicities, which in our case are the distances between tree clusters. 
Note that the geometry of the different patterns cannot be uncovered with the structure function, since it
involves a spherical average. In Fig.~\ref{maxstructure}, we show the transition to patterns using the
maximum of the structure function as a function of the competition parameter. A peak appears when 
there are spatial structures in the system, so $Max[S(k)]\neq0$. However, we do not have information about the values where the shapes of the patterns change.
Taking $R=8~m$, the peak is always at $\lambda_c=10 m$ for our deterministic savanna model, independently of the competition
and fire resistance parameters, provided that they take values that ensure the emergence of patterns in the system
(see the line labeled by $\Gamma=0$ in Fig.~\ref{strucboth}; 
for the definition of $\Gamma$ see Sec.~\ref{stochastic}).
This result is in good agreement with the theoretical one provided for the wavelength by the
linear stability analysis $\lambda=2\pi/k_{max}=9.78~m$, which is also independent of competition and resistance to fires.

\section{Stochastic model}\label{stochastic}

The perfectly periodic patterns emerging in Fig.~\ref{patterns} seem
to be far from the disordered ones usually observed in aerial photographs of mesic savannas and shown by
individual based models \citep{calabrese, jeltsch1999detecting, barot1999demography, caylor2003tree}. 
We have so far described a savanna system in terms of the density of trees with a deterministic
dynamics. The interpretation of the field $\rho(\mathbf{x},t)$ is the density of tree (active) 
sites in a small volume, $V$. If we think of trees as reacting particles
which are born and die probabilistically, then to provide a reasonable description of the underlying individual-based
birth and death dynamics,
we have to add a noise term to the standard deterministic equation. It  
will take into account the {\it intrinsic} stochasticity present at the individual level in the system.

\begin{figure}
\begin{center}
\includegraphics[width=0.55\textwidth]{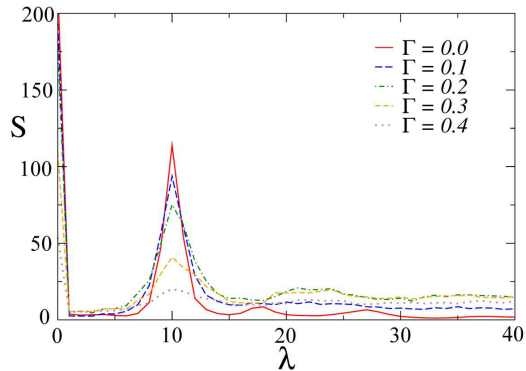}
\caption{Numerical computation of the structure function defined in Eq.~(\ref{strucfunc}) for different values
of the demographic noise intensity. $\delta=9.8$, $\sigma=2.9$, $R=8~m$, $\alpha=1.0$, $b=5.0$.}
\label{strucboth}
\end{center}
\end{figure}

If we take a small volume, $V$, the number of reactions taking place is proportional to the number of particles therein, $N$, with 
small deviations. If $N$ is large enough, the central limit theorem applies to the sum of $N$ independent random
variables and predicts that the amplitude of the deviation is of the order of $\sqrt{N}\propto\sqrt{\rho(\mathbf{x},t)}$ \citep{gardiner}.
This stochasticity is referred to as demographic noise. The macroscopic equation is now stochastic,
\begin{equation}\label{savsto}
   \frac{\partial \rho(\mathbf{x}, t)}{\partial t}=b_{eff}(\rho)[\rho(\mathbf{x}, t)-\rho^{2}(\mathbf{x}, t)]-
\alpha\rho(\mathbf{x}, t)+\Gamma\sqrt{\rho(\mathbf{x}, t)}\eta(\mathbf{x},t),
\end{equation}
where $\Gamma \propto \sqrt{b_{eff}}$ (but we take it as a constant, \citep{dickman}) modulates the intensity of $\eta(\mathbf{x},t)$,
 a Gaussian white noise term with zero mean and correlations given by Dirac delta distributions
\begin{equation}
 <\eta(\mathbf{x},t)\eta(\mathbf{x'},t')>=\delta(\mathbf{x}-\mathbf{x'})\delta(t-t').
\end{equation}
The complete description of the dynamics in Eq.(\ref{savsto})
should have the potential to describe more realistic patterns.

\begin{figure}
\begin{center} 
\includegraphics[width=0.6\textwidth]{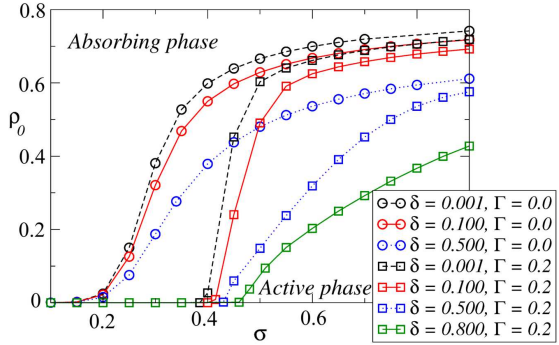}
\caption{Active-absorbing phase transition in the deterministic (Circles) and the stochastic
model (Squares). In the later case, we integrate the Eq.~(\ref{savsto}) with $\Gamma=0.2$ and
average the density of trees in the steady state.}
\label{noisy-transition}
\end{center}
\end{figure}

We first investigate the effect of demographic noise  on the persistence of trees in the system.
We show in (Fig. \ref{noisy-transition}) that the critical point, $\sigma^{*}$, depends on the value of the
competition parameter $\delta$. This effect is rather small, so that when $\delta$ increases the transition to the
grassland state appears only for a slightly larger $\sigma$ (i.e, less frequent fire). The reason seems to be that fire
frequency and intensity depend on grass biomass. Seasonally wet savannas support much more grass biomass that serves as 
fuel for fires during the dry season \citep{dodo2, hanan2008}. Dry savannas have much lower grass biomass, so they do not burn as often or as intensely.
The shift of the critical value of $\sigma$ when competition is stronger
is consistent with the one showed in \citep{calabrese}, as can be seen comparing Fig.~2 therein
with Fig.~\ref{noisy-transition} here. Besides, the values obtained for $\sigma^{*}$ are larger when we consider the demographic stochasticity \citep{stanley}
neglected in the deterministic field approach.

\begin{figure}
\centering
\includegraphics[width=0.6\textwidth]{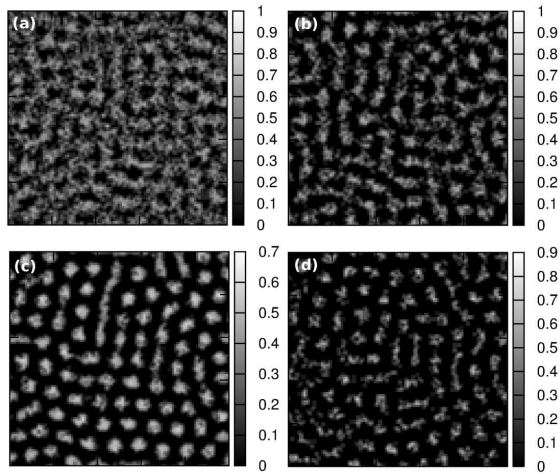}
\caption{Patterns of the stochastic model in a square patch of savanna of $1~ha$. $\sigma=2.9$, $R=8.0~m$, $b=5.0$ and $\alpha=1.0$ in all the plots.
 (a) $\Gamma=0.2$, $\delta=3.0$. (b) $\Gamma=0.2$, $\delta=5.0$. (c) $\Gamma=0.1$, $\delta=10.0$. (b) $\Gamma=0.2$, $\delta=10.0$.}
\label{stopatterns}
\end{figure}

We explore numerically the stochastic savanna model using an algorithm developed in
\cite{dickman} (See \ref{sec:numerical}).
Note that the noise makes the transition to pattern smoother so the change
from homogeneous to inhomogeneous spatial distributions is not as clear as it is in the limit
where the demographic noise vanishes (See Fig.~\ref{maxstructure}). The presence of
demographic noise in the model, as shown in Fig.~\ref{phspace} (red stars),
also decreases the value of the competition strength at which
patterns appear in the system, as has been observed in other systems. Mathematically, these new patterns appear since 
demographic noise maintains Fourier modes of the solution which, due to the value of the parameters,
would decay in a deterministic approach \citep{butler}. Biologically, exclusion zones are promoted by demographic noise, since it does not 
affect regions where there are not trees. On the other hand in vegetated areas fluctuations may enhance tree density, leading to stronger competition.
The presence of demographic noise in the model allows the existence of patterns under more humid conditions.
This result is highly relevant for mesic savannas, as we expect competition to be of low to intermediate strength in such systems.
We show two examples of these irregular patterns in Fig.~\ref{stopatterns}(a) and Fig.~\ref{stopatterns}(b). Unrealistic stripe-like patterns
no longer appear in the stochastic model. 

We have studied the dynamics of the system for some values of the fire and 
competition parameters. Demographic noise influences the spatial 
structures shown by the model. The deterministic approach shows a full spectrum of
patterns which are not visually realistic for mesic savannas (but for arid systems). The role of
the noise is to transform this spectrum of regular, unrealistic patterns into more irregular ones 
(Figures \ref{stopatterns}(a)-\ref{stopatterns}(d))
that remind the observed in aerial photographs of real mesic savannas. On the other hand, these patterns
are statistically equivalent to the deterministic ones, as it is shown with the structure function in Fig. \ref{strucboth}.
The dominant 
scale in the solution is given by the interaction radio, $R$, and it is independent of the 
amplitude of the noise (see the structure function in Fig.~\ref{strucboth},
peaked around $\lambda=10~m$ independently of the noise). Besides, over a certain treshold in the amplitude, demographic
noise destroys the population of trees. Therefore, the model presents
an active-absorbing transition with the noise strength, $\Gamma$, being the control parameter. 

\section{Discussion}
\label{sec:discussion}

Understanding the mechanisms that produce spatial patterns in savanna tree populations has long been an area of interest among savanna ecologists 
\citep{skarpe1991spatial, jeltsch1999detecting, barot1999demography, caylor2003tree, scanlon2007positive}.
A key step in such an analysis is defining the most parsimonious combination of mechanisms that will produce the pattern in question. 
In this chapter the combination of long-range competition for resources and the facilitation induced by fire  
are considered the responsible of the spatial structures, in the line of studies of vegetation pattern formation
in arid systems, where also a combination of long-range inhibition and short-range facilitation is introduced \citep{klausmeier, LefeverLejeune, TheAmNat2002, Hardenberg}.
The main difference is that the facilitation provided by the protection effect of adult trees against fires in our savanna model takes the short-range facilitation to 
its infinitesimally short limit (i.e, local limit). Under this assumption we have studied the conditions under which our model could account for patterns.
We have shown that nonlocal competition combined with local facilitation induces the full range of
observed spatial patterns, provided the competition term enters nonlinearly 
in the equation for the density of trees,
and that competition is strong enough.

The key technical requirement for this effect to occur is that the competition kernel must be an almost constant function in a given competition region, and decay
abruptly out of the region. We verify this condition working with supergaussian kernel functions.
In practice, this means that competition kernels whose Fourier transform takes 
negative values for some wavenumber values, will lead to competition driven clustering.

The other mechanism we have considered for a minimalistic but realistic
savanna model, fire, has been shown to be relevant
for the coexistence of trees and grass and for the shape of the patterns.
However, competition is the main ingredient allowing pattern solutions to exist in the model. If the shape of the kernel
allows these types of solutions, then the specific values of fire and competition
parameters determine the kind of spatial structure that develops.
It is also worth mentioning that one can observe the full spectrum of patterns in the limit where fires vanish 
($\sigma\rightarrow\infty$), so there is no facilitation at all, provided competition is strong enough. However,
when there is no competition, $\delta=0$, no patterns develop regardless of the value of the fire term.
Therefore, we conclude that the nonlocal competition term is responsible for the emergence of clustered distributions 
of trees in the model, with the fire term playing a relevant role only to fix the 
value of the competition parameter at which patterns appear. In other words, for a given competition strength, 
patterns appear more readily when fire is combined with competition. A similar mechanism of competitive interactions
between species has been shown to give rise to clusters of species in the context of classical ecological niche theory. Scheffer and 
van Nes \citep{scheffer2006} showed that species distribution in niche space was clustered, and Pigolotti et al. \citep{pigo} showed that 
this arises as an instability of the nonlocal nonlinear equation describing the competition of species.

Long-distance competition for resources in combination with the local facilitation due to the protection effect
of adult trees in the establishment of juvenile ones can explain the emergence of realistic structures of trees in mesic savannas. 
In these environmental conditions, competition is limited, so we should restrict to small to intermediate values of the parameter $\delta$, 
and the effect of fires is also worth to be taken into account. However, these two ingredients give a full range of patterns
observed in vegetated systems, but not in the particular case of savannas. It is necessary to consider the role of demographic noise, 
which is present in the system through the stochastic nature of the birth and death processes of individual trees. In this complete
framework our model shows irregular patterns of trees similar to the observed in real savannas.

The other important feature of savannas, the characteristic long-time coexistence of trees and grass is well
captured with our model (Figures \ref{transition} and \ref{noisy-transition}). Besides, the presence of 
demographic noise, as it is shown in Fig.~\ref{noisy-transition}, makes our approach much 
more realistic, since the persistence of trees in the face of fires is related to the water in the system.
On the other hand, demographic stochasticity causes tree extinction at lower fire frequencies (larger $\sigma$) than
in the deterministic case. This is because random fluctuations in tree density are of sufficient
magnitude that this can hit zero even if the deterministic stationary 
tree density (for a given fire frequency) is greater than zero. This effect vanishes if
we increase the system size. The demographic noise is proportional to the density of trees (proportional
to $(L_{x}\times L_{y})^{-1}$), so fluctuations are smaller if we study bigger patches of savannas.
As usually happens in the study of critical phenomena in Statistical Mechanics, the extinction times due to 
demographic noise increase exponentially with the size of the system for those intensities of competition
and fire that allow the presence of trees in the stationary state. Over the critical line, this time will follow
a power law scaling, and a logarithmic one when the stationary state of the deterministic model is already absorbing 
(without trees) \citep{marro}.

\section{Summary} \label{summary}

We have shown the formation of patterns in a minimal savanna model, that 
considers the combination of long-range competition and local facilitation mechanisms
as well as the transition from  
trees-grass coexistence to a grass only state. 
 
The salient feature of the model is that it only considers
nonlocal (and nonlinear) competition through a kernel function which defines the length
of the interaction, while the facilitation is considered to have an infinitesimally short influence range.
Our model thus differs from standard kernel-based savanna models that feature both
short-range facilitation and long-range competition. The same sequence of spatial patterns
appears in both approaches, confirming Rietkerk and van de Koppel's \citep{RietkerkTrends} suggestion
that short-range facilitation does not induce spatial pattern formation by itself, and long-distance competition
is also needed. It also suggests that long-range competition could be not only a necessary, but also a 
sufficient condition to the appearance of spatial structures of trees.

Inspired by \citep{calabrese}, we have proposed a nonlocal deterministic macroscopic equation 
for the evolution of the local density of trees where fire and
tree-tree competition are the dominant mechanisms.
If the kernel function falls off with distance very quickly (the Fourier
transform is always positive) the system only has homogenous solutions. 
In the opposite case, patterns may appear depending on the value of the parameters ($\delta$ and $\sigma$),
and in a sequence similar to the spatial
structures appearing in standard kernel-based models. Under less favorable environmental
conditions, trees tend to arrange in more robust structures to survive 
(Fig.~\ref{patterns}(d)). Biologically, trees are lumped
in dense groups, separated by empty regions. 
Entrance of new trees in these {\it exclusion zones} is impossible
due to the intense competition they experience there.

A great strength of our approach is that our deterministic analysis is formal, 
and we have shown the different spatial distributions of the
trees that occur as competition becomes more intense, concluding that self organization of trees is a good
mechanism to promote tree survival under adverse conditions \citep{TheAmNat2002}. 
Trees tend to cluster in the high competition (low resources) limit (Fig.~\ref{patterns}(d)), due to the 
formation of exclusion zones caused by nonlocal competition, and not as a result of facilitation.
However, because we are dealing with a deterministic model, the patterns
are too regular and the transition between the grass-only and a tree-populated
states is independent of tree competition. We therefore considered stochasticity
coming from the stochastic nature of individual birth and death events,
to provide a more realistic description of savanna dynamics. Calabrese et al. \citep{calabrese} also noted that savanna-to-grassland
transition was independent of competition intensity in the mean field approach, but not when demographic noise was included. In
the present model, both the grassland to savanna transition 
and the spatial structures that develop are influenced by demographic stochasticity. In the case of spatial 
structures, demographic noise is specially relevant, since it turns much of the unrealistic patterns of the deterministic 
model into more realistic ones, that remind the observed in real savannas. It also allows the existence of periodic arrangements 
of trees in more humid systems, which means environmental conditions closer to mesic savannas.

We have quantified the characteristic spacing of spatial patterns
through the structure function. The irregular patterns produced by the stochastic model still
have a dominant wavelength whose value is the same as in the
deterministic model and depends only on the value of the range 
of the interaction, $R$, in the kernel function. The match between the typical spatial scale of the
patterns and the characteristic distance over which nonlocal competition acts suggests
that it could be responsible for the presence of clustered spatial structures. 
In the next chapter we will propose a competition model, neglecting every facilitative interaction, to confirm this hypothesis
in arid to semiarid systems. 
   \chapter{Semiarid systems} \label{chap:semi}

Regular vegetation patterns in semiarid ecosystems have been traditionally believed to arise
from the interplay between long-range competition and facilitation processes
acting at smaller distances. In this chapter, it is shown that under rather general
conditions, long-range competition alone may be enough to shape these
patterns. To this end we propose three simple, general models for the dynamics of
vegetation, that include only long-range competition between plants 
through a nonlocal term, where the kernel function quantifies the
intensity and range of the interaction. Firstly, long-range competition
is introduced influencing the growth of vegetation, secondly it is assumed to 
affect its death. Finally, it is considered
as a term independent of the local birth-death vegetation dynamics, entering
linearly in the equation.
In all the situations, regardless of the way in which competition acts,
we recover the vegetation spatial structures
that account for facilitation in addition to competition. Models 
only have consider the finite range of the
competition among plants, given by the length of the roots.

\section{Introduction}
\label{sec:Introduction}

\begin{figure}
\centering
\includegraphics[width=0.75\textwidth, clip=true]{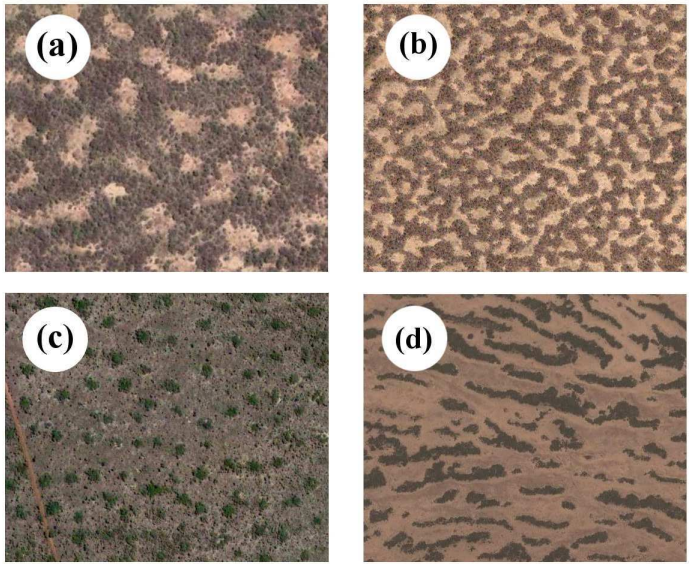}
\caption{Aerial photographies obtained from Google Earth. (a) Gapped pattern (12\textdegree22'44.44'' N; 2\textdegree24'03.00'' E).
(b) Laberynth pattern (12\textdegree38'15.70'' N; 3\textdegree13'05.25'' E). (c) Spot pattern (12\textdegree18'32.51'' N; 2\textdegree22'38.22'' E).
(d) Tiger bush (13\textdegree21'44.14'' N; 2\textdegree04'53.30'' E).}
\label{patrones}
\end{figure}

Regular patterns  and spatial
organization of vegetation have been observed in many arid and
semiarid ecosystems worldwide (Fig.~\ref{patrones}), covering a diverse range of
plant taxa and soil types \citep{klausmeier,
RietkerkTrends,thompson}. A key common ingredient in these
systems is that plant growth is severely limited by water
availability, and thus plants likely compete strongly for water
\citep{TheAmNat2002}. The study of such patterns is especially
interesting because their features may reveal much about the
underlying physical and biological processes that generated
them in addition to giving information on the characteristics of
the ecosystem. It is possible, for instance, to infer their
resilience against anthropogenic disturbances or climatic
changes that could cause abrupt shifts in the system and lead
it to a desert state
\citep{vandekoppel2002,D'Odoricoa,D'Odoricob}.

Much research has therefore focused on identifying the mechanisms
that can produce spatial patterning in water limited systems
\citep{LefeverLejeune,klausmeier,pueyo}. An important class of
deterministic vegetation models (i.e., those not considering
noise associated with random disturbances) that can produce
regular patterns are the kernel-based models
\citep{D'Odoricoa}. These models produce patterns via a
symmetry-breaking instability (i.e., a mechanism by which the
symmetric-homogeneous state loses stability and a periodic
pattern is created) that has its origins in the interplay
between short-range facilitation and long-range competition
\citep{D'Odoricob,RietkerkTrends,BorgognoRG}, with field
observations confirming this hypothesis in some landscapes
\citep{Dunkerley}. Therefore it has been long assumed that both
of these mechanisms must be present in semiarid systems to
account for observed vegetation patterns, although quantifying
the importance of each one has proven to be a difficult and
contentious task \citep{Barbier,Veblen}. A key role theory can
play here is to identify the minimal requirements for pattern
formation to occur. Some authors have speculated that
pattern formation, under certain conditions, could occur
without short-range facilitation \citep{RietkerkTrends}. In this line,
in the previous chapter we proposed a model for mesic savannas
that includes fire and plant-plant
competition as key ingredients. Fire
introduces a positive feedback so that this model considers
both competition and facilitation mechanisms. However, the
model still produces regular patterns even when the
facilitative interaction, fire, is considered at its very
short-range (in fact, local) limit. These considerations
suggest that local facilitation may be superfluous for pattern
formation, and that a deeper exploration of the range of
conditions under which pattern formation can occur in the
absence of facilitation is therefore warranted. This will 
be the major concern of this chapter.

We will study simple,
but quite general, single-variable models for
vegetation density in
water-limited regions. Only competitive interactions are considered, modeled
in different ways to analyse if patterns
depend on how competition enters in the dynamical equations \cite{Martinez-Garcia2013,Martinez-garcia2014}. The role
of nonlinearities is also investigated.
We show that when only a single broadly applicable
condition is met, that competitive interactions have a finite
range, the full set of regular patterns formerly attributed to
the interaction between short-range facilitation and
long-distance competition can be produced in the absence of
facilitation.

\section{Competition in a nonlocal nonlinear birth term}
\label{birthsec}

Arid and semiarid ecosystems are typified by patches of
vegetation interspersed with bare ground. Water is a very
limited resource for which juvenile plants must compete with
those that have already established. Logistic-type
population models have been used in a wide variety of
applications including semiarid systems and, as shown in the previous chapter, savannas
\citep{calabrese}. They thus form a reasonable and very general
starting point. Specifically, we consider the
large-scale long-time description of the model in terms of a
continuous-time evolution equation for the density of trees,
$\rho ({\bf x},t)$. Death occurs at a constant rate $\alpha$,
whereas population growth occurs via a sequence of seed
production, dispersal, and seed establishment processes. Seed
production occurs at a rate $\beta_0$ per plant. For simplicity
we consider dispersal to be purely local and then if all seeds
would give rise to new  plants the growth rate would be
$\beta_0\rho({\bf x},t)$. But once a seed lands, it will have
to overcome competition in order to establish as a
new plant. We consider two different competition mechanisms.
First, space availability alone limits density to a maximum
value given by $\rho_{max}$. Thus, $0\leq \rho ({\bf x},t) \leq
\rho_{max}$. The proportion of available space at site ${\bf
x}$ is $1-\rho({\bf x},t)/\rho_{max}$, so that the growth rate
given by seed production should be reduced by this factor.
Second, once the seed germinates, it has to overcome
competition for resources with other plants. This is included in the model by an additional factor
$r=r(\tilde{\rho},\delta)$, $0\leq r \leq 1$, which is the
probability of overcoming competition. This probability decreases with increasing average vegetation density within a
neighborhood $\tilde \rho$, and the strength of this decrease depends on the competition intensity parameter, $\delta$. Higher values of $\delta$
represent more arid lands, and thus stronger competition for
water. In the following, we measure density in units so that
$\rho_{max}=1$. Combining all processes, the evolution equation
for the density then takes the form:

\begin{equation} \label{model}
\frac{\partial \rho({\bf x},t)}{\partial t}=\beta_0 r(\tilde{\rho},\delta)\rho({\bf x},t)(1-\rho({\bf x},t))-\alpha\rho({\bf x},t).
\end{equation}

$\tilde{\rho}=\tilde{\rho}({\bf x},t)$ is the nonlocal density
of vegetation that is obtained by averaging (with a proper
weighting function) the density of plants in a neighborhood:

\begin{equation}\label{nonlocal}
\tilde{\rho}({\bf x},t)=\int G(|{\bf x}-{\bf x'}|)\rho({\bf x'},t)d{\bf x'},
\end{equation}
 where $G({\bf x})$ is a normalized kernel function, which
accounts for the weighted mean vegetation density, and defines
the neighborhood of the plant. A Laplacian term could be
included in the right-hand side of Eq.~(\ref{model}) as a way to model
seed dispersal, but doing so would not qualitatively change our
results, so we have left it out.

We have presented a phenomenological derivation of the model.
An open problem is to infer this type of description from a 
mechanistic one where the explicit interactive dynamics of
vegetation competing for water is considered. Preliminary results on this derivation are 
shown in the Appendix \ref{app:derivation}.

  \begin{figure}
  \centering
  \includegraphics[angle=-90,width=0.8\textwidth]{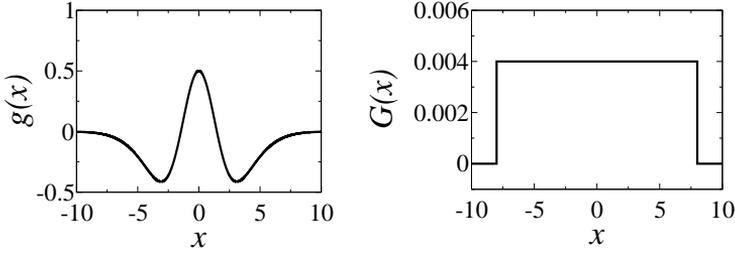}
  \caption{(Left) Kernel function of standard one-dimensional
kernel-based models 
considering both competitive and facilitative interactions.
It is built with a combination of positive and negative Gaussian functions, 
$g(x)=1.5\exp\left(-(x/2)^{2}\right)-\exp\left(-(x/4)^{2}\right)$. Notice that
it takes positive and negative values at different distances.
(Right) Competitive-only top-hat kernel with range $R=8$. $G(x)$ is always positive.}
  \label{fig:kernel}
  \end{figure} 

In previous kernel-based vegetation models \citep{LefeverLejeune,D'Odoricoa},
the kernel function contained
information on the class of interactions present in the system,
that were both facilitative and competitive.
That is, it can take positive and negative values either enhancing or inhibiting the vegetation growth.
On the contrary, we introduce purely competitive interactions
through the nonlocal function $r(\tilde{\rho},\delta)$, where
the kernel is always positive and thus defines the area of influence of a focal plant, and
how this influence decays with distance. These two kernels are compared in Fig.~\ref{fig:kernel}.
Competition is included
by assuming that the probability of establishment
$r$ decreases with increasing vegetation density in the
surroundings:

\begin{equation}
\frac{\partial r(\tilde{\rho},\delta)}{\partial\tilde{\rho}}\leq 0. \label{cond2}
\end{equation}
As $\delta$ modulates the strength of the competition, it must
be that $r(\tilde{\rho},\delta=0)=1$, and
$r(\tilde{\rho},\delta\rightarrow\infty)=0$. This means that
when water is abundant ($\delta=0$) competition for water is
not important  ($r=1$), whereas new plants cannot establish in
the limit of extremely arid systems, $\delta\rightarrow\infty$.

Note the generality of this vegetation competition model: a
spatially nonlocal population growth term of
logistic type with rate fulfilling Eq.~(\ref{cond2}), and a
linear death term. A complete description of our model should specify
both the kernel function $G$ as well as $r$, but we can go
further with the analysis in general terms.

The possible homogenous stationary values of the density for
Eq.~(\ref{model}) are: a) no
vegetation $\rho=0$,  and b) the  vegetated state $\rho=\rho_{0}$.
The system will show
either one or the other depending on the relationship between
the seed production and death rates, $\beta_{0}$ and $\alpha$ \citep{calabrese}.
The non-trivial homogeneous stationary solution, $\rho_{0}$, can be obtained by solving

\begin{equation}\label{homo}
\beta_{0}r(\rho_{0},\delta)(1-\rho_{0})-\alpha=0,
\end{equation}
that has only one solution in the interval
$\rho_{0}\in[0,1]$ because of the conditions imposed on the
function $r$ in equation(\ref{cond2}). We now ask if this
stationary solution gives rise to periodic structures via a
symmetry-breaking instability as happens in other models that
include not only competition but also facilitation mechanisms
in the interactions \citep{BorgognoRG}. To explore
this possibility in our model, we perform a linear stability
analysis \citep{cross} introducing a small perturbation to the
stationary solution, so $\rho({\bf
x},t)=\rho_{0}+\epsilon\psi({\bf x},t)$, with $\epsilon\ll 1$.
Technical details of this derivation may be found in Appendix
\ref{deriv}. The perturbation growth rate is

\begin{equation}
\label{condition}
 \lambda({\bf k})=-\alpha\rho_{0}\left[\frac{1}{1-\rho_{0}}-\frac{r'(\rho_{0},\delta)}{r(\rho_{0},\delta)}\hat{G}({\bf k})\right],
\end{equation}
where $\hat{G}({\bf k})$ is the Fourier transform of the
kernel, $\hat{G}({\bf k})=\int G({\bf x})\exp(i{\bf k}\cdot{\bf
x})d{\bf x}$, and
$r'(\rho_{0},\delta)\equiv\left(\frac{\partial r}{\partial
\tilde\rho}\right)_{\tilde\rho=\rho_{0}}$.

Patterns appear if the maximum of the growth rate (i.e., of the
most unstable mode), $\lambda(k_{c})$, is positive, which means
that the perturbation grows with time. From
Eq.~(\ref{condition}), this is only possible if the Fourier
transform of the kernel function, $\hat{G}({\bf k})$, takes
negative values, since  $r'(\rho_{0},\delta)<0$. As has been in the previous chapter, this happens,
for example, for all stretched exponentials $G(|{\bf
x}|)\propto \exp{(-|{\bf x}/R|^p})$ with $p>2$, where $R$ is a
typical interaction length \citep{pigo,pigolotti2010}.
Kernels satisfying this criterion have broader shoulders and
shorter tails (i.e., are more platykurtic) than the Gaussian
function, which is obtained for $p=2$. In reality, any
competitive interaction among plants will have finite range
because their roots, which mediate the interaction, have finite
length. The interaction range $R$ between two plants
will be twice the typical root length. Kernels with finite range can,
in general, be modeled by considering a truncated function such
that $G(|x|)=CF(|x|)\Pi(|x|)$, where $C$ is a normalization
constant, $\Pi(x)$ is a unit-step function defined as
$\Pi(x)=1$ if $|x| \leq R$ and $\Pi(x)=0$ if $|x| > R$, and
$F(|x|)$ is a function of the distance that models the
interactions among the plants. Because of the finite range in
the kernel function, the Fourier transform will show
oscillations and thus will always take negative values. The
functional form of the probability of surviving the
competition, $r(\tilde{\rho},\delta)$, changes only the
parameter regime where patterns first develop, but they will
appear in the system, regardless of its form, for
$r'(\rho_{0},\delta)/r(\rho_{0},\delta)$ large enough.

For the rest of our analysis, we will use $F(x)=1$, so the
kernel is given by $G(x)=1/\pi R^2$ if $|x| \leq R$ and
$G(x)=0$ if $|x| > R$, which defines an interaction area of
radius $R$ (that is, roots of typical length
$R/2$). Its Fourier transform (in two dimensions) is

\begin{equation}\label{tophat}
 \hat{G}({\bf k})=\frac{2J_{1}(|{\bf k}|R)}{|{\bf k}|R},
\end{equation}
where $J_{1}(|{\bf k}|R)$ is the first-order Bessel function.
We will further specify the model by assuming
particular forms for the growth rates.
Let us consider a probability of surviving competition given by

\begin{equation}\label{pldecay}
 r(\tilde{\rho},\delta)=\frac{1}{(1+\delta\tilde{\rho})^{q}},
\end{equation}
with $q>0$. In the particular case of $q=1$, the homogeneous
density, $\rho_{0}$, and the perturbation growth rate,
$\lambda$, can be obtained analytically. Numerical evaluations
must be done if $q\neq1$. In the following, for simplicity, we
consider the case $q=1$ and only briefly discuss other values.
The nontrivial stationary solution, $\rho_{0}\neq0$, can be
obtained analytically

\begin{equation}\label{rhostat}
 \rho_{0}=\frac{\beta_{0}-\alpha}{\beta_{0}+\alpha\delta},
\end{equation}
where $\beta_{0}\geq\alpha$. Eq.~(\ref{rhostat}) shows
that the homogeneous density of trees in the stationary state
decays as $\sim\delta^{-1}$ with increasing competition
strength (i.e., large $\delta$). It can be analytically shown
that the same dependence of $\rho_{0}$ on large $\delta$ occurs
for any value of $q$.

  \begin{figure}
  \centering
  \includegraphics[width=0.8\textwidth, trim= 1mm 1mm 1mm .5mm,clip=true]{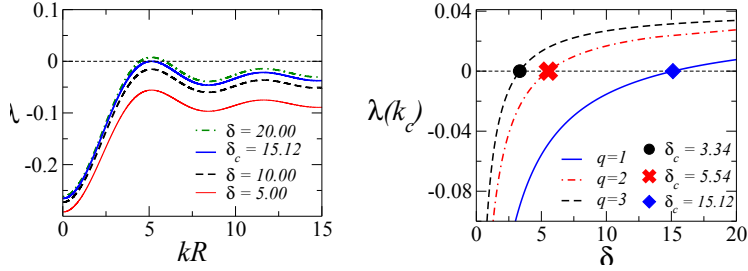}
  \caption{ (Left) Perturbation growth rate given by Eq.~(\ref{dispower}) using a
  unit-step kernel for different values of $\delta$.
  From bottom to top $\delta=5.00$, $\delta=10.00$, $\delta_{c}=15.12$, $\delta=20.00$. (Right) $\lambda(k_{c})$, as a function of $\delta$, using
  $r(\tilde{\rho},\delta)$ given by Eq.~(\ref{pldecay}). From right to left $q=1$, $q=2$, $q=3$. In both panels, other parameters: $\beta_{0}=1.0$ and $\alpha=0.5$.}
  \label{figpldisp}
  \end{figure}
 
From Eq.~(\ref{condition}), the growth rate of the
perturbations can also be calculated

\begin{equation}\label{dispower}
 \lambda({\bf k})=\frac{(\alpha-\beta_{0})(\beta_{0}+\alpha\delta\hat{G}({\bf k}))}{\beta_{0}(1+\delta)},
\end{equation}
and is shown in Fig.~\ref{figpldisp} (Left) for different
values of the competition strength. When the growth rate of the
most unstable mode (i.e. the maximum of $\lambda (k)$),
$k_{c}$, becomes positive, patterns emerge in the system
\citep{BorgognoRG}. To obtain the critical value of the
competition parameter at the transition to patterns,
$\delta_{c}$, we have to calculate the most unstable mode as
the first extreme of $\lambda (\bf k)$ at $k\neq0$, i.e., the
first zero of the derivative of $\hat{G}(\bf k)$. This was already 
done in the previous chapter, and gives $k_{c}=5.136/R$. It depends
only on the the range defining $G (\bf r)$). 
This value changes depending on the kernel, but in the case of
kernels with a finite range (i.e. truncated by a unit step
function of radius $R$) it is always of this order.
The critical wavenumber is determined mainly by the
contribution of the unit step function to the Fourier
transform, which is always the same. This result is also
independent of the other parameters of the system, and shows
that the nonlocal competition mechanism is responsible for the
formation of patterns in the system.

To identify the parameter values for the transition to
patterns, we solve $\lambda({\bf k}_{c})=0$ in
Eq.~(\ref{dispower}), which shows that patterns emerge when
competition strength exceeds
$\delta_{c}=-\beta_{0}/\alpha\hat{G}(\mathbf{k_{c}})$, which is
positive because $\hat{G}(\mathbf{k_{c}})<0$. Figure
\ref{figpldisp} (Right) shows the growth rate of the most
unstable mode as a function of competition strength for
different values of the exponent $q$ for fixed values
$\beta_{0}=1$, and $\alpha=0.5$. Note that the critical value
of the competition parameter depends on the functional form of
$r$. This dependence could be used to tune the value of $q$ to
have a realistic competition strength for the transition to
patterns, provided that one has sufficient data.

\begin{figure}[H]
  \centering
  \includegraphics[width=0.6\textwidth]{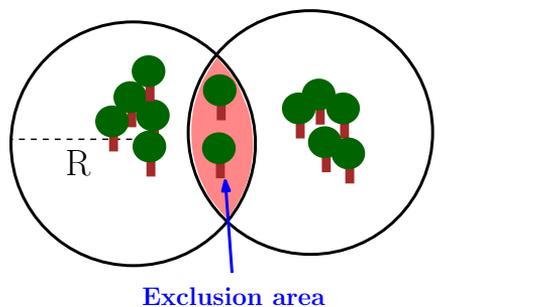}
  \caption{Schematic representation of the formation of exclusion areas, where plants have to compete with two different
  vegetation patches, whereas plants in each patch compete only with individuals in its own patch.}
  \label{exclusion}
  \end{figure}

We can also explain the separation length between clusters of
plants using ecological arguments adn expanding the concept of
\textit{exclusion areas} that was mentioned in the previous chapter.
Consider a random and
inhomogeneous distribution of plants. Maxima of this
distribution identify places with the highest plant density.
Imagine that two such maxima occur at a distance larger than
$R$ but smaller than $2R$ from each other. There will be no
direct interaction between the roots of plants in these
different patches because they are separated by a distance
larger than the interaction range $R$ (twice the
root extension as first order approximation). But there is an area in-between which is
simultaneously within the range of both patches. Compared with
plants occurring inside a cluster, which only have to compete
with plants in their own cluster, those that occur in-between
clusters will experience stronger competition and will
therefore tend to disappear (Fig.~\ref{exclusion}). We call
these regions {\it exclusion
areas}
\citep{Hernandez-Garcia,pigo,pigolotti2010}. The
disappearance of plants in these exclusion areas in turn
reduces competition on the two well-populated patches, so that
a positive feedback appears reinforcing the establishment of
plants in patches periodically separated with a distance
between $R$ and $2R$. We stress again that competition alone is
responsible for the symmetry breaking instability, and no
facilitative interactions are needed for pattern formation.

This mechanism does not work only in the particular case
of considering competition among plants modifying the birth 
rate. It also leads to the formation of patterns if nonlocal competition
is introduced in the death rate or entering linearly in the dynamics. This will be shown in Sections \ref{deathsec} and \ref{sec:linear}
respectively.

\begin{figure}[H]
  \centering
  \includegraphics[width=0.6\textwidth]{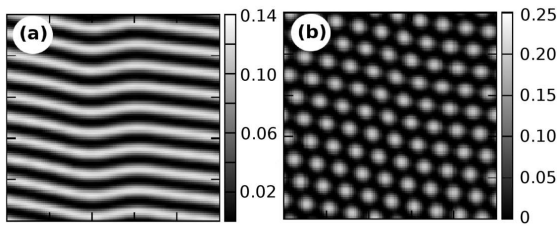}
  \caption{Steady spatial structures shown by the model using the $r(\tilde{\rho},\delta)$ given by Eq.~(\ref{pldecay})
  with $q=1$. Darker grey levels represent smaller densities. (a) Vegetation stripes, $\delta=16.0$.
  (b) Vegetation spots, $\delta=17.0$.  Other parameters: $\beta_{0}=1.0$ and $\alpha=0.5$}
  \label{patternspower}
  \end{figure}

Finally, we have numerically integrated Eq.~(\ref{model}) in a
patch of $10^{4}$~m$^{2}$ with periodic boundary conditions and
a competition range of $R=8$~m. Time stepping is done with an
Euler algorithm. The results (see Fig.~\ref{patternspower})
exhibit steady striped and spotted vegetation patterns
typically arising from symmetry breaking.

Similar results can be obtained for
different growth rates, for example considering a family of
stretched exponentials in the probability 
of overcoming long-range competition $r(\tilde{\rho},\delta)= \exp\left(-\delta\tilde{\rho}^{s}\right)$.

\begin{figure}
  \centering
  \includegraphics[width=0.4\textwidth]{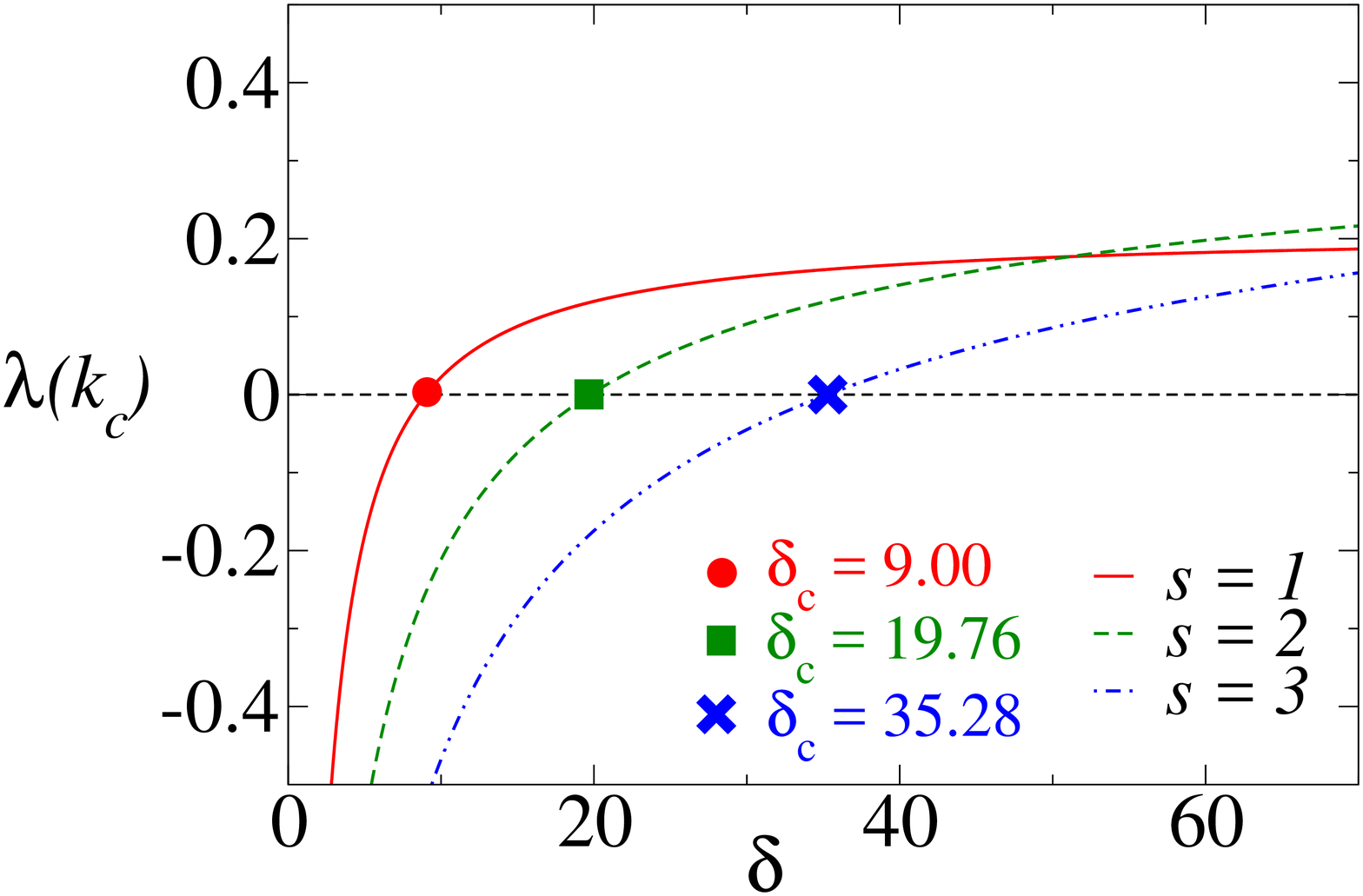}
  \caption{Perturbation growth rate of the most unstable mode as a function of the competition strength for different values
  of the exponent $s$. $s=1$ full line, $s=2$ dashed line, and $s=3$ dotted-dashed line. Other parameters: $\alpha=1$, $\beta_{0}=5$,
  and $R=8$.}
  \label{maxdisp-exp}
  \end{figure}

This gives a perturbation growth rate of the form
\begin{equation}
 \lambda({\bf k})=-\alpha\rho_{0}\left(\frac{1}{1-\rho_{0}}+\delta s\rho_{0}^{p-1}\hat{G}({\bf k})\right),
\end{equation}
where the stationary density, $\rho_{0}$ has to be obtained numerically from Eq.~(\ref{homo}).
The value of the competition parameter at the transition to patterns $\delta_{c}$ also
has to be obtained numerically for a given set of values of $s$, $\alpha$, and $\beta_{0}$.
This critical value is shown in  Fig.~\ref{maxdisp-exp} using a top-hat kernel
for different values of $s$. It is represented the value of the perturbation growth
rate of the most unstable mode $\lambda({\bf k_{c}})$ as a function of the competition strength
$\delta$. 

This further confirms our result that competition is the only
necessary ingredient for the formation of vegetation patterns
in the present framework, and that this does not depend on the
functional form of the probability of surviving competition
(growth rate) provided it verifies the requirements given by
Eq.~(\ref{cond2}).

\section{Competition in a nonlocal nonlinear death term}\label{deathsec}
In the previous section we have considered that plant death occurs at a constant rate
and the birth process takes place following a sequence of seed production, 
dispersal and establishment. The establishment of new 
seeds has been studied as a process depending on the density of vegetation in 
a given neighborhood. However, it is possible to consider
it affecting the death rate.

In this framework, the population growth has two stages, namely seed production at rate $\beta_{0}$
and local seed dispersal. This means that once a seed is produced by a plant, it still has to 
compete for available space, limiting the density of vegetation to a maximum value $\rho_{max}$. As it was
done before, we choose units so that $\rho_{max}=1$. On the other hand, the death of plants 
is influenced by the availability of resources and thus include nonlocal interactions.
When water is abundant the competition for it is not relevant and
plants die at a typical constant rate $\alpha_{0}$. However, the scarcity of
resources promotes the death of vegetation, and this is included in the model 
by an additional factor $h$ in the death term, the probability of dying because of competition, $0\leq h\leq 1$.
That is, the probability of not being able to overcome competition $h(\tilde{\rho},\delta)=1-r(\tilde{\rho},\delta)$,
where $r$ is the function introduced in Sec.~\ref{birthsec}. 
The probability of dying because of competition for resources has to increase with
the density of vegetation
\begin{equation}
 \frac{\partial h(\tilde{\rho},\delta)}{\partial \tilde{\rho}}\geq 0,
\end{equation}
and it must tend to its maximum value in the limit of extremely arid systems $(\delta\rightarrow\infty)$ and to 
vanish when water is not a constraint for vegetation growth $(\delta\rightarrow 0)$. These properties
can be derived from the properties of the function $r$ and its relationship with $h$.

Under these conditions, the model equation for the density of vegetation is
 \begin{equation}\label{sav11}
\frac{\partial \rho({\bf x},t)}{\partial t}=\beta\rho({\bf x},t)(1-\rho({\bf x},t))-\alpha_{0}h(\tilde{\rho}({\bf x},t),\delta)\rho({\bf x},t),
\end{equation}
where $\tilde{\rho}({\bf x},t)$ is the nonlocal density of vegetation at ${\bf x}$,
$\tilde{\rho}({\bf x},t)=\int\rho({\bf x'},t)G(|{\bf x}-{\bf x'}|)dx'$, and $G({\bf x})$ is the kernel
function that defines an interaction range and modulates its strength with the
distance from the focal plant, as it was in Sec.~\ref{birthsec}. Also following the step
of this previous section, we study the existence of patterns in this model.

 First of all, the stationary solutions of Eq.~(\ref{sav11}), $\rho_{0}$, are obtained solving 
 \begin{equation}\label{stadeath}
\beta\rho_{0}(1-\rho_{0})-\alpha_{0}h(\rho_{0},\delta)\rho_{0}=0,
 \end{equation}
 that has a trivial solution, $\rho_{0}=0$ referring to the desert state. The 
 vegetated state must be obtained from 
 \begin{equation}\label{vegsta}
 \beta(1-\rho_{0})-\alpha_{0}h(\rho_{0},\delta)=0,
  \end{equation}
 once the function $h$ has been chosen.
 
 Second of all, the formation of patterns in the system has to be studied through a linear stability analysis of 
 Eq.~(\ref{sav11}), introducing a small perturbation to the stationary homogeneous state, $\rho_{0}$.
 Considering that $\rho({\bf x},t)=\rho_{0}+\epsilon\psi({\bf x},t)$, with $\epsilon\ll1$, the perturbation
 evolves according to the following linear integro-differential equation
 \begin{equation}
  \frac{\partial \psi({\bf x},t)}{\partial t}=\beta(1-2\rho_{0})\psi({\bf x},t)-\alpha_{0}r(\rho_{0},\delta)\psi({\bf x},t)-\alpha_{0}\rho_{0}r'(\rho_{0},\delta)\int G(|{\bf x}-{\bf x'}|)\psi({\bf x'},t)dx',
 \end{equation}
that can be solved using the Fourier transform to obtain the growth rate of the perturbation,
\begin{equation}
 \lambda({\bf k})=\beta(1-2\rho_{0})-\alpha_{0} h(\rho_{0},\delta)-\alpha_{0}\rho_{0}h'(\rho_{0},\delta)\hat{G}({\bf k}),
\end{equation}
where, again, $\hat{G}({\bf k})$ is the Fourier transform of the Kernel function. Using Eq.~(\ref{stadeath}), one finally gets 
\begin{equation}
 \lambda({\bf k})=-\rho_{0}\left[\beta+\alpha_{0}h'(\rho_{0},\delta)\hat{G}({\bf k})\right],
\end{equation}
that shows that, as it was in the case of models with nonlocal interactions in the birth rate, that
the patterns appear only when the Fourier transform of the kernel takes negative values.
To investigate the particular behavior of this model in a simple situation, one needs to choose a function $h$.

\begin{figure}
\centering
\includegraphics[width=0.4\textwidth,angle=-90]{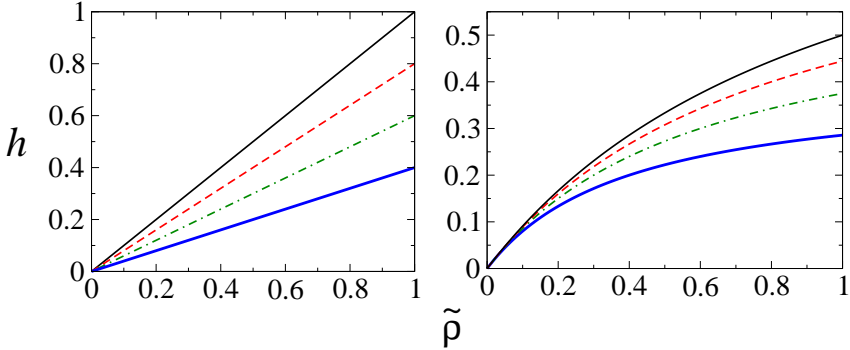}
\caption{Probability of dying because of long range competition as a function of the density of 
vegetation. Left panel, $h(\tilde{\rho}({\bf x},t),\delta)=\delta\tilde{\rho}$.
Right panel, $h(\tilde{\rho}({\bf x},t),\delta)=\frac{\delta\tilde{\rho}}{\delta+\tilde{\rho}}$. 
From top to bottom $\delta=1,\delta=0.8,\delta=0.6,\delta=0.4$ in both panels.}
\label{multicomp}
\end{figure}

The simplest case that allows a complete analysis is to choose a 
probability of dying because of competition growing linearly with the nonlocal density of vegetation,
\begin{equation}\label{rectah}
h(\tilde{\rho}({\bf x},t)=\delta\tilde{\rho}({\bf x},t).
\end{equation}

The behavior of this function $h$ is shown in the left panel of Fig.~\ref{multicomp} for different
values of $\delta$. The choice of this functional form in $h$ restricts the domain of the parameter $\delta$. 
It has to be $\delta\in[0,1]$ since it must be $h\leq 1$ to represent a probability.

\begin{figure}
\centering
\includegraphics[width=0.65\textwidth]{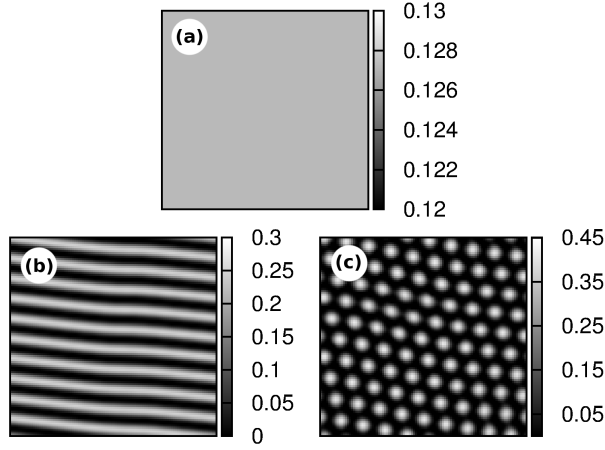}
\caption{Distribution of vegetation of the model in a patch of size $100\times100$
with a linear probability of dying because of long-range competition, Eq.~(\ref{rectah}). 
(a) $\delta=0.7$, (b) $\delta=0.8$, (c) $\delta=0.9$. $\alpha_{0}=1$, $\beta=0.1$, $R=8$.}
\label{patterns_recta}
\end{figure}

The nontrivial stationary solution, Eq.~(\ref{vegsta}), is
\begin{equation}
 \rho_{0}=\frac{\beta}{\beta+\alpha_{0}\delta}.
\end{equation}
This solution has the proper behavior in the limits $\delta\rightarrow 0$ and $\delta\rightarrow\infty$. 
The growth rate of the perturbation is
\begin{equation}
 \lambda({\bf k})=-\frac{\beta}{\beta+\alpha_{0}\delta}\left[\beta+\alpha_{0}\delta\hat{G}({\bf k})\right],
\end{equation}
from where we obtain a transition to pattern at,
\begin{equation}\label{critdeltarecta}
 \delta_{c}=-\frac{\beta}{\alpha_{0}\hat{G}({\bf k_{c}})},
\end{equation}
where $k_{c}$ is the most unstable mode, i.e., that with the highest growing rate.
Eq.~(\ref{critdeltarecta}) limits the possible value of the 
birth and the death rates for a given kernel. They have to take values such as $\delta_{c}\leq 1$. This is a new condition
that was not present in the model presented in Sec.~\ref{birthsec}, and appears because of the constraints imposed on the values of the competition parameter
by the linear function $h$. For $\alpha_{0}=1$, $\beta=0.1$ and a top-hat kernel of radius $R=8$, patterns emerge for a competition 
strength $\delta_{c}\approx0.747$. As it can be observed in Fig.~\ref{patterns_recta}, the homogeneous distribution is the stationary 
solution when $\delta<\delta_{c}$ (Panel a), while patterns (stripes and spots) appear otherwise (Panels b and c respectively).

A different function for the probability of dying 
might be chosen \footnote{It is important to justify that $\delta$ has no dimension, and there is a saturating
density, $\kappa$, that we have set to $\kappa=1$ so that $r=\frac{\delta\tilde{\rho}}{\delta\kappa+\tilde{\rho}}$.},
\begin{equation}
h(\tilde{\rho}({\bf x},t),\delta)=\frac{\delta\tilde{\rho}}{\delta+\tilde{\rho}},
\end{equation}
that is shown in the right panel of Fig.~\ref{multicomp} for different values of $\delta$.
Although this makes the analysis much more complicated, it allows the competition strength parameter
$\delta$ to take any positive real value. We present some results showing that the functional
form of $h$ is not relevant for pattern formation.

The nontrivial stationary solution is 
 \begin{equation}\label{rho02}
  \rho_{0}=\frac{-[(\alpha_{0}+\beta)\delta-\beta]+\sqrt{[(\alpha_{0}+\beta)\delta-\beta]^{2}+4\beta^{2}\delta}}{2\beta},
 \end{equation}
 where we have chosen the positive sign in the square root because it must be $\rho_{0}>0$. The growth rate of the perturbation is 
 \begin{equation}\label{lambda2}
  \lambda({\bf k})=-\rho_{0}\left[\beta+\frac{\alpha_{0}\delta^{2}\hat{G}({\bf k})}{(\delta+\rho_{0})^{2}}\right].
 \end{equation}
Plugging Eq.~(\ref{rho02}) into Eq.~(\ref{lambda2}) and considering the most unstable mode, ${\bf k_{c}}$,
we can obtain an expression for the value of the competition parameter at the transition to patterns,
\begin{equation}
 \delta_{c}=\frac{(\alpha_{0}-\beta)\hat{G}({\bf k_{c}})+\sqrt{\alpha_{0}\beta^{3}(\hat{G}({\bf k_{c}})-1)^{2}\hat{G}({\bf k_{c}})}}{(\alpha_{0}^{2}+\beta^{2})\hat{G}({\bf k_{c}})-\alpha_{0}\beta(\hat{G}({\bf k_{c}})+1)^{2}}.
 \end{equation}
Using $\alpha_{0}=1$, $\beta=0.1$ and a constant top-hat kernel of radius $R=8$ one obtains $\delta_{c}\approx0.663$. This result can be checked
integrating numerically Eq.~(\ref{sav11}) with these values of the rates and the interaction range. We do not show this patterns again because they
are equivalent as the ones in Fig.~\ref{patterns_recta}.

\section{Competition in a nonlocal linear death term}\label{sec:linear}
The study that we are going to present in this section is inspired by the 
kernel-based models reviewed in \citep{BorgognoRG}, where it is assumed that both
competitive and facilitative interactions between plants must come into play
for pattern formation. These models are not well posed, since the density
of vegetation can become negative at some spatial points because the long range interactions
entering linearly. This incovenience is often overcome assuming that whenever the density becomes 
negative, it takes a value $\rho({\bf x}, t)=0$. In this section, we extend these models to the situations where 
there is only competition among trees, and also study the effect of seed dispersal through a diffusive term.
The vegetation density changes in time because of its local dynamics (logistic growth) and the spatial
interactions (competition) with other points of the space,
\begin{eqnarray}\label{dodomodel}
 \frac{\partial \rho(x,t)}{\partial t}=D\nabla^{2}\rho(x,t)+\rho(x,t)\left[1-\kappa^{-1}\rho(x,t)\right]-\Omega\int G(|x-x'|)\rho(x',t)dx', \nonumber \\
\end{eqnarray}
where $\kappa$ is the {\it carrying capacity} of the system (i.e, the maximum density of vegetation in the
absence of competition) and $\Omega$ is the {\it interaction strength parameter}. To account for purely competitive interactions
$\Omega$ and the Kernel function $G({\bf x})$ are positive.
It is important to remark, once more, that the dynamics of this model is not bounded at $\rho(x,t)=0$, 
so negative values must be avoided by imposing that $\rho=0$ when it becomes negative 
in the numerical simulations.

The stationary solutions are, $\rho_{0}=0$ (no vegetated state), and 
\begin{equation}
\rho_{0}=(1-\Omega)\kappa,
\end{equation}
that limits the values of the interaction strength parameter. It must be $\Omega \leq 1$ to verify that $\rho_{0}\geq 0$,
while in the limit $\Omega = 0$, where the are not long range interactions, the population of plants grows logistically 
until it reaches its carrying capacity.

\begin{figure}
\begin{center} 
\includegraphics[width=0.6\textwidth]{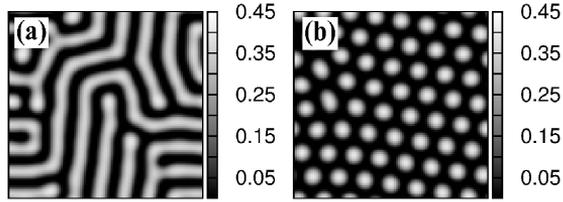}
\caption{Spatial distribution of vegetation. (a) $D=1$, $\Omega=0.7$, and (b) $D=1$, $\Omega=0.9$. $R=8$ in all the panels.}
\label{diff}
\end{center}
\end{figure}

The linear stability analysis of Eq.~(\ref{dodomodel}) gives a peturbation gorwth rate,
\begin{equation}\label{dododisp}
 \lambda({\bf k})=-Dk^{2}+(1-2\kappa^{-1}\rho_{0})+\Omega \hat{G}(k),
\end{equation}
that, in the absence of diffusion, $D=0$, allows the existence of patterns if
\begin{equation}\label{critomega}
 \Omega\ge\Omega_{c}=\frac{1}{2-\hat{G}(k_{c})},
\end{equation}
where $k_{c}$ is the most unstable mode.
Provided that we choose, again, a normalized constant kernel of width $2R$
the transition to patterns appears at $\Omega_{c}\approx0.47$.
Integrating the full Eq.~(\ref{dodomodel}) 
including long-range seed dispersal $(D\neq0)$ laberynthic and spotted patterns emerge.
This result is shown in Figures \ref{diff}(a) and \ref{diff}(b) for different values of $\Omega$.

\section{Summary and conclusions}\label{sec:summary}

In this chapter we have presented several  
models of vegetation in water-limited regions where, 
despite the absence of facilitative
interactions, regular structures may still appear: stripes 
and spots of vegetation interspersed on the bare soil forming a hexagonal lattice. 
Patterns consisting on spots of bare soil characteristic of models with competition and 
facilitation are not observed when positive interactions are neglected.
In fact, in the previous chapter, where we studied a model for mesic savannas with local
facilitation, the gapped distributions appear only when the system is 
on the transition line to patterns. This result is different from what is observed in models including nonlocal facilitation,
that show the whole sequence of patterns (gaps, stripes and spots), with each type of structure in a wide
region of the parameter space. This suggests that facilitative interactions, although not 
indispensable for the formation of patterns in arid systems, could be relevant to substain 
some of the structures that have been reported in field observations. Future work is planned to address 
this problem.

From a mathematical point of view, nonlocality enters through an {\it influence}
function that determines the number of plants competing within a range with any given plant.
A first-order approximation of this distance can be given by (twice) the typical length of the roots,
but field measurements are needed in order to determine the resources uptaking range of a given plant species.
A necessary condition for pattern transitions, common to all the presented models,
is the existence of negative values of the 
Fourier transform of the kernel function, which always happens,
but not only,
for kernel functions with finite range. 

From a biological point of view, competitive interactions alone may give
rise to spatial structures because of the 
development of regions (typically located between maxima
of the plant density) where
competition is stronger, preventing the growth of more vegetation.

An unfortunate consequence of the universal character of these models is that
the information that is possible to 
gain on the underlying biophysical mechanisms operating in the system
just by studying the spatial structures of the vegetation is limited. 
Many different mechanisms lead to the same patterns.
Although they are universal models should be specific to each system,
emphasizing the importance of empirical studies. Field work may help
theoretical efforts by placing biologically 
reasonable bounds on the shape and extent of 
the kernel functions used in the models, and 
also by approximations to the probablity of overcoming competition, $r(\tilde{\rho},\delta)$, or of 
dying because of it $h(\tilde{\rho},\delta)$.

Finally, with this work, we aim at showing that
competitive interactions alone may be responsible 
of patterns in arid to semiarid systems, regardless 
of how they are introduced in the different modeling approaches.
Under certain conditions, nonlocal competition alone may be responsible
of the formation of patterns regardless of entering in the equations either linearly or nonlinearly.
We hope that our results 
shed light on the task of 
understanding the fundamental biological mechanisms
-and the possible absence of facilitation- that
could be at the origin of pattern formation in semiarid systems. 

   \begin{appendices}
   \chapter{Linear stability analysis}\label{deriv} 

In chapters \ref{chap:mesic} and \ref{chap:semi}, we have referred many times to the linear
stability analysis of the equations in order to gain
insight into the possible emergence of patterns.
This is a broadly used technique. Its
objective is to obtain the temporal evolution of
small perturbations to the stationary homogeneous state of the
system. To this aim we assume idealized conditions that can be summarized in three points:
choosing a uniform
geometry for the walls that confine the medium; replacing 
the lateral boundaries with ones consistent with translational symmetry;
and choosing to study the linear instability towards time-independent patterns
that are consistent with translational symmetry.
We impose periodic boundary conditions and study the 
formation of stationary patterns.

Our starting point is Eq.~(\ref{model}),
\begin{equation} \label{model2}
\frac{\partial \rho({\bf x},t)}{\partial t}=\beta_0 r(\tilde{\rho},\delta)\rho({\bf x},t)(1-\rho({\bf x},t))-\alpha\rho({\bf x},t).
\end{equation}

Considering that the system is in the homogeneous stationary state, $\rho({\bf x},t)=\rho_{0}$,
how will perturbations to this state evolve with time? Will they vanish, taking the system back
to the homogeneous state, or will they grow up allowing the formation of stationary spatial structures?
The key quantity will be the growth rate of the perturbation, 
restricted to the linear regime, so that only small perturbations in the short time limit
will be considered. If a small perturbation is done to the stationary homogeneous state, the density is
\begin{equation}\label{perturbation}
\rho({\bf x},t)=\rho_{0}+\epsilon\psi({\bf x},t),
\end{equation}
with $\epsilon \ll 1$. Plugging Eq.~(\ref{perturbation}) into Eq.~(\ref{model2}), it is
\begin{equation}\label{eqdiff:pert}
 \epsilon\frac{\partial\psi}{\partial t}=\beta_0 r\left(\rho_{0}+\epsilon\int\psi({\bf x'},t)G(|{\bf x}-{\bf x'}|)dx',\delta\right)(\rho_{0}+\epsilon\psi)(1-\rho_{0}-\epsilon\psi)-\alpha(\rho_{0}+\epsilon\psi),
\end{equation}
where the spatiotemporal dependence in the perturbation $\psi$ is omitted to simplify the notation. 
We have also used that $\rho_{0}$ is stationary and thus its time derivative equals to zero together
with the normalization of the kernel to simplify
the dependence on $\rho_{0}$ in the function $r$.

Eq.~(\ref{eqdiff:pert}) can be simplified because we are interested in
small perturbations. The function $r$ can be expanded in a Taylor
series and higher order terms in $\epsilon$ neglected. This gives a 
linearized integro-differential equation for the evolution of $\psi$,
\begin{eqnarray}\label{pert_lin}
\frac{\partial \psi}{\partial t}=\beta_{0}r(\rho_{0},\delta)(1-2\rho_{0}) \psi-\alpha\rho_{0} \psi+\beta_{0}r'(\rho_{0},\delta)\rho_{0}(1-\rho_{0})\int G(|{\bf x}-{\bf x'}|)\psi({\bf x'},t)d{\bf x'}, \nonumber \\
\end{eqnarray}
that can be solved using Fourier transform.
Notice that the stationary solution of Eq.~(\ref{model2}), 
\begin{equation}\label{homo:app}
\beta_{0}r(\rho_{0},\delta)\rho_{0}(1-\rho_{0})-\alpha\rho_{0}=0,
\end{equation}
has been used to obtain Eq.~(\ref{pert_lin}). We apply the Fourier
transform to Eq.~(\ref{pert_lin}),
\begin{eqnarray}\label{transformed}
\frac{\partial \hat{\psi}({\bf k},t)}{\partial t}&=&\beta_{0}r(\rho_{0},\delta)(1-2\rho_{0}) \hat{\psi}({\bf k},t)-\alpha\rho_{0} \hat{\psi}({\bf k},t) \nonumber \\
&+&\beta_{0}r'(\rho_{0},\delta)\rho_{0}(1-\rho_{0})\hat{G}({\bf k})\hat{\psi}({\bf k},t),
\end{eqnarray}
where $\hat{\psi}({\bf k},t)=\int{\rm e}^{i k\cdot x}\psi({\bf
x},t)d{\bf x}$ is the Fourier transform of the perturbation,
and equivalently, $\hat{G}(k)$ is the Fourier transform of the
kernel.

Finally, Eq.~(\ref{transformed}) is solved considering $\hat{\psi}({\bf
k},t)\propto \exp(\lambda({\bf k})t)$, so that a linear growth rate of the perturbation is obtained
\begin{equation}\label{reldis}
 \lambda({\bf k})=\beta_{0}\left[r(\rho_{0},\delta)(1-2\rho_{0})+(1-\rho_{0})\rho_{0}r'(\rho_{0},\delta)\hat{G}({\bf k})\right]-\alpha.
\end{equation}

   \chapter{Numerical integration of Eq.~(2.19)}
\label{sec:numerical}

The integration of stochastic equations where the noise amplitude depends on the square root of the 
variable, $\rho$, and there are absorbing states (i.e, states where the system stays indefinitely),
has awaken a great interest, specially in the study of critical phenomena (i.e, properties of the system that appear
when it is close to the critical point). The 
amplitude of the fluctuations tends to zero there, and thus numerical instabilities may appear. Recently \citep{sto1, sto2}
a very efficient method has been developed, but we have used in this work an older one, presented in \citep{dickman}, since its 
implementation is easier and it gives precise results working far from the transition point. It consists on
discretizing the Langevin equation, taking a step size $\Delta\rho$ in the variable.

First of all, to integrate the Eq.~(\ref{savsto}) we discretize the space, whose area is $L_{x}\times L_{y}$, in a $N_{x}\times N_{y}$ regular grid
and compute the integral in the exponential term. It is approximated by a sum of the field evaluated in the nodes
\begin{equation}
 \int\rho(\bx,t)G(\bx-\bx')dx\approx\sum_{i=1}^{N_{x}}\sum_{j=1}^{N_{y}}\rho_{i,j}G_{i,j;i',j'}\Delta x\Delta y,
\end{equation}
where $\Delta x$ and $\Delta y$ are the spatial steps in both directions. They are given by $\Delta_{x,y}=L_{x,y}/(N_{x,y}-1)$.

Then, we integrate the temporal dependence. The key of the algorithm is to prevent $\rho+\Delta\rho$ to take negative values. From a general equation
\begin{equation}
 \frac{d\rho}{dt}=f(\rho)+\sqrt{\rho}\eta(t),
\end{equation}
where $\eta(t)$ is a Gaussian white noise with zero mean and delta correlated, it is
\begin{equation}\label{discrho}
 \Delta\rho=f(\rho)\Delta t+\sqrt{\rho}\Delta W,
\end{equation}
where $\Delta W=\sqrt{\Delta t}Y$. $Y$ is a Gaussian number with zero mean and unit variance.
To avoid negative values of $\rho+\Delta\rho$, \cite{dickman}
proposes to dicretize the density setting $\rho=n\rho_{min}$ and to simetrically truncate the Gaussian distribution from
where $Y$ is drawn, so that $|Y|\leq Y_{max}$. The negative values are avoided requiring 
\begin{equation}\label{constraint}
 Y_{max}\sqrt{\Delta t}\leq\rho_{min}.
\end{equation} 
This can be seen adding $\rho$ in both sides of Eq.~(\ref{discrho}), considering the discretization of the density and assuming that $Y$ takes its most
negative value according to the constraint (\ref{constraint}), so
\begin{equation}\label{disc5}
 n\rho_{min}+\Delta\rho=f(n\rho_{min})\Delta t+n\rho_{min}-\sqrt{n}\rho_{min},
\end{equation}
that is always positive for $n$ integer.

Eq.~(\ref{constraint}) can be verified in many ways but, following \cite{dickman} again, we use
\begin{eqnarray}
 Y_{max}&=&\frac{|\ln \Delta t|}{3}, \nonumber \\
 \rho_{min}&=&\frac{(\ln \Delta t)^{2}\Delta t}{9}.
\end{eqnarray}
Finally, rescaling the equation, we can achieve a discretized version in which positive
and zero-mean noise are ensured at the cost of a ``quantized'' density.

   \chapter{Derivation of the effective nonlocal description from tree-water dynamics}\label{app:derivation}

In this section we present a preliminar (and not fully 
satisfactory) attempt to derive the model 
presented in Section \ref{birthsec} of Chapter \ref{chap:semi} 
(the derivation corresponding to the nonlocal death model in \ref{deathsec}
is a straightforward extension of this calculation).

Let us consider a system involving dimensionless vegetation density, $\rho(\bx,t)$, and soil-water $w(\bx,t)$.
The dynamics is purely local and competitive and takes the form:   
\begin{eqnarray}
 \frac{\partial\rho}{\partial t}&=&\beta\rho(1-\rho)w-\alpha\rho, \label{biomass}\\
 \frac{\partial w}{\partial t}&=&-\mu\rho w-\gamma w+I+D_w\nabla^{2}w, \label{water}
\end{eqnarray}
where the nondimensional positive parameters are: 
the seed production rate $\beta$; the vegetation death rate 
$\alpha$; the consumption rate of water by vegetation, $\mu$;
the evaporation rate $\gamma$, 
and the rainfall, $I$. 
Water percolation in the ground is modeled by a diffusion 
constant 
$D_w$.
Note that this model is a simplified version, which only includes competitive interactions,
of the model presented in \cite{Gilad2004}.

Since the characteristic time scale of the water is much faster than the one of the biomass 
we can do an adiabatic
elimination of the variable $w$ (i.e. $\frac{\partial w}{\partial t}=0$) so that 
\begin{equation}
-\mu\rho w-\gamma w+I+D_w\nabla^{2}w=0,
\end{equation}
and thus
\begin{equation}\label{eqwat}
  \left(D_w\nabla^{2}-\gamma\right)w=\mu\rho w-I,
\end{equation}
whose formal solution can be obtained using Green's functions, $G_d$,
\begin{equation}
 w(\bx)=\int G_d(\bx-{\bf s})(\mu\rho({\bf s})w({\bf s})-I)d{\bf s},
\label{eqw}
\end{equation}
with the boundary conditions $w(x\rightarrow \pm \infty)=0$.
For simplicity we now consider a one-dimensional situation, although
analogous calculations can be done in two dimensions. 
The Green's function is the solution
of
 \begin{equation}\label{deltas}
 \left(D_w\nabla^{2} -\gamma \right)G_d=\delta({\bx}-{\bs}),
\end{equation}
and it is given by
\begin{equation}\label{kernelgreen}
 G_d({\bx},{\bs})=-\frac{1}{2}\exp\left(-\sqrt{\frac{\gamma}{D_w}}|{\bx}-{\bs}|\right)
\end{equation}
Taking the nondimensional small number
 $\mu$ as the perturbative parameter, we can further obtain
an approximate expression for $w$ from Eq. (\ref{eqw}) 
\begin{equation}
 w({\bx})=-IG_{d0}\left[1+\mu\int G_d({\bx}-{\bs})\rho({\bs})d{\bs}+{\cal O}({\mu^2})\right],
\label{eqw2}
\end{equation}
where $G_{d0}=\int G_d({\bx}-{\bs})d{\bs}<0$, since the Green's function is always negative.
 Plugging this in the equation for the biomass density  (\ref{biomass}), we obtain
the closed expression:
\begin{equation} \label{eq}
 \frac{\partial\rho}{\partial t}=\beta\rho(1-\rho)\left\lbrace-IG_{d0}
\left[\mu\int G_d({\bx}-{\bs})\rho({\bs})d{\bs}+1\right]\right\rbrace-\alpha\rho.
\end{equation}
Defining the positive nonlocal density
 $\tilde{\rho}=\int G_c({\bx}-{\bs})\rho({\bs})d{\bs}$, where $G_c=-G_d$,
 we can write Eq.~(\ref{eq}) as
\begin{equation} \label{eq2}
 \frac{\partial\rho}{\partial t}=\beta \bar r(\tilde{\rho})\rho(1-\rho)-\alpha\rho,
\end{equation}
where we have defined $\bar r(\tilde{\rho})=I|G_{d0}|\left(1-\mu\tilde{\rho}\right)$.

To have a good agreement with the effective nonlocal dynamics 
Eq.~(\ref{model}), $\bar r>0$ since it
 represents a probability. This is certainly the case
for small $\mu$. 
Note that some additional conditions on the normalization of the Green's function have to be imposed to
limit $r$ to values less than $1$. Also
$\bar r' (\tilde \rho)=-I\mu|G_{d0}|$  is always negative,
as we expected. 

In this particular example we obtained the exponential kernel of Eq.~(\ref{kernelgreen}), 
which does not have the finite-range support that would
be associated to the finite root extent. As a 
consequence, the Fourier transform of this kernel has
no negative components and then does not lead to pattern formation.  
The simple modeling of water dispersion
by means of a diffusion constant does not 
contain the additional spatial scale 
associated to root size, and should be replaced 
by some mechanism implementing root effects.
On the other side, the finite-range of the kernel 
 is a sufficient but
not a necessary condition
for its Fourier transform to have negative values. 
It is well-known the existence
of infinite-range kernels whose Fourier transform has negative values.
This is the case of all stretched exponentials 
$G(x) \propto \exp ({-|\bx|^p})$ with $p>2$ 
as has been already mentioned several times before.
Kernels satisfying this are more platykurtic than the Gaussian function. 
Work is in progress along this line.  

    \end{appendices}

\part[\textit{\textsc{Animal mobility}}]{\textbf{\textit{\textsc{Animal mobility \\}}}
         }   
 
 \label{part:move}
      \chapter{Optimal search in interacting populations}\label{chap:search}

In this chapter we investigate the relationships between search efficiency, movement strategy, and nonlocal
communication in the biological context of animal foraging. We consider situations where the members of a population of foragers perform either Gaussian jumps or L\'evy
flights, and show that the search time is minimized when communication among individuals occurs at intermediate ranges, independently of the type of movement \cite{Martinez-Garcia2014a,Martinez-Garcia2013b}. 
While Brownian strategies are more strongly influenced by the communication mechanism, L\'evy flights still result in shorter overall
search durations.

\section{Introduction}

Situations where a single individual or a group of searchers must find an object (target)
appear in many different fields including chemistry \citep{rmp_benichou},
information theory \citep{pirolli},
and animal foraging \citep{libroforaging,mendez2014random}. The study of these searching problems
has generated an increasing amount of work in the last years, many of them
oriented towards the identification of efficient strategies  
\citep{rmp_benichou,libroforaging,nature-vergassola,benPRE2006}.
Many remarkable examples can be found in the context of biological
encounters,
such as proteins searching for targets on DNA \citep{bio-taylor}, or animals searching
for a mate, shelter or food 
\citep{Campos2013,rmp_benichou, Viswanathan2008, mejiamonasterio,bartomeus}. 
In these cases, the search time is generally limiting and minimizing it
can increase individual fitness or reaction rates.

The optimality of a search strategy depends 
strongly on the nature of both the targets and the searchers
\citep{bartumeus2005animal,BartumeusPRL}. In the context of animal foraging, which is our focus here, searchers 
may move randomly, may use memory and experience to locate dispersed targets or they may also combine random search with memory-based search.
In highly social species, groups of searchers may share 
information when no single individual is sufficiently 
knowledgeable. 
This is based on \textit{the many wrong 
hypothesis} \citep{hoare, torneypnas}, that states that error in sensing of individuals
can be reduced by interacting with the rest of the group, where all individuals can act as sensors.

It is well known that individual movement plays a central role in search efficiency, and many studies have focused
on the comparative efficiency of L\'evy and  
Brownian movement strategies \citep{rmp_benichou,Viswanathan2008,bartumeus2005animal,BartumeusPRL}.
L\'evy flights are more efficient in some random search scenarios \citep{visna1999,viswaphysica}, but whether or not they
are used in real animal search strategies is still an open and contentious topic \citep{edwards2011overturning,edwards2007}.
Much less effort, however,
has been spent on trying to understand the long-range (i.e. nonlocal) interaction
mechanisms among social searchers. While diverse
observations suggest that such interactions occur in many taxa, including 
bacteria \citep{LiuPassino}, insects, and mammals \citep{dianamonkey,mccomb},
previous studies have focused almost exclusively on how the collective movements of a group of animals can emerge from
local interactions among individuals \citep{PhysRevE.86.011901,Kolpas2013,Couzin2002}.
The distance at which communication can be maintained strongly depends on the species. 
A variety of mammalian species are known to communicate acoustically over distances of up to several kilometers \citep{mccomb,callhyenas,calllions}, 
but while group formation via vocalizations has been well studied \citep{mccomb,whales,hyenas}, incidental benefits such as increased foraging
efficiency have received little research attention. Therefore many open questions remain,
particularly on the interrelation between communication and optimal search for resources. 
How can communication facilitate group formation and identification of areas of high quality resources? Does a communication range exist
that optimizes foraging efficiency? To what degree does search efficiency depend on the communication
mechanism? Finally, how does communication affect individual space use
in a heterogeneous environment?

Here, we present a theoretical model to study the effect of nonlocal communication on the 
search efficiency of a group of individuals. Communication and the amount of shared information are
directly connected through the distance travelled by the signals emitted by successful searchers (communication range).
To introduce the model, we start from an individual based description and compare
the situations where the individuals employ either L\'evy or Brownian random search strategies.
In the first section we study many properties of the model such as the existence of an optimal communication range
and the influence of the distance between targets and the movement strategy on its value.
For tractability, we will next consider a simplified, one-dimensional version of 
the model and compute analytically the search time for both
Brownian and L\'evy searchers as a function of the communication length scale. This simplified model
allows us to unveil the dependence of 
this time on both the parameters governing individual mobility, and on 
the distance between targets. Finally, a continuum description in terms of an equation for 
the density of searchers is derived. The chapter ends with a discussion comparing Brownian and L\'evy strategies.

\section{The Individual Based Model for Brownian searchers} \label{2dbrown}
We consider $N$ particles which undergo a $2D$ Brownian random walk. Correlated random walks,
often more appropriate to model directional persistence in  animal movement, reduce to Brownian motion for large spatiotemporal scales \citep{turchin}.
The movement is biased by the gradients of the landscape quality (local information), and
by the interaction among individuals through a communication mechanism that is
activated when good resources are found, thus providing information on habitat quality in other areas (nonlocal information).
The dynamics of any of the particles $i = 1,..., N$ is 
\begin{equation}\label{lange}
\dot{\mathbf{r}_i}(t)=B_{g}\nabla g(\mathbf{r}_i)+
B_{C}\nabla S(\mathbf{r}_i)+\mathbf{\eta}_{i}(t),
\end{equation}
where $\mathbf{\eta}_{i}(t)$ is a Gaussian white noise term characterized by
$\langle\eta_{i}(t)\rangle=0$, and $\langle\eta_{i}(t)\eta_{j}(t')\rangle=2D\delta_{ij}\delta(t-t')$, 
with  $D$  the diffusion coefficient.
The term $B_{g}\nabla g(\mathbf{r}_i)$ refers to the local search, where $g(\mathbf{r})$ is the
environmental quality function (amount of grass, prey, etc...)
and  $B_{g}$ is the local search bias parameter.
$B_{C}\nabla S(\mathbf{r}_i)$ is the nonlocal 
search term, with 
$B_{C}$ the nonlocal search bias parameter and $S(\mathbf{r}_i)$
is the {\it available information function} of the individual $i$. 
It represents the information arriving at the spatial position of the animal $i$ as a result
of the communication with the rest of the population. 
This term makes the individuals move along the gradients of the information received. 
It is a function of the superposition
of pairwise interactions between the individual $i$ and each one of its conspecifics,
\begin{equation}\label{colective}
 S(\mathbf{r}_i)=F\left(\sum_{j=1, j\neq i}^{N} A[g(\mathbf{r}_j)]V(\mathbf{r_i},\mathbf{r}_j)\right).
\end{equation}
$F$ is an arbitrary {\it perception function} that must be set in each application of the model,
$V(\mathbf{r}_i,\mathbf{r}_j)$ is the 
interaction between the receptor particle $i$ depending on its position 
$\mathbf{r}_i$ and the emitting particle fixed at $\mathbf{r}_j$, 
and $A[g(\mathbf{r}_j)]$ is the activation function (typically, a Heaviside function) that
indicates that the individual at $\mathbf{r}_j$ calls the others if it is
in a good habitat. 

\begin{figure}
\centering
\includegraphics[width=0.5\textwidth, clip=true]{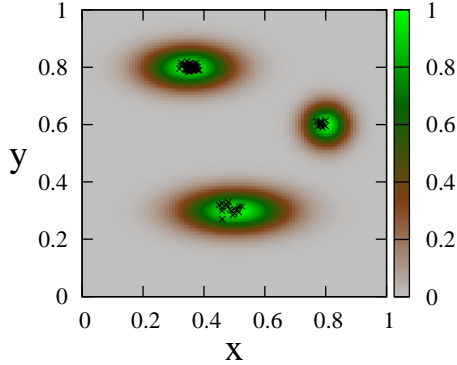}
\caption{Distribution of searchers in the long limit using as theoretical resources landscape
formed by three Gaussian patches.}
\label{distgausres}
\end{figure}

We begin with Monte Carlo simulations of the individual based dynamics in Eq.~(\ref{lange})
using a square system ($L_{x}=L_{y}=1$) with periodic boundary conditions, and a population
of $N=100$ individuals. We use a theoretical landscape quality function, 
$g(\mathbf{r})$, consisting of three 
non-normalized Gaussian functions, to ensure that $g(\mathbf{r})\in[0,1]$, centered at different spatial points (Fig.~\ref{distgausres}).
As a first approach, we consider a Gaussian-like interaction 
kernel. Manipulating its typical range via the standard deviation, $\sigma$, the information sent
by a successful searcher will travel farther and the number of individuals with which the communication
is established will be larger. 
The available information function of the individual $i$ depending on its position is
\begin{equation}\label{perteo}
S(\mathbf{r}_i)=\sum_{j=1, j\neq i}^{N}A[g(\mathbf{r}_j)]\frac{\exp\left(-\frac{(\mathbf{r}_i-\mathbf{r}_j)^{2}}
{2\sigma^{2}}\right)}{2\pi\sigma^{2}},
\end{equation}
where, as mentioned before, $A[g(\mathbf{r})]$ is a theta Heaviside function that
activates the interaction when the quality is over a certain threshold $\kappa$,
$A[g(\mathbf{r})]=\Theta(g(\mathbf{r})-\kappa)$. The perception function has been chosen as 
the identity for simplicity $F=I$. 

\begin{figure}
\centering
\includegraphics[width=0.45\textwidth, clip=true]{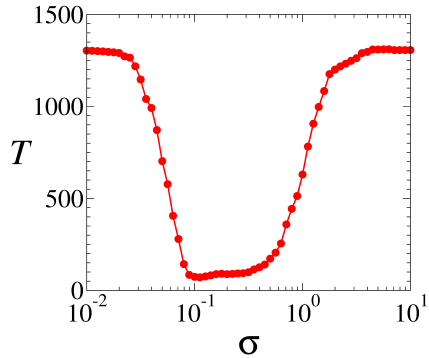}
\caption{Search time using the individual based description with $B_g=0.50$, $B_c=0.75$, $D=0.05$ and $\kappa=0.85 g_{max}$.}
\label{both}
\end{figure}

The question is
how the typical communication distance affects the average efficiency of individuals 
searching for targets in space (areas of high-quality forage). We give an answer in terms of
spatial distributions of individuals at long times starting from a random initial condition
and the mean first arrival time to the targets, $T$, as it is done in related works \citep{Benichou2005}.
This quantity (Fig.~\ref{both}) may be optimized with a 
communication range parameter, $\sigma$, of intermediate scale. 
As it was said before, the number of individuals from which a given animal receives a signal will typically
increase with the interaction scale. 
When this scale is too small, individuals receive too little information (no information when $\sigma=0$),  
and thus exhibit low search efficiency (Fig.~\ref{both}). Similarly, interaction scales that are too large lead to 
individuals being overwhelmed with information from all directions, also resulting in inefficient search (Fig.~\ref{both}). In this case, the information
received by any individual is constant over the whole space, 
so that it does not have gradients to follow. Only intermediate communication scales supply
the receiving individual with an optimal amount of information with which to 
efficiently locate the callers and the high-quality habitat areas they occupy.
The values of the threshold $\kappa$, as long as they fall within a reasonable range, 
only change the absolute time scales of the searching process.

\section{L\'evy flights}

In the case of L\'evy flights individual trajectories show sequences of short displacements interspersed with long
straight displacements. Alternatively, L\'evy flights have been shown as a good searching strategy that
may be used by some species. However, empirical studies have generated controversy, since many of 
the statistical methods used to support the presence of L\'evy flights in nature have been questioned,
and the issue remains unresolved \citep{nature_hum,edwards2007,edwards2011overturning}.
In this section the case of L\'evy searchers is considered. The results will show that neither the behavior of the model, 
nor the existence of an intermediate optimal communication scale, depend on the characteristics of the motion of the individuals.

L\'evy flights do not have a typical length scale and thus
the searcher can, in principle, make jumps as larger as the size of the system. The lengths of the jumps, $l>0$, are sorted from a 
probability distribution with a long tail \citep{metzler,klages}
\begin{equation}\label{levyasymp}
P_{\mu}(l)\approx \tilde{l}^{\mu}l^{-(\mu+1)}, \ \ l\rightarrow\infty,
\end{equation}
with $l\gg\tilde{l}$, and $0<\mu<2$, where $\tilde{l}$ is a characteristic length scale of the system.
This distribution is
not defined for $\mu<0$, its mean and variance are unbounded for $0<\mu\leq1$, and it has a mean but no variance for $1<\mu<2$. 
Finally, for $\mu\geq2$, the two first moments exist and thus it obeys the central limit theorem. The Brownian motion 
limit is recovered in this latter case, while very long jumps are more frequent when $\mu\rightarrow 0$.
This extreme is usually referred as the ballistic limit, with a high abundance of straight-line long displacements \citep{libroforaging,mendez2014random}.
The cumulative distribution corresponding to Eq.~(\ref{levyasymp}) is 
\begin{equation}\label{cumnonor}
 \Psi_{\mu}(l)\approx\mu^{-1}\left(\frac{l}{\tilde{l}}\right)^{-\mu}, \ \ l\rightarrow\infty.
\end{equation}

As a simple normalizable cumulative distribution function, with the
asymptotic behavior of Eq.~(\ref{cumnonor}), we will use \citep{Elsref}
\begin{equation}\label{cumnor}
 \Psi_{\mu}(l)=\frac{1}{\tilde{l}\left(1+\frac{l}{\tilde{l}} b^{1/\mu}\right)^{\mu}},
\end{equation}
whose probability distribution, $P_{\mu}(l)=\Psi_{\mu}'(l)$, is given by
\begin{equation}\label{levydist}
 P_{\mu}(l)=\frac{\mu b^{1/\mu}}{\tilde{l}\left(1+\frac{l}{\tilde{l}} b^{1/\mu }\right)^{\mu+1}},
\end{equation}
with $0<\mu<2$, and $b=[\Gamma(1-\mu/2)\Gamma(\mu/2)]/\Gamma(\mu)$. 
We fix $\tilde{l}=h=1$. In addition, individuals will stop if they find a target during a displacement of length $l$.
This naturally introduces a cutoff in the length of the jumps, which becomes more important as target density increases \citep{libroforaging}. 
However, as we will focus on a situation where target density is low, we introduce an exponential cutoff of the order of the system size in the jump length
probability distribution to ensure that very long jumps 
without physical meaning (they imply very high velocities)
do not occur
\begin{equation}\label{cutoffdist}
\varphi_{\mu}(l)=C\frac{\exp(-l/L)\mu b^{1/\mu}}{\left(1+l b^{1/\mu} \right)^{\mu+1}},
\end{equation}
where $C=\int_{0}^{\infty}\varphi_{\mu}(l)dl$ is the normalization constant, and $\tilde{l}=h$. 

Finally for the pairwise interaction we choose a family of functions given by 
\begin{equation}\label{twobody}
V({\bf r}_{i},{\bf r})=\exp\left(-\frac{|{\bf r}_{i}-{\bf r}_{j}|^{p}}{\sigma}\right),
\end{equation}
 where $\sigma^{1/p}$ gives the
typical communication scale. For simplicity, and without loss of generality, we will consider only the case $p=2$.
Indeed, the choice of the function $V$ is not relevant for the behavior of the model, provided that
it defines an interaction length scale through the parameter $\sigma$. This scale
must tend to zero in the limit $\sigma\rightarrow 0$ and to
infinity in the limit $\sigma\rightarrow \infty$. This assures that the 
gradient of the calling function vanishes in these limits. 

\begin{figure}
\begin{center} 
\includegraphics[width=0.65\textwidth]{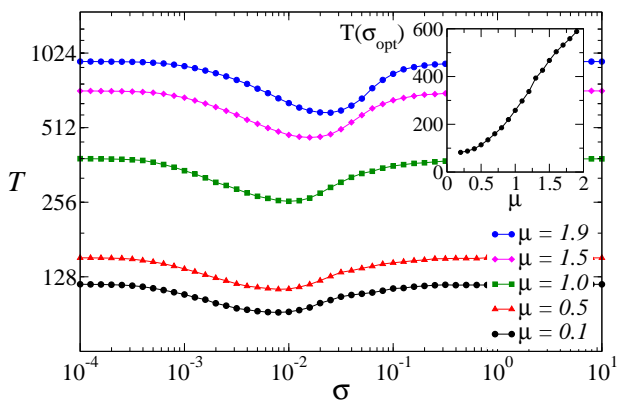}
\caption{Mean first arrival time for L\'evy flights with different values of $\mu$ in the $2D$ model.
$B_{g}=1$, $B_{c}=1$, $\tau_{0}=50$. (Inset) Mean time at the optimal communication range as a function of the L\'evy exponent, $\mu$. Lines are interpolations. }
\label{comp}
\end{center}
\end{figure}

Generally, the search is faster when long displacements occur more frequently.
Fig.~\ref{comp} shows search time for different values of $\mu$ and the 
mean searching time at the optimal communication range as a function of the L\'evy exponent (inset).
As the frequence of long displacements decreases (increasing $\mu$) the search is slower.
Again, the effect of the communication mechanism is more important when
we approach the Brownian limit ($\mu\rightarrow 2$), as will be 
explained in Sec.~\ref{sec:discussion_search}. 

\section{One-dimensional analytical approximations}\label{sec:analytics_search}
\begin{figure}
\begin{center} 
\includegraphics[width=.8\textwidth]{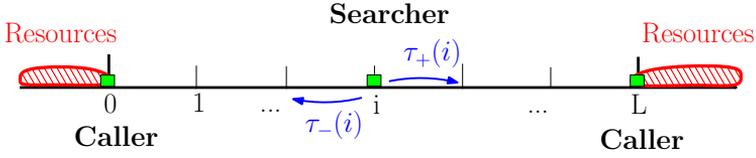}
\caption{Scheme of the simple version of the model.}
\label{scheme}
\end{center}
\end{figure}

 To gain clearer insight and provide analytical 
arguments, we study a minimalistic version of the model.
Consider a single individual in a one-dimensional space of length $L$, so
that the highest quality areas are located beyond the limits of the 
system, i.e. at $x=-1$ and $x=L+1$ (see Fig.~\ref{scheme}).
Note that this would correspond to the ideal situation where all
the members of the population but one -the searcher- have
already reached one of the targets. 
A landscape quality function, $g(x)$ must also be defined.
Provided it is a smooth, well-behaved function, its particular shape 
is not relevant. We therefore assume a Gaussian-like quality landscape,
\begin{equation}\label{qlandscape}
g(x)={\rm e}^{-\frac{(x+1)^{2}}{\sigma_{r}}}+{\rm e}^{-\frac{(x-L-1)^{2}}{\sigma_{r}}},
\end{equation}
where $\sigma_{r}$ gives the characteristic width of a high quality region.
Notice that $g(x)$ is defined so that 
highest quality areas are located, as mentioned,  at $x=-1$ and $x=L+1$.
This ensures that the gradient of the function does not vanish at the extremes of the system
(Fig.~\ref{scheme}), and it is equivalent to setting the value of the threshold $\kappa$ such that the targets
start at $x=0$ and $x=L$. We assume that a foraging area is good enough when its quality 
is higher than $80\%$ of the ideal environment, which means $\kappa=0.8$. As we center the patches of resources
at $x=-1$ and $x=L+1$, fixing a good quality treshold at $\kappa=0.8$ is equivalent to fix the width of the environmental quality function at $\sigma_{r}=4.5$,
to ensure that $g(0)=g(L)\approx0.80$. However, the qualitative behavior of the model is independent of this choice.

Finally, for the pairwise communication function we choose the functions defined in Eq.~(\ref{twobody}) with $p=2$ again.
The combination of local and nonlocal information gives the total available information for the searcher,
$R(x)=B_{g}g(x)+B_{c}S(x)$. 

To obtain analytical results, we work in the following on a discrete space.
The stochastic particle dynamics equivalent to  Eq.~(\ref{lange}) considers 
left and right jumping rates which are  defined for every individual using the total information function,
\begin{equation}\label{rates}
 \tau_{\pm}(x)=\max\left(\tau_{0}+\frac{R(x\pm h)-R(x)}{h},\alpha\right),
\end{equation}
where $\alpha$ is a small positive constant to avoid negative rates that has been given 
an arbitrary value $(\alpha=10^{-4})$, and $h$ is the spatial discretization ($h=1$).
Finally, $\tau_{0}$ is the jumping rate of an individual in absence of information, and it is related 
to the diffusion component of the dynamics of Eq.~(\ref{lange}).
Given the transition rates of Eq.~(\ref{rates}), the movement with a higher gain of information has a higher rate, and therefore a larger probability of taking 
place. 

The simplest situation, which allows an analytical treatment of the problem, is to consider only $N=3$ individuals. Two of them
are located in the top quality areas just beyond the frontiers of the system limit, $x=-1$ and $x=L+1$,
and the other one is still searching for a target. 
Under these considerations, using the environmental quality function defined in Eq.~(\ref{qlandscape}), and the 
pairwise potential of Eq.~(\ref{twobody}), the total available information for the searcher is
\begin{equation}\label{totalinfo}
 R(x;\sigma,L)=B_{g}\left({\rm e}^{-\frac{(x+1)^{2}}{\sigma_{r}}}+{\rm e}^{-\frac{(x-L-1)^{2}}
{\sigma_{r}}}\right)+B_{C}\left({\rm e}^{-\frac{(x-L-1)^{2}}{\sigma}}+{\rm e}^{-\frac{(x+1)^{2}}{\sigma}}\right).
\end{equation}

As it was done in Sec.~\ref{2dbrown}, the efficiency of the search process is measured in terms of the first arrival
time at one of the high quality areas, either at $x=0$ or $x=L$,  starting from $x_{0}=L/2$. From the definition of the transition
rates in Eq.~(\ref{rates}), $\tau_{+}(L-1)\gg\tau_{-}(L)$,
and equivalently $\tau_{-}(1)\gg\tau_{+}(0)$. This means that at both extremes of the system, the 
rate at which particles arrive is much higher than the rate at which they leave, so particles do not move
when they arrive in the top quality areas. This allows us to consider  
both extremes $x=0$ and $x=L$ of the system as absorbing, 
and the first arrival time may be obtained from the flux of presence probability of the searcher there \citep{redner}
\begin{equation}\label{meantime}
 \langle T(\sigma)\rangle=\int_{0}^{\infty}t\left(\frac{\partial P(0,t)}{\partial t}+\frac{\partial P(L,t)}{\partial t}\right)dt.
\end{equation}
This definition will be used in the following sections to investigate
the influence of sharing information (i.e., of the interaction mechanism) on search times. The results will be compared with
those obtained using a deterministic approximation of the movement of the searcher.
We study two different random strategies: Brownian and L\'evy.

\subsection{Brownian motion}
\begin{figure}
\begin{center} 
\centering
\includegraphics[scale=0.35]{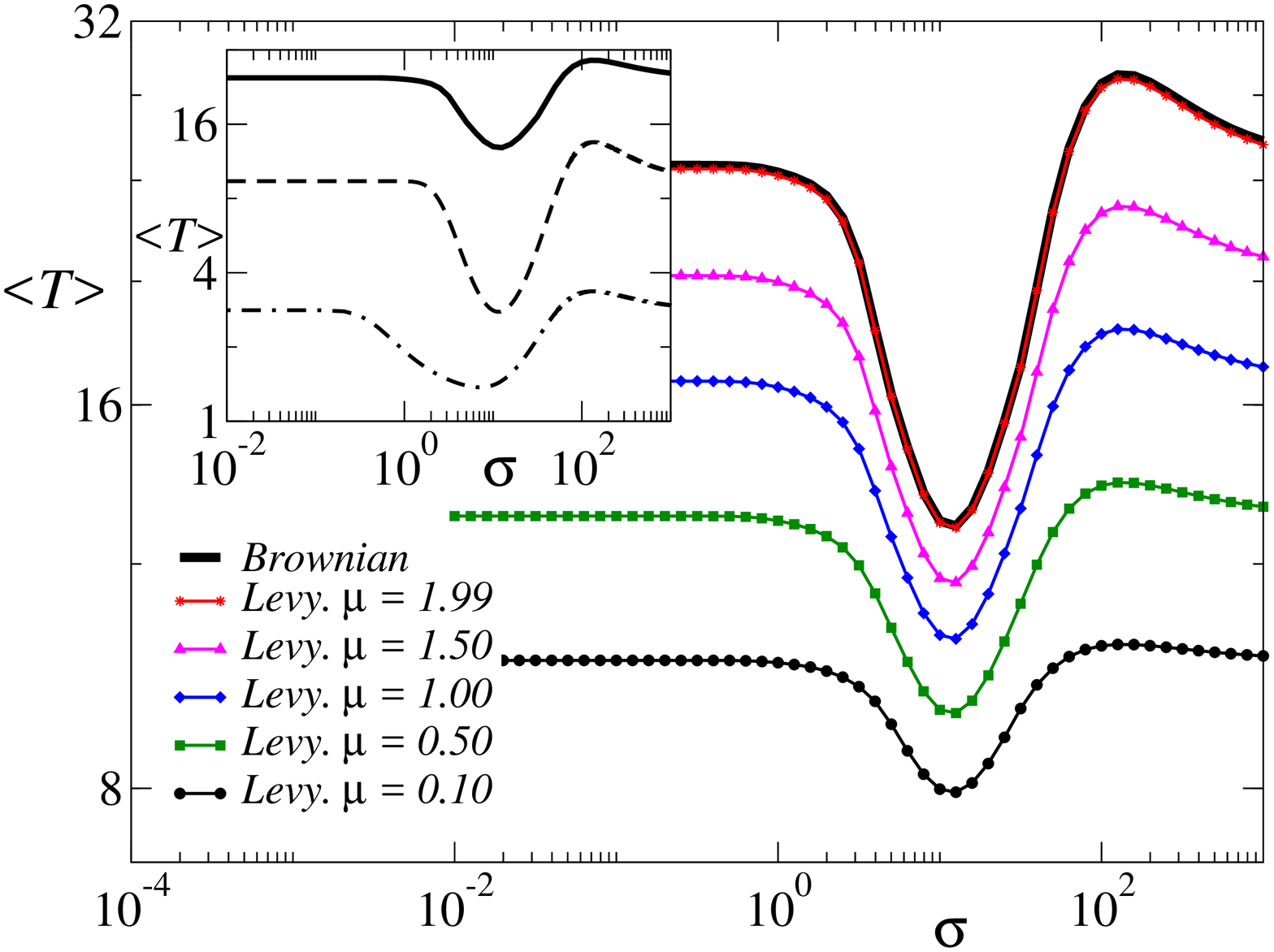}
\caption{First arrival time solving Eq.~(\ref{diffset}) for the Brownian jumps and Eq.~(\ref{diffsetlevy}) in the case of L\'evy flights for different values of $\mu$. Lines are 
interpolations. Inset: First arrival time using its definition Eq.~(\ref{meantime}) (full line) and Eq.~(\ref{time_search}) with $\epsilon=2$ (dashed line) and $\epsilon=0$ (dotted dashed line) for a Brownian searcher. 
In both panels: $L=9$, $\sigma_{r}=4.5$, $B_{g}=1$, and $B_{c}=1$.}
\label{timesana}
\end{center}
\end{figure}

We start studying the Brownian case, where the searcher only
jumps -with a given rate- to its nearest neighbors. Therefore the coupling of the set of differential equations describing the occupancy probability 
of every site of the system is (notice that lattice spacing $h=1$),
\begin{eqnarray}\label{diffset}
 \frac{\partial P(0,t)}{\partial t}&=&-\tau_{+}(0)P(0,t)+\tau_{-}(1)P(1,t), \nonumber \\
 \frac{\partial P(i,t)}{\partial t}&=&-(\tau_{+}(i)+\tau_{-}(i))P(i,t)+\tau_{+}(i-1)P(i-1,t)+\tau_{-}(i+1)P(i+1,t), \nonumber \\
 \frac{\partial P(L,t)}{\partial t}&=&-\tau_{-}(L)P(L,t)+\tau_{+}(L-1)P(L-1,t).
\end{eqnarray}
with $i=1,...,L-1$. 
If the initial position of the particle is known, it
is possible to solve Eq.~(\ref{diffset}) using the Laplace transform.
Once the probability distribution of each point has been obtained, it is possible to obtain
the mean first arrival time using Eq.~(\ref{meantime}).
The thick line in Fig.~\ref{timesana} shows this result, indicating that the searching process is optimal
(minimal time to arrive to one of the good quality areas) 
for intermediate values of $\sigma$.
A particularly simple limit in Eq.~(\ref{diffset}) appears when 
$\tau_{+}>>\tau_{-}$ when $x>L/2$ (and the contrary on the 
other half of the system). The search time is  $T(\sigma)=\frac{L}{2\tau_{+}}$.
This is the expected result since the movement is mainly in one direction
and at a constant rate.

In biological terms this means 
that the optimal situation for the individuals is to deal with intermediate amounts of information.
Extreme situations, where too much ($\sigma\rightarrow\infty$) or too little ($\sigma\rightarrow0$) information is provided by the population,
have the same effect on the mean first arrival time, which tends to the same asymptotic value in both limits.
In both cases, the search is driven only by the local perception of the environment.

This calculation gives exact results, but it implies
fixing the system size, solving a set of equations of dimension $L$, and finally
obtaining the inverse Laplace transform of the solutions. The main disadvantage of this approach
is that it is not possible to study the influence of the distance
between targets on the optimal communication length. To circumvent this we
use a deterministic approach in 
the continuum  limit $h\rightarrow 0$ 
and define, using the symmetry of the system, 
a mean drift velocity towards one of the high quality areas, $x=L$,
\begin{equation}\label{average}
\langle v_{d}(\sigma, L)\rangle =\int_{L/2}^{L}(\tau_{+}(x)-\tau_{-}(x))dx,
\end{equation}
Substituting the definition of the transition rates Eq.~(\ref{rates}), the drift velocity is,
\begin{equation}\label{drift}
 \langle v_{d}(\sigma, L)\rangle=2\left[R\left(L\right)-R\left(\frac{L}{2}\right)\right],
\end{equation}
and therefore the search time is
\begin{equation}\label{time_search}
\langle T(\sigma,L)\rangle=\frac{N/2}{\langle v_{d}(\sigma,L)\rangle}.
\end{equation}

 We compute the searching time using Eq.~(\ref{time_search})
with the same values of the parameters used before ($\sigma_{r}=4.5$, $B_{g}=1$, and $B_{c}=1$, $L=9$) 
to compare it with the results given by Eq.~(\ref{meantime})
(inset of Fig.~\ref{timesana}). The approach in Eq.~(\ref{time_search}) (dotted-dashed line) reproduces the qualitative behavior
of the searching time although underestimates the value of the optimal communication range ($\sigma_{opt}=7.2$
while Eq.~(\ref{meantime}) produces $\sigma_{opt}=12.5$). This can be fixed excluding from the average in Eq.~(\ref{average})
the boundary of the system introducing a parameter $\epsilon$ in the limits of the integration.
To estimate the value of $\epsilon$ it is useful to plot
$\tau_{+}(x)-\tau_{-}(x)$ versus $x$ (not shown). The difference between rates, although 
depending on $\sigma$, starts increasing quickly when $x\geq L-2$, so one can estimate $\epsilon=2$. 
The inset of Fig.~\ref{timesana} shows the exit time as a function of the communication range computed with this approach (dashed line).
Its optimal value is in good agreement with the result obtained using the definition
of the search time (thick line), with $\sigma_{opt}\approx12.5$ for both approaches.
However the temporal scale of the problem (the absolute values of the times), although higher
than with $\epsilon=0$,  is still lower in this calculation. Results for $\epsilon=2$ 
correspond to the dashed line in the inset of Fig.~\ref{timesana}.

Regardless of 
the value of $\epsilon$ used in the average, this approximation underestimates the 
temporal scale of the problem (the absolute values of the times).
This is because it is assumed that the searcher follows a deterministic
movement to the target neglecting any fluctuation that may slow the process.

Finally, increasing $\sigma$ beyond its optimal value, 
there is a maximum for the search time for any of the approaches.
For these values of the communication range, the nonlocal information
at the middle of the system coming from both targets is higher than in the extremes and
thus there is a bias to the middle in the movement of the searcher.
This small effect, that vanishes when $\sigma$ increases and the information tends to be constant
in the whole system, seems to be an artifact of the particular arrangement 
of the simplified $1D$ system, and does not seem relevant for any real-world consideration of this kind of model.
In addition, it does not substantially affect the dynamics because local perception of the environment
pushes the individual towards one of the targets.

\begin{figure}
\begin{center} 
\centering
\includegraphics[scale=0.30]{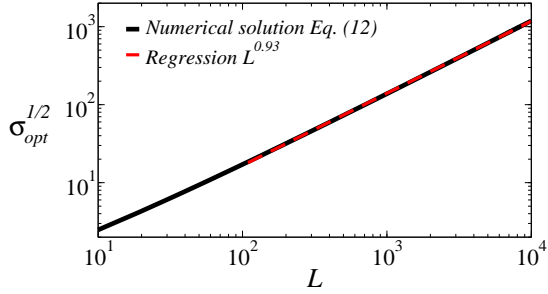}
\caption{Scaling of the optimal communication range parameter with the distance between targets (system size in the $1D$ simple model).}
\label{scaling}
\end{center}
\end{figure}

Finally, within this deterministic approximation, besides studying larger systems with no additional computational cost,
it is possible to obtain the optimal value of the interaction range parameter, $\sigma_{opt}$: 
\begin{equation}\label{optsigma}
\left(\frac{\partial T}{\partial \sigma}\right)_{\sigma=\sigma_{opt}}=0, 
\end{equation}
which has to be solved numerically for different sizes of the system. 
 The typical optimal communication scale defined by $\sigma^{1/p}$, (i.e., by $\sigma^{1/2}$ since $p=2$)
grows approximately linearly with the distance between targets in the asymptotic limit. Using a regression of the results obtained from the integration 
of Eq.~(\ref{optsigma}) yields an exponent $\sigma_{opt}^{1/2}\propto L^{0.93}$ for $L\gg1$ (Fig.~\ref{scaling}). 

\subsection{L\'evy flights.}
Proceeding as in the case of Brownian motion, considering a L\'evy searcher the 
set of equations for the probability of occupancy is 
\begin{eqnarray}\label{diffsetlevy}
 \frac{\partial P(0,t)}{\partial t}&=&\sum_{j=1}^{L}\tau_{-}(j)B_{j}P(j,t)-\tau_{+}(0)P(0,t)\left(B_{L}+\sum_{j=1}^{L-1}A_{j}\right), \nonumber \\
  \frac{\partial P(i,t)}{\partial t}&=&\sum_{j=0}^{i-1}\tau_{+}(j)A_{i-j}P(j,t)+\sum_{j=i+1}^{L}\tau_{-}(j)A_{j-i}P(j,t)- \nonumber \\
  &&\tau_{-}(i)P(i,t)\left(B_{i}+\sum_{j=1}^{i-1}A_{j}\right)-\tau_{+}(i)P(i,t)\left(B_{L-i}+\sum_{j=1}^{L-i-1}A_{j}\right),  \nonumber \\
 \frac{\partial P(L,t)}{\partial t}&=&\sum_{j=0}^{L-1}\tau_{+}(j)B_{L-j}P(j,t)-\tau_{-}(L)P(L,t)\left(B_{L}+\sum_{j=1}^{L-1}A_{j}\right), \nonumber \\
\end{eqnarray}
with $i=1,\ldots,L-1$.

We assume that if a jump of length in between $j-1$ and $j$ takes place, the individual gets the position $j$.
To this aim, the coefficients $A_{j}$ enter in the set of equations~(\ref{diffsetlevy}) and are defined as $A_{j}=\int_{j-1}^{j}\Psi_{\mu}(l)dl$. They give the probability
of a jump of length between $j-1$ and $j$ to happen. The coefficients $B_{j}$ are defined as $B_{j}=\int_{j-1}^{\infty}\Psi_{\mu}(l)dl$, to 
take into account that the searcher stops if it arrives to a target. This introduces a cutoff in the jumping length distribution Eq.~(\ref{levydist}). 

Given the size of the system, $L$, which fixes the dimension of the system of equations~(\ref{diffsetlevy}), it is possible
to obtain an analytical solution for the occupancy probabilities and the mean arrival time to the
targets using Eq.~(\ref{meantime}).
This is shown in Fig.~\ref{timesana}, where the Brownian 
limit is recovered when $\mu\rightarrow 2$. It is also observed that when long jumps are frequent the search is much faster, although the gain in
search efficiency due to the communication mechanisms is lower close to the ballistic limit (i.e., $\mu\rightarrow 0$).
This will be explained later in Sec.~\ref{sec:discussion_search}.

Similarly to the Brownian case, a particularly simple limit in Eq.~(\ref{diffsetlevy})
appears when 
$\tau_{+}>>\tau_{-}$ for $x>L/2$ (and the contrary on the 
other half of the system). The search time is 
\begin{equation}
T(x=L,\sigma)\propto\frac{1}{\tau_{+}}, \nonumber \\
\end{equation}
where the proportionality constant is a combination of the coefficients $A_i$ that depends on the size of the system.

\section{Continuum approximation}
From the Langevin Eq.~(\ref{lange}), and following the standard
arguments presented in \cite{Dean1996} and \cite{Marconi1999} it is possible to
write an equation for the evolution of the
density of individuals, $\rho (\mathbf{r},t)$ \footnote{In Appendix \ref{app:macroder}
we show this derivation in detail.}. This approach will allow us to
fix the parameters of the problem having a better understanding 
of the role they are playing in the dynamics through a dimensional analysis.
 However, in the case of the large grazing mammals we are going to study later,
it is not very suitable to describe a population as a continuum since 
the number of individuals is not very high and the typical distances among them is large.
Neglecting fluctuations the continuum equation for the density is 
\small
\begin{eqnarray}\label{macro}
\frac{\partial\rho(\mathbf{r},t)}{\partial t}=D\nabla^{2}\rho(\mathbf{r},t)+B_{g}\nabla \left[\rho(\mathbf{r},t)\nabla g(\mathbf{r})\right]+B_{c}\nabla\left[\rho(\mathbf{r},t)\nabla F \left(\int d\mathbf{r'}\rho(\mathbf{r},t) A[g(\mathbf{r'})]
V(\mathbf{r},\mathbf{r'})\right)\right], \nonumber \\
\end{eqnarray}
\normalsize
which is quite similar to the one derived in \cite{Savelev2005} to study the 
transport of interacting particles on a substrate.

\begin{figure}
\centering
\includegraphics[width=0.45\textwidth, clip=true]{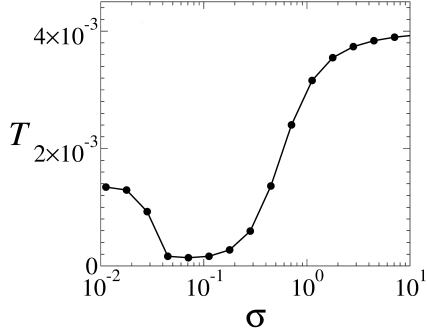}
\caption{Search time using the macroscopic equation with $B_g=0.50$, $B_c=50$, $D=0.75$ and $\kappa=0.85 g_{max}$.}
\label{macrotime}
\end{figure}

\begin{figure}
\begin{center} 
\centering
\includegraphics[trim=0 0 0 0.4, clip,scale=0.30]{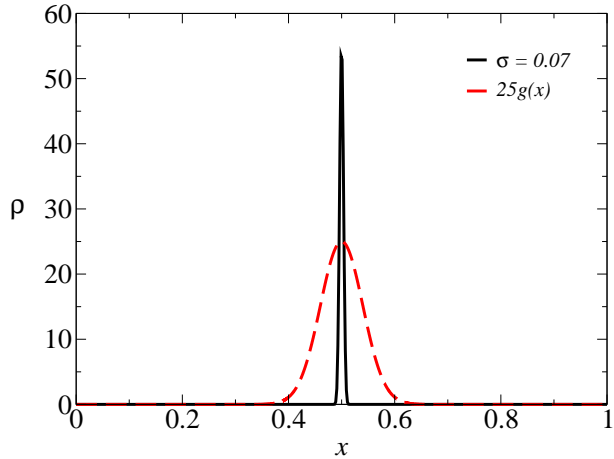}
\caption{Density distribution in the long time limit (black full line). The resources distribution (red dashed line) 
fas been multiplied by a factor $25$ for clarity.}
\label{macrodistributions}
\end{center}
\end{figure}

The same  behavior is also shown by the macroscopic 
Eq.~(\ref{macro}) (Fig.~\ref{macrotime}). 
Now $T$ is defined as the time that passes until
half of the population has found a target, that is
$\int_{g(\mathbf{r})\geq\kappa}\rho(\mathbf{r},t)d\mathbf{r} \geq N/2$.
We have integrated the Eq.~(\ref{macro}) in $1D$ system of length $L=1$, using a single Gaussian patch of 
resources centered at $L/2$ and periodic boundary conditions
for a random initial condition. This is equivalent
to the case of an infinite system with equidistant high quality areas.
We have taken the calling bias as being much stronger than the resource 
bias to make the nonlocal mechanism much more important in the
search process and thus easier to see how the communication 
range parameter affects the search time. The differences between 
the $2D$ individual-based and the $1D$ deterministic density equation description, coupled with the parameter choices
(stronger bias in the density equation), explain the different observed time-scales 
in Figs.~\ref{both} and \ref{macrotime}. Due to the simplicity of resources landscape, the stationary distribution of
individuals is Dirac delta peaked at the maximum of resources (the center of the system). In Fig.~\ref{macrodistributions}
we show the density of individuals in the long time limit (not stationary).
All the animals are in good habitats, i.e., in areas where the maxima of the $g$ function occur.

\section{Brownian jumps vs. L\'evy flights}\label{sec:discussion_search}

As a general result of the model, searching is faster when individuals have intermediate 
amounts of information, regardless of the kind of movement strategy
followed by the population (Brownian or L\'evy). However,
communication has a larger impact on Brownian motion, i.e.,
the depth of the well at $\sigma_{opt}$ is larger (Figs.~\ref{comp} and \ref{timesana}).

\begin{figure}
\begin{center} 
\includegraphics[width=0.45\textwidth]{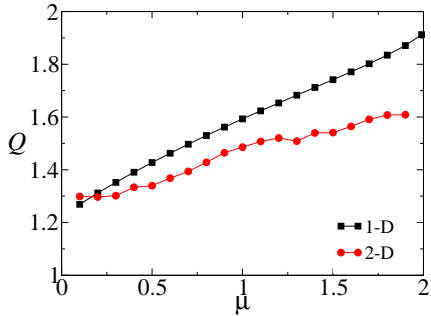}
\caption{Improvement of the searching process because of the communication mechanism. Circles correspond 
to the $2D$ model and squares to $1D$. Lines are interpolations.}
\label{mejora}
\end{center}
\end{figure}

A measure of the improvement in search performance at the optimal communication range is given
by the ratio between the search time without communication and that at 
the optimal communication range, $Q=T_{\sigma\rightarrow0}/T_{\sigma_{opt}}$. 
This quantity is plotted in Fig.~\ref{mejora} for different L\'evy exponents.
As previously mentioned, Brownian searchers that are not able to perform long displacements benefit more
from communication than L\'evy searchers. This is because introducing an additional source of information
increases the directionality of the random motion and prevents the searcher from revisiting the same place many times, which
is the key problem with Brownian search strategies \citep{libroforaging}. A Brownian walker
has no directionality in the movement, so provided with sources of information (communication together with the local 
quality of the landscape) it can search much more efficiently. This effect is less important
for L\'evy searchers due to the presence of long, straight-line moves that, by themselves, decrease the number
of times that a particular area is revisited.
In addition the return probability to a given point is much higher in $1D$ than in $2D$. Therefore the directionality 
introduced by the communication has a stronger effect in the simple $1D$ scenario that we have studied. It is also important 
to remark that in this case the walker only can move either to the right or to the left at each step. This will make the influence of the
bias due to communication much stronger in the jumping probabilities.

In summary, the communication mechanism is less important in L\'evy strategies, so that its effect is less noticeable
as it is shown in Fig.~\ref{mejora} both in $1D$ and $2D$. 

\begin{figure}
\begin{center} 
\includegraphics[width=0.6\textwidth]{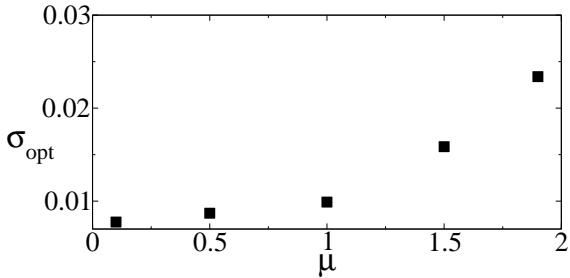}
\caption{Optimal communication range as a function of the L\'evy exponent.}
\label{scalingopt}
\end{center}
\end{figure}

However, the value of the optimal interaction range
changes with the kind of motion. This is shown in the $2D$ model by the dependence of the mean
search time on the communication range for different L\'evy exponents (Fig.~\ref{comp}).
The value of $\sigma_{opt}$ increases with the L\'evy exponent, 
so Brownian searchers $(\mu\rightarrow2)$ need
to spread the information farther (a larger value of $\sigma_{opt}$)
than L\'evy $(\mu=1)$ walkers to obtain the maximum benefit.
In Fig.~\ref{scalingopt} we show the value of the optimal communication range, $\sigma_{opt}$, as a function of the L\'evy exponent.
L\'evy trajectories show clusters of short displacements with frequent turns occasionally broken up by long linear displacements,
which account for most of the target encounters. However, because these steps are often much longer than the average distance between targets
they are not positively influenced by communication, so any benefit a L\'evy strategy gains from communication occurs during the series
of short displacements. The time that an individual spends doing short movements is limited by the interarrival time of the large steps, so 
unless an individual is already relatively close to a target, it will not have time to reach a target before the next big step comes and
moves it far away from that original target. Therefore the
optimal communication range decreases with decreasing L\'evy exponent, $\mu$, as longer displacements become more frequent at lower $\mu$ values. 

In addition, the value of $\sigma_{opt}$ depends on both the number of targets and their spatial distribution,
as was shown in Sec.~\ref{sec:analytics_search} for a simple $1D$ situation where $\sigma_{opt}\sim L$.

\section{Summary and conclusions}\label{search:summmary}

In this chapter we compared
Brownian and L\'evy search strategies using a population of individuals 
that exchange information about the location of spatially distributed targets. 
Using a simple $1D$ model we have provided analytical results on both cases, concluding that
frequent long jumps ($\mu\rightarrow 0$, ballistic limit) minimize the searching times.

 However the effect of a communication mechanism is more pronounced in the limit of short 
jumps i.e., Brownian motion. This means that a population of individuals employing Brownian motion
gains proportionally more benefit from communicating and sharing information than does a population of
L\'evy walkers, where long jumps are more or less frequent depending on the value of the L\'evy exponent $\mu$.
When messages are exchanged in a range that minimizes search
duration, communication is the driving force in the Brownian limit, but 
occasional long jumps are still responsible for most of the encounters with targets in the case of long-tailed
step-length distributions.

The  main result of this work is rather general: 
independently of the kind of communication performed by the population, and of the spatial
distribution of the targets, a population of individuals with the ability to communicate will find the 
targets in a shorter time if the information is spread at intermediate ranges. Both an excess and a lack of information increase the search time.
However, the communication mechanism does not have the same quantitative effect on the different moving strategies (i.e., ballistic, L\'evy or Brownian).
Uninformed Brownian individuals perform a random movement revisiting the same position many times, so having an external source of information
introduces directionality on the movement, decreasing the number of times that a point in the space is visited. In the case of L\'evy and ballistic 
strategies $(\mu\rightarrow0)$, communication is less noticeable because individuals are able to do long jumps.
This is already a source of directionality that prevents individuals from revisiting 
the same points in space many times, and thus weakening the effect of the directionality introduced by communication.
      \chapter{Foraging in \textit{ Procapra gutturosa}}

In this chapter we show an application of the model presented in Chapter \ref{chap:search} to the particular case of acoustic
communication among Mongolian gazelles ({\it Procapra gutturosa}) \cite{Martinez-Garcia2013b},
for which data are available, searching for good habitat areas. Using Monte Carlo simulations,
our results point out that the search is optimal (i.e. the mean first hitting time among searchers is minimum)
at intermediate scales of communication. We also present this result in terms of the frequency of the sounds, showing
a good agreement with field measurements of the sounds emitted by these gazelles in the wild.
The formation of groups in the populations is also studied.

\section{Introduction}
Many living organisms, including bacteria \citep{LiuPassino}, insects, and mammals \citep{dianamonkey,mccomb}
communicate for a variety of reasons including facilitation of social cohesion \citep{cap, pfefferle}, 
defense against predators \citep{predator}, maintenance of territories \citep{stamps,frey2}, and to pool information
 on resource locations when no single individual is sufficiently knowledgeable \citep{bee,hoare,berdahl,simons,torneypnas}.
 Communication among individuals frequently leads to group formation \citep{eftimie}, which often has clear direct benefits such as
 reducing individual vulnerability to predators. Such strategies may, however, also have important incidental benefits.
 For example, an individual that has found a good foraging patch might try to attract conspecifics to reduce its risk of
 predation, but also provides its conspecifics with information on the location of good forage, thus increasing the foraging
 efficiency of those responding to the call.

\begin{figure}
\centering
\includegraphics[width=0.80\textwidth, clip=true]{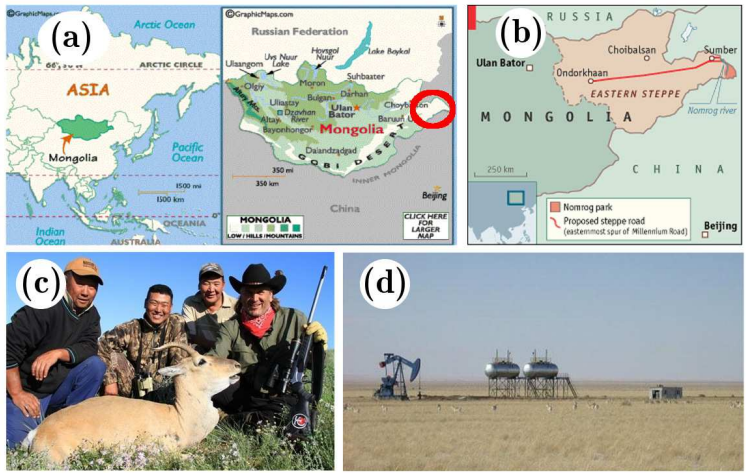}
\caption{(a) Location of the Eastern Steppe. (b) Construction of roads in the step has caused a habitat fragmentation. (c) Hunting is one of
the major threats to Mongolian gazelles. (d) Oil and mineral explorations in the Steppe.}
\label{gazelle}
\end{figure}

We apply a specialized version of the model introduced in Chapter \ref{chap:search} to the particular case of acoustic communication among
Mongolian gazelles, the dominant wild herbivore in the Eastern Steppe of Mongolia (Fig.~\ref{gazelle}a) among 14 species of ungulates. 
A population of about one milion of animals is estimated, but it is difficult to give a good measurement
because of large fluctuations due to the extreme conditions in the steppe that cause periods with important population losses.
In addition, the nomadism of the species, travelling long distances during the year, makes more difficult a demographic control.
In any case, the species is still recognized as \textit{one of Asia's largest remaining wildlife populations} \citep{Olson2005}, although
it has experienced a major reduction in range during the past century, and is further threatened by excessive hunting and continued habitat loss and fragmentation
(grassland steppes are increasingly being carved up by fences, roads, agriculture, and densely settled areas while 
oil fields and pipelines are being developed in the region) (Fig.~\ref{gazelle} b-d).
In fact, Mongolian gazelles were formerly distributed across the whole area of the Republic of Mongolia but the range of this species, between the 1940s and 1960s,
was reduced by 70\% owing to excessive official hunting and poaching. Nowadays, although individuals or small groups are found across a wider geographical range,
higher concentrations
of this gazelle species are now limited to the Eastern Steppe (Fig.~\ref{gazelle}a) where they avoid narrow valleys, forest, sand dunes or cultivated fields
unless driven there by exceptional circumstances. The plant cover of the dry Steppes of Eastern Mongolia is extremely sparse, generally 5–20\%  (Fig.~\ref{gazelle}, right panel), 
rarely reaching 30–40\% \citep{freygebler,Mueller2008}.

\begin{figure}
\centering
\includegraphics[width=0.80\textwidth, clip=true]{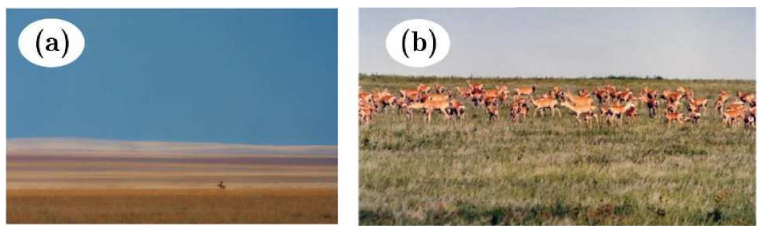}
\caption{(a) Typical landscape in the steppe. (b) Group of gazelles grazing in the Eastern Steppe.}
\label{fig2}
\end{figure}

In summary, gazelles must find each
other to form grazing groups less susceptible to preadator's attacks (Fig.~\ref{gazelle}), and relatively small areas of good forage
in a vast landscape where sound can travel substantial distances \citep{frey}(Fig.~\ref{fig2}a) .
We aim to explore whether acoustic communication in the steppe could lead to the formation of observed large aggregations of animals (Fig.~\ref{fig2}b)
\citep{olson2009}, and how search efficiency depends on the distances over which calls can be perceived. We wonder
if the frequency of the voice of the gazelles is optimal to communicate in the steppe, and
if the call length-scales that optimize search in real landscapes are biologically and physically plausible. 
To do this, we couple an individual-based representation of our model with remotely-sensed data on resource quality in the Eastern Steppe.

\section{The model for acoustic communication}

A detailed analysis of gazelle relocation data has shown that, over the spatiotemporal scales relevant 
to searching for resources (days to weeks), the movement of Mongolian gazelles can be closely approximated
by simple Brownian motion for the spatiotemporal scales involved in foraging. Therefore, the starting point of the modeling is
\begin{equation}\label{lange2}
\dot{\mathbf{r}_i}(t)=B_{g}\nabla g(\mathbf{r}_i)+B_{C}\nabla S(\mathbf{r}_i)+\mathbf{\eta}_{i}(t),
\end{equation}
where $\mathbf{\eta}_{i}(t)$ is a Gaussian white noise term to characterize Brownian motion. The function $\mathbf{\eta}_{i}(t)$
is defined by its statistical properties:
its mean value $\langle\eta_{i}(t)\rangle=0$, and the correlation $\langle\eta_{i}(t)\eta_{j}(t')\rangle=2D\delta_{ij}\delta(t-t')$. 
$D$ is  the diffusion coefficient. The terms $B_{g}\nabla g(\mathbf{r}_i)$ and $B_{C}\nabla S(\mathbf{r}_i)$ are referred to the local
and the nonlocal search, as it was defined in Chapter \ref{chap:search}. These terms drive the movement to
the best grazing areas.

The function $g(\mathbf{r})$ quantifies the habitat quality in the Eastern Steppe of Mongolia through the Normalized
Difference Vegetation Index (NDVI), one of the most widely used vegetation quality estimators. It can be calculated from satellite imagery, 
and has been already applied to gazelle habitat associations in the Mongolian Steppe \citep{Mueller2008}. 
NDVI is characterized by the function $g_{d}({\bf r})$, a
continuous function taking values between $0$ (no vegetation)  and $1$ (fully vegetated).
As the vegetation at low NDVI is too sparse, and at high NDVI is too mature and indigestible,
gazelles typically seek forage patches characterized by intermediate NDVI values \citep{Mueller2008}.
To make gradients of resources drive the movement of the individuals to
regions with intermediate NDVI values, we apply to the data a linear transformation:
\begin{eqnarray}
f(x)= \left\{ \begin{array}{lcc}
  g_{d}({\bf r}) & {\rm if} & g_{d}({\bf r})<0.5 \\ 
 1-g_{d}({\bf r}) & {\rm if} & g_{d}({\bf r})\geq 0.5    
              \end{array}
   \right.
\end{eqnarray}

The new function $g({\bf r})$ defines a resources landscape with values between $0$ and  $0.5$, where $0$ represents both
fully vegetated and no vegetation (i.e., low quality forage).
We study a subregion with an extension of around $23000$~km$^{2}$ (Coordinates: $46.50\text{\textdegree}$N, $114.15\text{\textdegree}$E; 
$46.50\text{\textdegree}$N, $116.80\text{\textdegree}$E; $47.65\text{\textdegree}$N, $114.15\text{\textdegree}$E;
$47.65\text{\textdegree}$N, $116.80\text{\textdegree}$E) and assume that the resources remain constant in time during the search.
It is crucial now to properly choose the perception function in order to 
realistically model the case of gazelles performing {\it acoustic communication}.
It is well known that the sensitivity of the response of the ear does not follow a linear scale, but approximately a logarithmic one. That is 
why the bel and the decibel are quite suitable to describe the acoustic perception of a listener.
Therefore we choose an acoustic perception function of the form
\begin{equation}
 S(\mathbf{r}_i)=10\log_{10}\left(\frac{\sum_{j=1, j\neq i}^{N}A[g(\mathbf{r}_j)]V(\mathbf{r}_i,\mathbf{r}_j)}{I_{0}}\right),
\end{equation}
where $F$ is taken as a logarithmic function. The sound calling of $j$, $V(\mathbf{r},\mathbf{r}_j)$, plays the role of a two body interaction potential, and
$I_{0}$ is the low perception threshold. We take the value of a human ear, $I_{0}=10^{-12}~W~m^{-2}$, which
is similar for most other mammals \citep{fletcher}, and in any case, is just a reference value on which our results will not depend.
The interaction potential mimicking acoustic communication is
\begin{equation}
 V(\mathbf{r}_i,\mathbf{r}_j)=\frac{P_{0}}{4\pi}\frac{{\rm e}^{-\gamma|\mathbf{r}_i-\mathbf{r}_j|}}{|\mathbf{r}-\mathbf{r}_j|^{2}},
\end{equation}
considering that sound from an acoustic source
attenuates in space mainly due to the atmospheric absorption (exponential term), and the spherical spreading
of the intensity ($4\pi r^{-2}$ contribution), and neglecting secondary effects \citep{naguib}.
 $P_{0}$ may be understood as the power of the sound at a distance
of $1~m$ from the source. According to Stoke's law of sound attenuation \citep{fletcher}, the absorption coefficient, $\gamma$, is given by 
\begin{equation}
\gamma=\frac{16\pi^{2}\eta\nu^{2}}{3\rho v^{3}},      
\end{equation}
where $\eta$ is the viscosity of the air, $\rho$ its density, $v$ the propagation velocity of
the acoustic signal (which depends on the 
temperature and the humidity), and $\nu$ its frequency. We work under environmental conditions of $T=20^{o}C$, and
relative humidity of 
$HR=50\%$, which are quite close to the corresponding empirical values for the summer months from the Baruun-Urt (Mongolia) weather station,
averaged over the last 4 years. These values give an absorption coefficient of $\gamma\approx10^{-10}\nu^{2}~m^{-1}$. The inverse
of the absorption coefficient, $\gamma^{-1}$, gives the typical length scale for the communication at each frequency, and thus plays the same
role as the standard deviation, $\sigma$, did in the Gaussian interaction used in the general model. From its functional dependence, different values of the frequency
will modify the value of the absorption coefficient, and consequently, will lead to different communication ranges. Therefore, we will use 
sound frequency, $\nu$, as the control parameter of the interaction range.

\begin{figure}
\centering
\includegraphics[width=0.50\textwidth, clip=true]{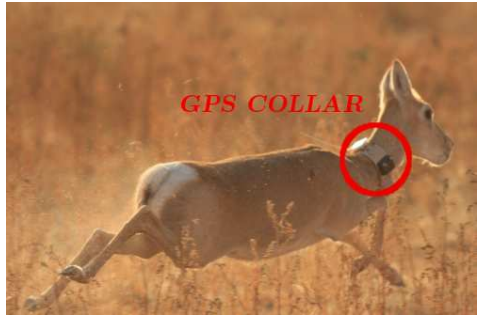}
\caption{Gazelle with a GPS collar.}
\label{gps}
\end{figure}

From a statistical analysis of GPS data tracking the positions of $36$ gazelles between 2007 and 2011,
we estimate a diffusion constant of $D=74$~km$^{2}$~day$^{-1}$ (Fig.~\ref{gps}). To give empirically-based values to the bias parameters, we define a 
drift velocity, and based on previous field work \citep{Mueller2008} we set $ v_{drift}=B_{g}\nabla g(\mathbf{r})+B_{c}\nabla S(\mathbf{r})=10~{\rm km~day}^{-1}$.
The local search mechanism is responsible for short-range slow movements, while nonlocal communication gives rise to long and faster movements, and
thus we require $B_{g}\nabla g(\mathbf{r})\ll B_{c}\nabla S(\mathbf{r})$.

\section{Results and discussion}

\begin{figure}
\begin{center} 
\includegraphics[width=0.5\textwidth]{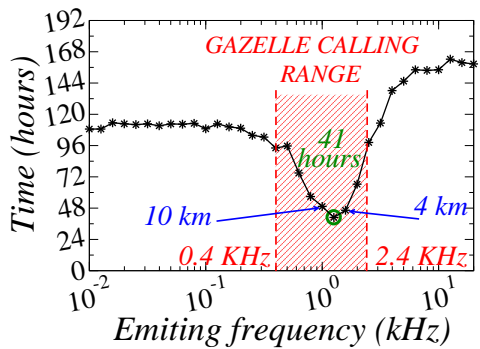}
\caption{ Mean arrival time for $500$ gazelles (averaged over $50$ realizations with different initial conditions).
Parameter values: $D=74$~km$^{2}$~day$^{-1}$, $B_{g}=2.6\times10^{-3}$~km$^{3}$~day$^{-1}$, $B_{c}=13$~km$^{2}$~day$^{-1}$, $\kappa=0.70g_{max}$.}
\label{kernels}
\end{center}
\end{figure}

\begin{figure}
\begin{center} 
\includegraphics[width=0.85\textwidth]{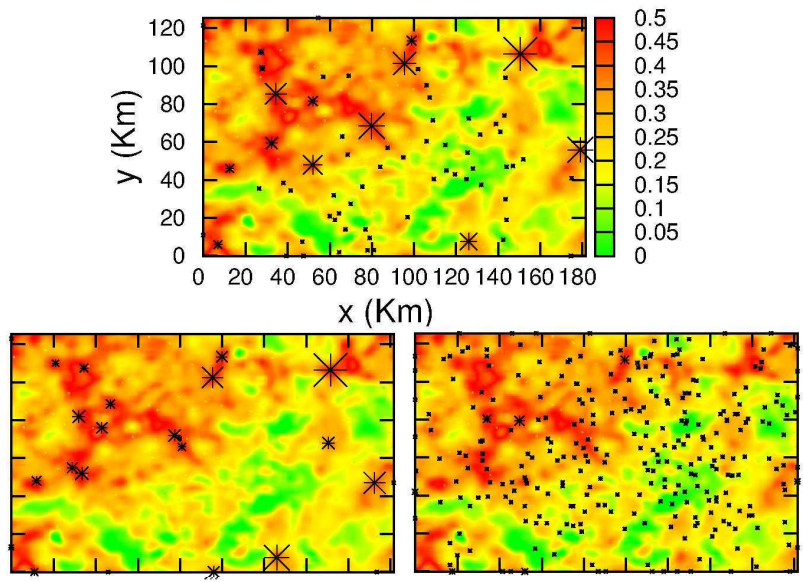}
\caption{Spatial distribution of $500$ gazelles after $1$ month (reflecting boundary conditions). $\nu=0.1$~kHz (top), $\nu=1$~kHz (bottom left), $\nu=15.8$~kHz (bottom right).
The size of the star is related to the size of the group at a position. Real data resources landscape.}
\label{100}
\end{center}
\end{figure}

We explore the dependence of this metric on the communication length, $\gamma^{-1}$,
or equivalently the frequency, $\nu$ (Fig.~\ref{kernels}). To this aim,
we couple an individual-based model following the dynamics of Eq.~(\ref{lange2}), 
with a data-based resources landscape sampled every $500$~m (shown in Fig.~\ref{100}),
and quantify the efficiency of the search for areas of high quality resources in terms of
the mean first arrival time of the population.
Similarly to other species, such as lions \citep{calllions} or hyenas \citep{callhyenas}, the optimal
foraging time ($41$~hours) is obtained for $\gamma^{-1}$ of the order of kilometers (around $6$~km).
This result cannot currently be checked with data. However, switching to frequencies, the optimal search is 
obtained when gazelles communicate at a frequency of $1.25$~kHz,
which lies inside the measured interval of frequencies of the sounds emitted by gazelles, $[0.4, 2.4]$~kHz \citep{freygebler,frey}.
This means that the search is optimal when the receiving individual has an intermediate 
amount of information. A lack of information leads to a slow, inefficient search, while an overabundance of
information makes the individual to get lost in the landscape.
These different regimes are also observed in the long time
spatial distributions (i.e. efficiency of the search in terms of quality) of the Fig.~\ref{100}.
For frequencies out of the optimal range, either smaller (Fig.~\ref{100} top) or
larger (Fig.~\ref{100} bottom right), some animals are still in low-quality areas at the end of the simulation period.
At intermediate communication scales, $\nu=1$~kHz, (Fig.~\ref{100} bottom left) all of the animals end up in
regions with the best resources, regardless of where they started from.

In summary, communication over intermediate scales results in faster search, 
and all the individuals form groups in areas of good resources.
While this has obvious advantages in terms of group defense and predator swamping,
it will also lead to rapid degradation of the forage (and thus habitat quality) at those locations. 
This is the problem of foraging influencing the patterns
of vegetation, which could be treated in future investigations.
Shorter-scale communication implies an almost individual search, which helps preserve local forage quality, 
but has clear disadvantages in terms of group defense strategies. On the other hand,
longer scales lead to the formation of big groups (faster degradation of foraging), 
and animals need more time to join a group,
which has negative consequences against predation. 
Furthermore, acoustic communication scales
significantly larger than the optimal scale for foraging efficiency identified here would be
biologically implausible, even if ultimate group size (and not rate of group formation) was
the most important aspect of an antipredation strategy.

Our study clarifies some questions on the relationship between 
communication and optimal search for resources. Our key result is that,
in general, intermediate communication distances optimize search
efficiency in terms of time and quality. Individuals are able to find the best quality
resource patches regardless of where they start from,  opening new questions about the distribution 
of individuals in heterogeneous landscapes.
The existence of maximum search efficiency at intermediate communication ranges 
is robust to the choice of functional form of $V(\mathbf{r})$, allowing 
the model to be generalized to many different ways of sharing information.
Also considering different species on the model (preys and predators or males and females  to studying
the case of mating) would be interesting extensions of this work.
Finally, regarding to the formation of groups because of communication among individuals, exploring tradeoffs between group 
defense and individual foraging efficiency in highly dynamic landscapes may be a promising avenue for future research.

     \begin{appendices}
    \chapter{Derivation of the macroscopic Eq.~(4.3)}\label{app:macroder}

In this appendix we will show the derivation of the macroscopic equation (\ref{macro})
in Chapter \ref{chap:search}, starting 
from the Langevin equation for the movement of a singles individual.

Considering a single individual, the Langevin equation is
\begin{equation}
 \dot{\mathbf{r}}_{i}(t)=B_{g}\nabla g(\mathbf{r})+B_{C}\nabla S_{i}(\mathbf{r})+\eta_{i}(t),
\end{equation}
where $\eta_{i}(t)$ is a Gaussian white noise with zero mean and correlation
delta-correlated in space and time. The available information function,
$S_{i}$ is given by 
\begin{equation}
 S_{i}(\mathbf{r})=F\left(\sum_{j=1, j\neq i}^{N} A[g(\mathbf{r}_j)]V(\mathbf{r},\mathbf{r_{j}})\right).
\end{equation}

To obtain the equation for the density of individuals, we will derive a density equation 
for the case of a single particle and then extend the result to a population 
with $N$ individuals \citep{Dean1996}.

In the case of a single particle the density is given by
\begin{equation}\label{deltadef}
 \rho(\mathbf{r},t)=\delta(\mathbf{r}-\mathbf{X_{i}}(t)),
\end{equation}
where $\mathbf{X_{i}}(t)$ is the stochastic trajectory of the particle.
Then, using this equation \ref{deltadef} for the density of searchers and the definition of the Dirac delta,
\begin{equation}
 f(\mathbf{X_{i}}(t))=\int d\mathbf{r}\rho_{i}(\mathbf{r},t)f(x)=\int d\mathbf{r}\delta(\mathbf{r}-\mathbf{X_{i}}(t))f(x),
\end{equation}
where $f(\mathbf{r})$ is an arbitrary function.
Its time derivative is
\begin{equation}\label{eq4}
 \frac{df(\mathbf{X_{i}}(t))}{dt}=\int d\mathbf{r}\delta(\mathbf{r}-\mathbf{X_{i}}(t))\frac{df(x)}{dt}.
\end{equation}

Using the It\^{o}'s formula Eq.~(\ref{eq4}) can be expanded,
\small
\begin{equation}
 \frac{df(\mathbf{X_{i}}(t))}{dt}=\int d\mathbf{r}\rho_{i}(\mathbf{r},t)\left[D\nabla^{2}f(\mathbf{r})
 +B_{g}\nabla g(\mathbf{r})\nabla f(\mathbf{r})+B_{c}\nabla S_{i}(\mathbf{r})\nabla f(\mathbf{r})+\nabla f(\mathbf{r})\eta_{i}(t)\right],
\end{equation}
\normalsize
and then, rearranging and integrating by parts each term\footnote{It is straightforward just choosing $\nabla f(x)=dV$ in the integration by parts. In the case 
of the laplacian term, we must integrate twice.},
\begin{eqnarray}\label{comp1}
 \frac{df(\mathbf{X_{i}}(t))}{dt}&=&\int d\mathbf{r}f(\mathbf{r})\left[D\nabla^{2}\rho_{i}(\mathbf{r},t)+B_{g}\nabla \left(\rho_{i}(\mathbf{r},t) \nabla g(\mathbf{r})\right)\right] \nonumber \\
 &+&\int d\mathbf{r}f(\mathbf{r})\left[B_{c}\nabla \left(\rho_{i}(\mathbf{r},t)\nabla S_{i}(\mathbf{r})\right)+\nabla (\rho_{i}(\mathbf{r},t)\eta_{i}(t))\right].
\end{eqnarray}

On the other hand the time derivative of $f(\mathbf{r})$ can be written as
\begin{equation}\label{comp2}
 \frac{df(\mathbf{X_{i}}(t))}{dt}=\int d\mathbf{r}f(\mathbf{r})\partial_{t}[\rho_{i}(\mathbf{r},t)].
\end{equation}

As both Eq.~(\ref{comp1}) and (\ref{comp2}) are true in the case of an arbitrary function, $f(\mathbf{r})$, it is possible
to write:
\begin{equation}\label{rhoalpha}
 \frac{\partial\rho_{i}(\mathbf{r},t)}{\partial t}=D\nabla^{2}\rho_{i}(\mathbf{r},t)+B_{g}\nabla \left[\rho_{i}(\mathbf{r},t)\nabla g(\mathbf{r})\right]+B_{c}\nabla \left[\rho_{i}(\mathbf{r},t)\nabla S_{i}(\mathbf{r})\right]+\nabla [\rho_{i}(\mathbf{r},t)\eta_{i}(t)].
\end{equation}

Finally, we neglect the last stochastic term to get a deterministic approximation as Eq.~(\ref{macro}).

    \chapter{Voronoi diagrams of the model.} \label{app_voronoi}
\begin{figure}
\begin{center} 
\includegraphics[width=.75\textwidth]{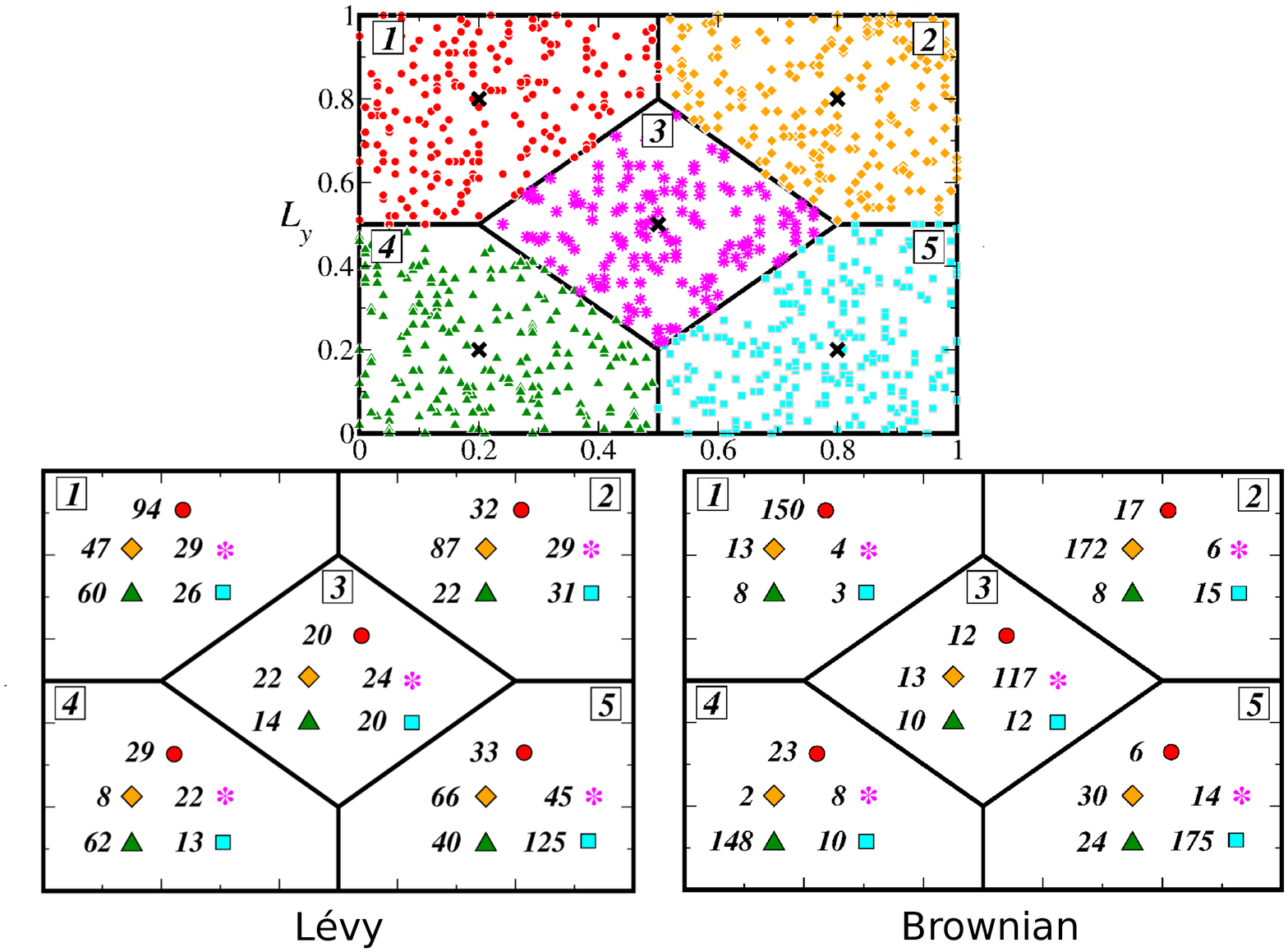}
\caption{(Color online). (Top) Initial random distribution of individuals, the symbol refers to the Voronoi cell at which every individual 
belongs initially.  (Bottom left) Number of individuals coming from each cell at each target at the end of the search using L\'evy flights.
(Right) Number of individuals coming from each cell
 at each target at the end of the search using Brownian motion. Parameters: $\sigma=0.01$ (optimal communication
range), $B_{g}=1$, $B_{c}=1$, $\tau_{0}=50$. The black crosses represent the location of the $5$ targets.}
\label{voronoi}
\end{center}
\end{figure}

 The behavior of the model, resulting in optimal searches at intermediate communication ranges,
can be explained in terms of Voronoi diagrams \cite{voronoi}. Consider every target as a seed that has 
associated a Voronoi cell formed by those points whose distance to that seed is less than or equal to
its distance to any other one (See Figure \ref{voronoi} (top) for a distribution
of the space in $5$ Voronoi cells for an initial distribution of particles with five targets (crosses)).
The searching time will be minimized when the information coming from the individuals located on one
target covers the full associated Voronoi cell, but only that cell.
In this situation, the searchers within that cell will receive information coming only from that target
 and move towards it. $\sigma_{opt}$ is the communication range that maximizes the gradient 
 (approximately the smallest value of $\sigma$ that makes
 the calling function not vanishing) of 
 the calling function at the frontiers of the Voronoi cells.
 Increasing the communication range provides individuals with information 
 coming from different targets, and makes them get overwhelmed in the limit $\sigma\rightarrow\infty$.
This Voronoi construction may also help to explain
the improvement of the searching strategies because of sharing
information.
The difference between Brownian and L\'evy strategies can be seen 
 in Figure \ref{voronoi} (Bottom).
They show the origin of the individuals that are at each target at the end of a L\'evy (Left) and a 
Brownian search (Right) (i.e., in which Voronoi cell they were at the beginning).
In the case of Brownian individuals most of the particles at every target were initially in its Voronoi cell. For
L\'evy flights the long displacements mix the population in the stationary state
(i.e., individuals at a target come from different cells). The 
communication mechanism is less important in L\'evy strategies, so that its effect is less noticeable and the encounters
of individuals with targets are caused mainly by the long displacements.

     \end{appendices}

  \part[\textit{\textsc{Temporal fluctuations}}]{\textbf{\textit{\textsc{Temporal fluctuations \\}}}\label{part:tempfluc}
         }

 \label{part:fluctuations}
     \chapter{Temporal disorder in up-down symmetric systems}\label{chap:tempdis}

In this chapter we study the effect of temporal fluctuations on systems with up-down symmetry 
through the behavior of the first-passage times \cite{Martinez-Garcia2012,Martinez-Garcia2011}. This is a relevant question in the modeling of ecosystems,
since they are subject to environmental changing conditions. Therefore, it is important to have models that include temporal disorder.
We analyze two well-known families of phase transitions in statistical physics —the Ising
and the generalized voter universality classes— and scrutinize the consequences of
placing them under fluctuating global conditions. It is observed that the variability
of the control parameter induces in both classes “temporal Griffiths phases” (TGPs),
characterized by broad regions in the parameter
space in which the mean first-passage times scale algebraically with system size.
In an ecological context, first-passage times are related to typical extinction
times, and studying how they are affected by the size of the system
(e.g. habitat fragmentation) is a problem of outmost relevance.

\section{Introduction}

Systems with up-down $Z_{2}$ symmetry --including the Ising model-- are
paradigmatic in mathematical ecology. They allow to address a big variety of problems
ranging from species competition \citep{spcomp} and neutral theories
of biodiversity \citep{Durret} to allele frequency in genetics \citep{allele}.
Some of them, such as the voter model, exhibit absorbing states,
a distinctive feature of nonequilibrium dynamics. Once these particular configurations
are reached, the system cannot escape from them so they imply the presence of currents
\citep{hinri,odor,GG,marro}. 

Phase transitions into absorbing states are quite universal and they depend on few
general properties of the system, such as symmetries and its dimensionality, and are insensitive to the 
underlying microscopic properties. 
This universality makes possible to establish a classification of the phase transitions into different classes.
Those systems exhibiting one absorbing state belong generically to the so called
Directed Percolation (DP) universality class and share the same set of
critical exponents and scaling functions. 
However, when there is some additional symmetry or conservation,
the phase transition exhibits critical scaling differing from DP. This
is the case of systems with two symmetric absorbing
states \citep{hinri,odor,GG,marro}, that show a phase transition usually referred to as Generalized voter (GV).
In an ecological context, this is a relevant class of models that can be used in many
situations with two equivalent species.

Analytical and numerical studies
\citep{dornic,droz,alhammal1,Vazquez-2008,Blythe} have shown that,
depending on some details such GV transition can split
into two separate ones: an Ising-like transition in which the up-down
symmetry is broken, and a second DP-like transition below which the
broken-symmetry phase collapses into the corresponding absorbing
state.  In particular, a general stochastic theory, aimed at capturing
the phenomenology of these systems, was proposed in \cite{alhammal1};
depending on general features they may exhibit a DP, an Ising, or a GV
transition.

When they try to mimick ecological systems, these models should not be isolated but,
instead, affected by external conditions or by environmental
fluctuations. The question of how external variability affects
diversity, robustness, and evolution of complex systems, is of outmost
relevance in ecology. Take, for instance, the example of
the neutral theory of biodiversity: if there are two $Z_2$-symmetric
(or neutral) species competing, what happens if depending on
environmental conditions one of the two species is favored at each
time step in a symmetric way?  Does such environmental variability
enhance species coexistence or does it hinder it?
\citep{Leigh,Vazquez-2010,BorgognoRG}.

Motivated by these questions, we study how basic properties of
up-down symmetric systems, such as response functions and
first-passage times, are affected by the presence of temporal
disorder.

Some previous works have explored from a theoretical point of view the effects of fluctuating global
conditions in simple models that exhibit phase transitions
\citep{jensen,alonso,kamenev}. Recently, a modified version of the simplest
representative of the DP class --i.e. the Contact Process-- equipped
with temporal disorder was studied in \cite{Vazquez-2011}. In this
model, the control parameter (birth probability) was taken to be a
random variable, varying at each time unit. As the control parameter
is allowed to take values above and below the transition point of the
pure contact process, the system alternates between the tendencies to
be active or absorbing. As shown in \cite{Vazquez-2011} this dynamical
frustration induces a logarithmic type of finite-size scaling at the
transition point and generates a subregion in the active phase
characterized by a generic algebraic scaling of the extinction times with system size. More
strikingly, this subregion is also characterized by generic
divergences in the system susceptibility, a property which is reserved
for critical points in pure systems.  This phenomenology is akin to the
one in systems with quenched ``spatial'' disorder \citep{Votja-2006},
which show algebraic relaxation of the order parameter, and
singularities in thermodynamic potentials in broad regions of
parameter space: the so-called, Griffiths Phases
\citep{griffiths}. 

In order to investigate whether the anomalous behavior that leads to
TGPs around absorbing state (DP) phase transitions is a universal
property of systems in other universality classes --and in particular,
in up-down symmetric systems-- we study the possibility of having TGPs
around Ising and GV transitions. We scrutinize simple models in these
two classes and assume that the corresponding control parameter
changes randomly in time, fluctuating around the transition point of
the corresponding pure model, and
study mean-first passage times. 
    
\section{Spatial disorder. Rare regions and the Griffiths Phase.}\label{qdis}

The presence of noise is an intrinsic property of natural system and it may change its behavior 
when compared with an ideal situation. Knowing whether and how the critical behavior changes when introducing 
a small amount of impurities is important in order to apply criticality to real systems. This is the case of the brain, where
Griffiths Phases and Lifshitz tails could play a relevant role \citep{moretti2013griffiths}. 

\begin{figure}
\centering
\includegraphics[width=0.6\textwidth]{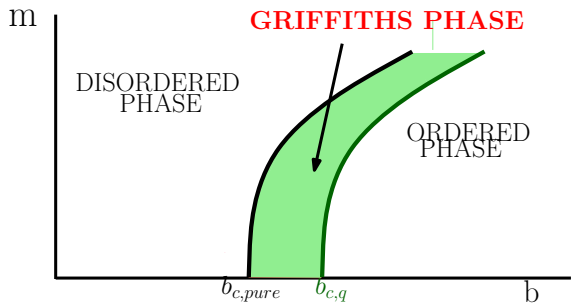}
\caption{The Griffiths Phase.}
\label{gp}
\end{figure}

Will the phase transition remain only at one point in presence of disorder or will the critical point split?
If so, will the critical behavior change quantitatively, giving new universality classes with new critical exponents, or even qualitatively with new
non-power law scalings at criticality? In this section we will review some of these questions in systems with quenched disorder 
i.e. depending on the spatial variables but that does not evolve in time \citep{odor,Votja-2006}.
One of the most common ways of introducing quenched disorder in a system is the {\it dilution}, that is,
the absence of spins in some fixed places of the lattice. 
The dilution reduces the tendency
towards magnetic long-range order in the system. Therefore, the critical value of the control parameter 
(typically the inverse of the temperature) for the pure model
(without noise), $b_{c,pure}$, moves into the ordered phase, $b_{c,q}$ (Fig. \ref{gp}).
That is, the transition to an ordered system is at a lower temperature.

On the other hand,
in the case of infinite systems, as happens in the thermodynamic limit, it is possible to find regions without vacancies of
an arbitrary size, regardless of the impurity concentration, that is the number of vacancies in the whole lattice.
When the value of the control parameter is between $b_{c,pure}$ and $b_{c,q}$, although the whole system is in the disordered
phase these pure regions can exhibit some
local order, which means a local non-vanishing value of the magnetization. These pure spatial regions are called {\it rare regions}
and the probability
of finding them decreases exponentially with its size $V_{RR}$ and the impurity concentration, $p$.
In addition, the dynamics in these regions is very slow since a coherent change (fluctuation) is needed in order to flip all the spins therein.

\begin{figure}
\centering
\includegraphics[width=0.7\textwidth]{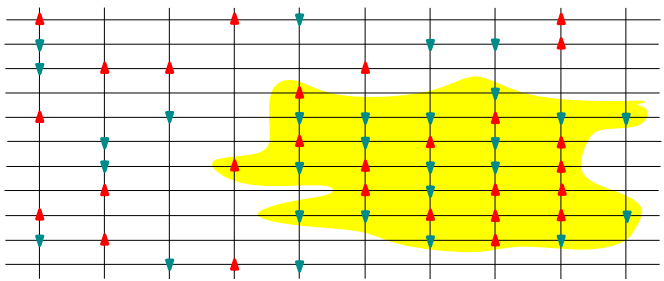}
\caption{Pure region, without vacancies, in a diluted spin interacting model. Red up triangles represent up spins. Dark cyan down triangles represent down spins.}
\label{rare}
\end{figure}

The interval between $b_{c,pure}$ and $b_{c,q}$, in the disordered phase, (Fig.~\ref{gp})
is the so-called {\it Griffiths Phase}, because it was Robert B. Griffiths the first who showed the possible existence of a singularity in
the free energy in this region \citep{griffiths2}. Its main characteristics are the 
generic divergences of the susceptibility, as a consequence of
the singularity in the free energy, and an anomalously slow relaxation to zero of the order parameter. Other time-dependent 
quantities also relax specially slow, mainly as a power law or a stretched exponential,
in contrast with the fast decay typical of pure systems, usually exponential.

In a given system with impurity concentration $p$, the probability of finding a rare region 
decreases exponentially with its d-dimensional volume, $V_{RR}$.
Calling $P_{RR}$ the probability of finding a rare region, up to constant factors, it is
\begin{equation}\label{rareprob}
 P_{RR}\propto {\rm e}^{-pV_{RR}},
\end{equation}
so it is very unlikely to find large rare region.
In contrast, the long-time dynamics inside the Griffiths phase is dominated by these regions.
Consider for instance the temporal evolution of the order parameter, typically magnetization in magnetic systems, $m$.
As its long time behavior is governed by rare regions, it is
\begin{equation}\label{mgp}
 m(t)\propto\int dV_{RR}P_{RR}{\rm e}^{-t/\xi_{t}(V_{RR})},
\end{equation}
where $\xi_{t}$ is the flipping time of a rare region, and increases exponentially with its size, 
so it is $\xi_{t}(V_{RR})\propto\exp(bV_{RR})$. The integral in Eq.~(\ref{mgp}) can be solved using
a saddle-node integration to obtain the slow relaxation of the magnetization typical of Griffiths
Phases
\begin{equation}
 m(t)\propto t^{-\phi},
\end{equation}
where $\phi$ is a non-universal exponent.

\section{Mean-field theory of $Z_2$-symmetric models with temporal disorder}
\label{MFmodels}

Interacting particle models, such as ecosystems, evolve stochastically over time. A useful technique
to study such systems is the mean-field (MF) approach, which implicitly
assumes a well-mixed situation, where each particle can interact with any
other,  providing a sound approximation in high dimensional systems.
One way in which the mean-field limit can be seen at work is by analyzing
a fully connected network (FCN), where each node
(particle) is directly connected to any other else, mimicking an infinite
dimensional system.
 
In the models that we study here, states can be labeled with occupation-number
variables $\rho_i$ taking a value $1$ if node $i$ is occupied or $0$
if it is empty, or alternatively by spin variables $S_{i}=2 \rho_i-1$,
with $S_{i}=\pm 1$. Using these latter, the natural order parameter is
the magnetization per spin, defined as
\begin{equation}
\label{magdef}
 m=\frac{1}{N}\sum_{i=1}^{N}S_{i},
\end{equation} 
where $N$ is the total number of particles in the system. Its value varies between $-1$ an $1$, both of them corresponding
to a fully ordered state, with the intermediate value $m=0$ reflecting a disordered system with the same amount of spins up and down.
The rate
equation for the probability $P(m,t)$ of having magnetization $m$ at a given time $t$, is 
\begin{eqnarray}
\label{prob} 
P(m,t+1/N) &=& \frac{\omega_{+}\left(m-2/N,b\right)}{N}
P\left(m-2/N,t\right) \\
&+&\frac{\omega_{-}\left(m+2/N,b\right)}{N}P\left(m+2/N,t\right) \nonumber \\ 
&+&\left[1-\frac{\omega_{-}(m,b)}{N}-\frac{\omega_{+}(m,b)}{N}\right]P(m,t)+\mathcal{O}(N^{-2}), \nonumber 
\end{eqnarray} 
where $\omega_{\pm}(m,b)$ are the transition probabilities from a
state with magnetization $m$ to a state with magnetization $m \pm
2/N$.

In the limit of $N\rightarrow\infty$ ($dt\rightarrow 0$) Eq.~(\ref{prob}) becomes a Master equation (Sec.~\ref{sec:masterq}),
\begin{eqnarray}\label{mastereqfluc}
\frac{\partial P(m,t)}{\partial t}&=& \omega_{+}\left(m-2/N,b\right)
P\left(m-2/N,t\right) \\
&+&\omega_{-}\left(m+2/N,b\right)P\left(m+2/N,t\right) \nonumber \\ 
&-&\left[\omega_{-}(m,b)+\omega_{+}(m,b)\right]P(m,t)+\mathcal{O}(N^{-2}). \nonumber 
\end{eqnarray}

This describes a process in which a ``spin'' is randomly
selected at every time-step (of length $dt = 1/N$), and inverted with
a probability that depends on $m$ and the control parameter $b$.  The
allowed magnetization changes in an individual update, $\Delta m=\pm
2/N$, are infinitesimally small in the $N\rightarrow\infty$ limit.
In this limit one can perform a Taylor expansion of the rates around $m$, leading to the Fokker-Planck equation 
\begin{equation} 
\label{fokker} 
\frac{\partial P(m,t)}{\partial
t}=-\frac{\partial}{\partial m}\left[f(m,b)P(m,t)\right]
+\frac{1}{2}\frac{\partial^{2}}{\partial m^{2}}\left[g(m,b)P(m,t)\right],
\end{equation} 
with drift and diffusion terms given, respectively, by
\begin{eqnarray} 
f(m,b)&=&2\left[\omega_{+}(m,b)-\omega_{-}(m,b)\right],
\label{langedrift} \\ \nonumber \\
g(m,b)&=&\frac{4\left[\omega_{+}(m,b)+\omega_{-}(m,b)\right]}{N}. 
\label{langediff}
\end{eqnarray}
From Eq.~(\ref{fokker}), and working in the It\^{o} scheme (as
justified by the fact that it comes from a discrete in time equation
\citep{Horsthemke1984}), its equivalent Langevin equation is
\citep{gardiner}
\begin{equation}
\label{langevin}
 \dot{m}=f(m,b)+\sqrt{g(m,b)} \, \eta(t),
\end{equation}
where the dot stands for time derivative, and $\eta(t)$ is a Gaussian
white noise of zero-mean and correlations
$\langle\eta(t)\eta(t')\rangle=\delta(t-t')$. The diffusion term is
proportional to $1/\sqrt{N}$, and therefore, it vanishes in the
thermodynamic limit ($N \to \infty$), leading to a deterministic
equation for $m$ (Sec.~\ref{sec:metmean}).

\begin{figure}
\centering 
\includegraphics[width=0.7\textwidth]{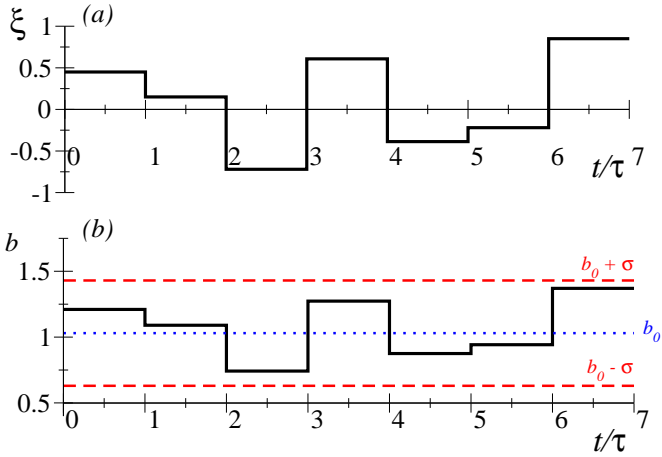}
\caption{(a) Typical realization of the colored noise $\xi(t)$, a step
  like function that takes values between $+1$ and $-1$. (b)
  Stochastic control parameter $b(t)=b_{0}+\sigma\xi(t)$ according to
  the values of the noise in (a), $b_{0}=1.03$ and $\sigma=0.4$.}
\label{noise}
\end{figure}

\begin{figure}[H]
\centering 
\includegraphics[width=0.7\textwidth]{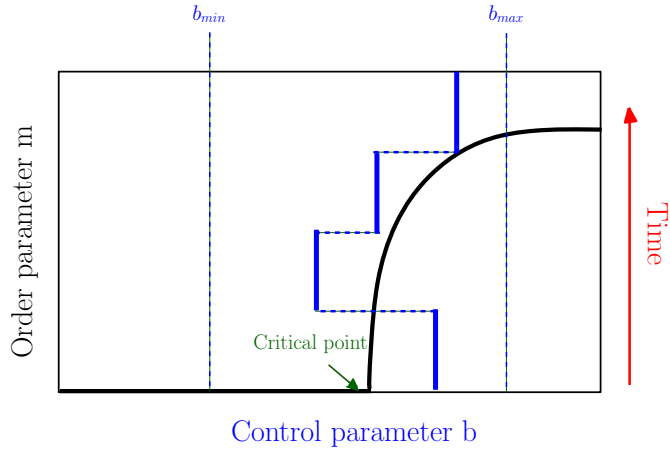}
\caption{Schematic representation of the fluctuating control parameter
  in the Ising model with Glauber dynamics. The system shifts
  between the ordered and the disordered phases.}
\label{noiseising}
\end{figure}

The drift and diffusion coefficients in Eq.~(\ref{langevin}) depend
not only on the magnetization, but also on the parameter $b$. To
analyze the behavior of the system when $b$ changes randomly over
time, and following previous works \citep{Vazquez-2010,Vazquez-2011},
we allow $b$ to take a new random value, extracted from a uniform
distribution, in the interval $(b_0-\sigma,b_0+\sigma)$ at each MC
step, i.e. every time interval $\tau = 1$.
Thus, we assume that the dynamics
of $b(t)$ obeys an Ornstein-Uhlenbeck process \begin{equation}
 b(t)=b_{0}+\sigma \, \xi(t),
\label{controlrandom}
\end{equation}
where $\xi(t)$ is a step-like function that randomly fluctuates
between $-1$ and $1$, as depicted in Fig.~\ref{noise}a. Its average correlation is
\begin{eqnarray}
\label{corr}
 \overline{<\xi(t)\xi(t+\Delta t)>} = \left\{ \begin{array}{ll}
\frac{1}{3}(1-|\Delta t|/\tau) & \mbox{for $|\Delta t|<\tau$}\\ \\
0 & \mbox{for $|\Delta t| >\tau$,}
\end{array}
\right.
\end{eqnarray}
where the bar stands for time averaging.  The parameters $b_0$ and
$\sigma$ are chosen with the requirement that $b$ takes values at both
sides of the transition point of the \emph{pure model} (see
Fig.~\ref{noise}b), that is, the model with constant $b$.  Thus, the
system randomly shifts between the tendencies to be in one phase or
the other (see Fig.~\ref{noiseising}).

The model presents both \emph{intrinsic} and \emph{extrinsic}
fluctuations, as represented by the white noise $\eta(t)$ and the
colored noise $\xi(t)$, respectively.  Plugging Eq.~(\ref{controlrandom}) for $b(t)$ into Eq.~(\ref{langevin}), and
retaining only linear terms in the noise one readily obtains
\begin{equation} 
\label{2nois}
\dot{m}=f_{0}(m)+\sqrt{g_{0}(m)} \, \eta(t)+j_{0}(m) \, \xi(t),
\end{equation}
where $f_{0}(m) \equiv f(m,b_0)$, $g_{0}(m) \equiv g(m,b_0)$ and
$j_{0}(m)$ is a function determined by the functional form of
$f(m,b)$, that might also depend on $b_{0}$.  To simplify the
analysis, we assume that relaxation times are much longer than the
autocorrelation time $\tau$, and thus take the limit $\tau \to 0$ in
the correlation function Eq.~(\ref{corr}), and transform the external
colored noise $\xi$ into a Gaussian white noise with effective
amplitude $K
\equiv\int_{-\infty}^{+\infty}\overline{<\xi(t)\xi(t+\Delta t)>}
\,d\Delta t = \tau/3$.  Then, we combine the two white noises into an
effective Gaussian white noise, whose square amplitude is the sum of
the squared amplitudes of both noises \citep{gardiner}, and finally
arrive at
\begin{equation}
\label{finallange}
\dot{m}=f_{0}(m)+\sqrt{g_{0}(m)+Kj_{0}^{2}(m)} \, \gamma(t),
\end{equation}
where $\langle \gamma(t) \rangle=0$ and $\langle \gamma(t) \gamma(t')
\rangle=\delta(t-t')$.

In the next two sections we analyze the dynamics of the kinetic Ising
model with Glauber dynamics and a variation of the voter model (the,
so-called, q-voter model) --which are representative of the Ising and
GV transitions respectively-- in the presence of external noise. For
that we follow the strategy developed in this section to derive
mean-field Langevin equations and present also results of numerical
simulations (for both finite and infinite dimensional systems), as
well as analytical calculations.

\section{Ising transition with temporal disorder}
\label{secising}

We consider the kinetic Ising model with Glauber dynamics
\citep{Glauber-1969}, as defined by the following transition rates
\begin{equation}\label{transgl}
 \Omega_{i}(S_{i}\rightarrow-S_{i})=\frac{1}{2}\left[1-S_{i}\,
{\rm \tanh}\left(\frac{b}{2d}\sum_{j\in \langle i \rangle}S_{j}\right)\right].
\end{equation}
The sum extends over the $2d$ nearest neighbors of a given spin $i$ on
a $d$-dimensional hypercubic lattice, and $b=J\beta$ is the control
parameter. $J$ is the coupling constant between spins, which we set to
$1$ from now on, and $\beta=(k_{B}T)^{-1}$. Note that $b$ in this case
is proportional to the inverse temperature.

\subsection{The Langevin equation}
\label{Lang-Eq}

In the mean-field case, the cubic lattice is replaced by a fully-connected network in
which the number of neighbors $2d$ of a given site is simply $N-1$.  Then, the 
transition rates of Eq.~(\ref{transgl}) can be expressed as
\begin{equation}
\label{transglmf}
 \Omega_{\pm}(m,b) \equiv \Omega(\mp \to \pm) = 
\frac{1}{2}\left[1\pm{\rm tanh}\left(b \, m\right)\right].
\end{equation}
which implies 
$ \omega_{\pm}(m,b)=\frac{1 \mp m}{2} \, \Omega_{\pm}(m,b)$
for jumps in the magnetization. Following the steps in the previous
section, and expanding $\Omega_{\pm}$ to third order in $m$, we obtain
\begin{equation} \label{isinglangevin}
 \dot{m}=a_{0}m-c_{0}m^{3}+\sqrt{\frac{1-b_{0} m^{2}}{N}+
K\sigma^{2}m^{2}(1-b_{0}^{2}m^{2})^{2}}\,\,\gamma(t),
\end{equation}
where $b_{0}$ is the mean value of the stochastic control parameter,
$a_{0}\equiv b_{0}-1$, and $c_{0}\equiv b_{0}^{3}/3$.

\begin{figure}[H]
\centering
  \includegraphics[width=0.4\textwidth]{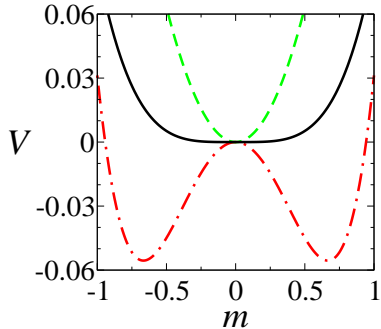}
  \caption{Potential for the Ising transition in a mean field approach.
    The dashed, solid and dot-dashed lines correspond to the
    paramagnetic phase, critical point, and the ferromagnetic phase,
    respectively.}
\label{potentialising}
\end{figure}

The potential $V(m) = -\frac{a_{0}}{2} m^2 +\frac{c_{0}}{4}m^{4}$
associated with the deterministic term of Eq.~(\ref{isinglangevin})
has the standard shape of the Ising class, that is, of systems
exhibiting a spontaneous breaking of the $Z_2$ symmetry (Fig.~\ref{potentialising}). A single
minimum at $m=0$ exists in the disordered phase, while two symmetric
ones, at $\pm\sqrt{a/c}$ exist below the critical point.

\subsection{Numerical results}

\begin{figure} \centering
  \includegraphics[width=0.7\textwidth]{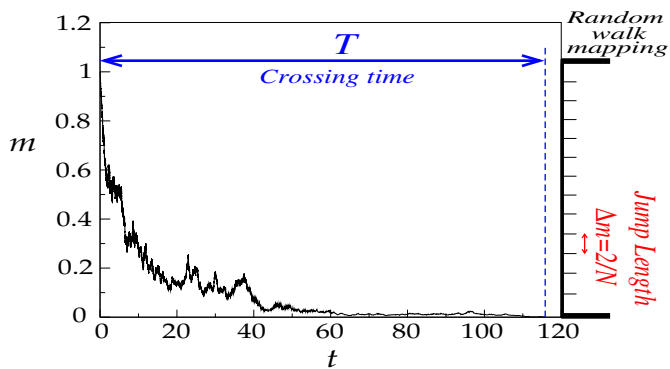} \caption{
    Single realization of the stochastic process. The system starts with all
    the spins in the same state $(m=1)$ and the dynamics is stopped
    when it crosses $m=0$, which defines the crossing time in the
    Ising model. We take $\sigma=0.4$, $b_{0}=0.98$ and system size
    $N=10^{6}$. On the right margin we sketch the mapping of the
    problem to a Random Walk with jump length $|\Delta m|=2/N$.}
\label{single}
\end{figure}

\begin{figure} \centering
  \includegraphics[width=0.65\textwidth]{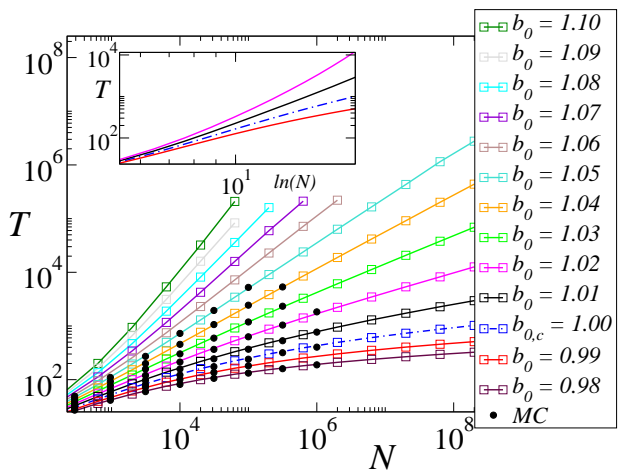} \caption{
    Main:
    Log-log plot of the crossing time $T(N)$ for the Ising Model with
    Glauber dynamics in mean field. $b_{0}=0.98$ (bottom) to $b_{0}=1.10$ (top).
    Monte Carlo simulations on a FCN
    (circles) and numerical integration of the Langevin equation~(\ref{isinglangevin})
    with $\sigma=0.4$ (squares and interpolation with solid lines). There is a
    region with generic algebraic scaling of $T(N)$ and continuously
    varying exponents, $b_{0}\in[1.01,1.10]$. Inset: log-log plot 
    of $T(N)$ vs. $\ln N$.
    At criticality (dotted-dashed line) the scaling is fitted to a 
    quadratic function in $\ln N$.}
\label{passtisingmf}
\end{figure}

In this section we study the behavior of the mean \emph{crossing time},
that has been shown to be relevant in the presence of temporal disorder \citep{Vazquez-2011}.
The crossing time is the time employed by the system
to reach the disordered zero-magnetization state for the first time,
starting from a fully ordered state with $|m|=1$ (see
Fig.~\ref{single}).  Crossing times were calculated by numerically
integrating Eq.~(\ref{isinglangevin}) for different realizations of
the noise $\gamma$ and averaging over many independent realizations.
These integrations were performed using a standard stochastic
Runge-Kutta scheme (note that, the noise term does not have any
pathological behavior at $m=0$ as occurs in systems with absorbing
states, for which more refined integration techniques are required
\citep{sto1}) .  Results are shown in Fig.~\ref{passtisingmf}.


To estimate the critical point, we calculated the time evolution of
the average magnetization $\langle m \rangle(t)$ by integrating the
Langevin equation~(\ref{isinglangevin}), and also by performing
Monte Carlo simulations of the particle system on a fully connected
network.  At the critical point $b_{0,c}$ the magnetization decays to
zero as $\langle m \rangle \sim t^{-\beta}$.  We have estimated
$b_{0,c} = 1$, which coincides with the pure case critical point
$b_{c,pure}=1$: the critical point in the presence of disorder in
mean-field is not shifted with respect to the pure system, in
agreement with the analytical calculation in appendix \ref{App-A}.  At
this critical point, as it is characteristic of TGPs
\citep{Vazquez-2011}, a scaling of the form $T \sim [\ln N]^{\alpha}$
is expected.  The numerically determined exponent value $\alpha \simeq
2.81$ for $\sigma=0.4$ is higher than the exponent $\alpha=2$ of the
asymptotic analytical prediction Eq.~(\ref{apA:final}), probably
because of the asymptotic regime in $\ln N$ has not been reached.
Instead, the behavior for arbitrary values of $N$ appears to be a
second order polynomial in $\ln N$, as we can see in
Eq.~(\ref{quadraticpol}).  Indeed, the numerical data is well fitted
by the quadratic function $a\,(\ln N)^2 + b \, \ln N + c $ (see inset
of Fig.~\ref{passtisingmf}).This is to be compared with the standard
power-law scaling $T\sim N^{\beta}$ characteristic of pure systems,
i.e. for $\sigma=0$. Moreover, a broad region showing algebraic
scaling $T \sim N^{\delta}$ with a continuously varying exponent
$\delta(b_0)$ ($\delta \rightarrow 0$ as $b_{0} \rightarrow
b^{+}_{0,c}$) appears in the ordered phase $b_0 > b_{0,c}$. Both
$\alpha$ and $\delta$ are not universal and depend on the noise
strength $\sigma$. Finally, in the disordered phase the scaling of $T$
is observed to be logarithmic, $T \sim \ln N$.

\begin{figure}
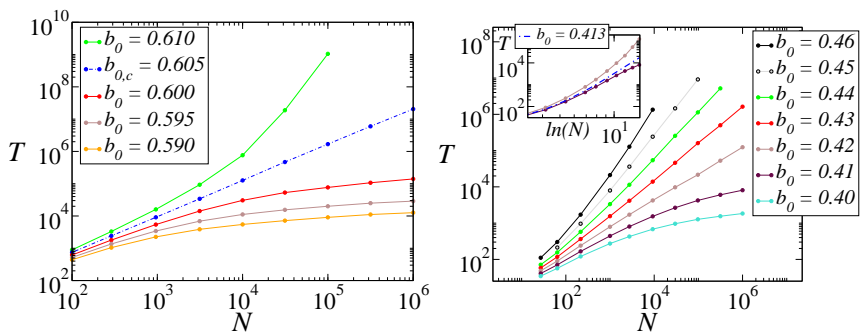

   \subfigure{\includegraphics[width=0.45\textwidth]{ising2d.eps}}
   \subfigure{\includegraphics[width=0.45\textwidth]{escape.eps}}
   \caption{ Log-log plot of the crossing time $T(N)$ for the Ising Model with
    Glauber dynamics on a regular lattice. (Left) $d=2$. $b_{0}=0.590$ (bottom) to $b_{0}=0.610$ (top). $\sigma=0.4$ (lines are interpolations).
    Power law scaling at the critical point (dotted-dashed line). TGP are not observed, 
    crossing time scales exponentially in the ordered phase (light green, upper, line). (Right) 
    $d=3$. $b_{0}=0.40$ (bottom) to $b_{0}=0.46$ (top). $\sigma=0.4$ (lines are interpolations).
    In a region $b \in [0.42, 0.46]$, generic algebraic scaling of $T(N)$ with continuously
    varying exponents. Inset: log-log plot of $T(N)$ vs. $\ln (N)$. It
    is estimated at criticality (dotted-dashed line) $T \sim (\ln N)^{5.29}$.}
   \label{isingfinite}                
\end{figure}

We have also performed Monte Carlo simulations of the time-disordered
Glauber model on two- and three-dimensional cubic lattices with
nearest neighbor interactions.  The critical point was computed
following standard methods, that is, by looking for a power law
scaling of $\langle m \rangle $ versus time, as we mentioned above.
In $d=2$, a shift in the critical point was found: from
$b_{c,pure}=0.441(1)$ in the pure model to $b_{0,c}=0.605(1)$ for
$\sigma=0.4$. However, the scaling behavior of $T$ with $N$ resembles
that of the pure model, with $T\sim N^{\beta}$ at criticality (with an
exponent numerically close to that of the pure model \citep{marro}),
and an exponential growth $T \sim \exp(c N)$, where $c$ is a positive
constant, in the ordered phase (Arrhenius law) ) { (see
  Fig.~\ref{isingfinite} (Left))}.  Thus, no region of generic algebraic scaling
appears in this low-dimensional system. On the contrary, in $d=3$,
results qualitatively similar to mean-field ones are recovered (see
Fig.~\ref{isingfinite} (Right)). The critical point is shifted from
$b_{c,pure}=0.222(1)$ (calculated in \cite{Heuer1993}) to
$b_{0,c}=0.413(2)$, with a critical exponent $\alpha(d=3)=5.29$ for
$\sigma=0.4$, and generic algebraic scaling in the ordered phase. In
conclusion, our numerical studies suggest that the lower critical
dimension for the TGPs in the Ising transition is $d_c=3$. This is in
agreement with the analytical finding in \cite{alonso}, establishing
that temporal disorder is irrelevant in Ising-like systems below three
dimensions. This result is to be compared with $d_c=2$ numerically
reported for the existence of TGPs in DP-like transitions
\citep{Vazquez-2011} (observe, however, that temporal disorder, in this
case, affects the value of critical exponents at criticality in all
spatial dimensions). Further studies are needed to clarify the
relation between disorder-relevance at criticality and the existence
or not of TGPs.
  
\subsection{Analytical results} 
\label{isinganal}

Let us consider the Langevin equation~(\ref{finallange}) in the
thermodynamic limit ($g_{0}(m)=0$). Given that the remaining intrinsic
noise comes from a transformation of a colored noise into a white
noise, the Stratonovich interpretation is to be used to obtain its
associated Fokker-Planck equation (see e.g.  \citep{Horsthemke1984})
\begin{eqnarray} \label{fpst} \frac{\partial P(m,t)}{\partial
    t}=-\frac{\partial}{\partial m} \Bigg\{ \left[
    f_{0}(m)+\frac{K}{2}j_{0}(m)j'_{0}(m)\right] P(m,t)\Bigg\}
  +\frac{1}{2}\frac{\partial^{2}}{\partial m^{2}} \Big\{ K
  j^{2}_{0}(m)P(m,t)\Big\}. \nonumber \\ \end{eqnarray} 

\begin{figure} \centering
  \includegraphics[width=0.55\textwidth]{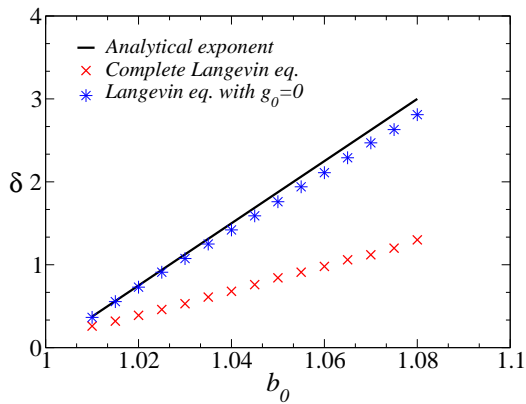}
  \caption{Comparison between anlytical and numerical values for the power law exponents in the mean crossing time.}
\label{exponente}
\end{figure}

Starting
from this N-independent Fokker-Planck equation~(\ref{fpst}), we will next provide
analytical results on the mean crossing times. An
effective dependence on $N$ is implemented by calculating the
first-passage time to the state $m=|2/N|$ rather than $m=0$. This is
equivalent to the assumption that the system reaches the zero
magnetization state with an equal number $N_+=N_-=N/2$ of up and down
spins when $|m|<2/N$, that is, when $N/2-1<N_+<N/2+1$. The mean-first
passage time $T$ associated with the Fokker-Planck
equation~(\ref{fpst}) obeys the differential equation (Chapter \ref{methods}) \citep{redner}
\begin{equation}\label{meanesc}
 \frac{K}{2}j^{2}_{0}(m)T''(m)+\left[ f_{0}(m)+\frac{K}{2}j_{0}(m)j'_{0}(m)\right]T'(m)=-1,
\end{equation}
with absorbing and reflecting boundaries at $|m|=2/N$ and $|m|=1$,
respectively.  The solution, starting at time $t=0$ from $m=1$ is given by 
\begin{equation}
\label{time}
T(m=1)=
2\int_{2/N}^{1}\frac{dy}{\psi(y)}\int_{y}^{1}\frac{\psi(z)}{Kj_{0}^{2}(z)}dz,
\end{equation}
where
\begin{equation}\label{psi1}
\psi(x)=\exp \Bigg\{ \int_{2/N}^{x} \frac{2f_{0}(x')+
Kj_{0}(x')j'_{0}(x')}{Kj_{0}^{2}(x')} dx' \Bigg\}. 
\end{equation}
Computing these integrals (see Appendix \ref{App-A}) we obtain
\begin{eqnarray} 
\label{timeana}
T \sim \left\{ \begin{array}{ll}
{\ln N}/(b_{0}-1) & \mbox{for $b_{0} < 1$} \\
3(\ln N)^{2} /\sigma^{2} & \mbox{for $b_{0} = 1$} \\
N^{\frac{6(b_{0}-1)}{\sigma^{2}}} & \mbox{for $b_{0} > 1$.}
\end{array}
\right.
\end{eqnarray}
These expressions qualitatively agree with the numerical results of
Fig.~\ref{passtisingmf}, showing that $T$ grows logarithmically with
$N$ in the absorbing phase $b_0<1$, as a power law in the active phase
$b_0>1$, and as a power of $\ln N$ (i.e. poly-logarithmically) at the
transition point $b_{0,c}=1$. The exponents $\delta=6(b_{0}-1)/\sigma^{2}$ 
do not agree well
with the numerically determined exponents. This is due to
the fact that we have neglected the $1/\sqrt{N}$ term by taking
$g_0=0$, which becomes of the same magnitude as the $j_0$ term when
$|m|$ approaches $2/N$. This was confirmed by
testing that analytical expressions Eq.~(\ref{timeana}) agree very
well with numerical integrations of Eq.~(\ref{isinglangevin})
performed for $g_{0}=0$, and setting the crossing point at $m=2/N$ (See Fig.~\ref{exponente}). In
summary, this analytical approach reproduces qualitatively --and in some
cases quantitatively-- the above reported non-trivial phenomenology.

\section{Generalized Voter transition with temporal disorder}
\label{secGV}

We study in this section the GV transition \citep{dornic}, which
appears when a $Z_2$-symmetry system simultaneously breaks the
symmetry and reaches one of the two absorbing states. A model
presenting this type of transition is the nonlinear $q$-voter model,
introduced in \cite{qvoter}. The microscopic dynamics of this
nonlinear version of the voter model consists in randomly picking a
spin $S_i$ and flipping it with a probability that depends on the
state of $q$ randomly chosen neighbors of $S_i$ (with possible
repetitions). If all neighbors are at the same state, then $S_i$
adopts it with probability $1$ (which implies, in particular, that the two
completely ordered configurations are absorbing). Otherwise, $S_i$
flips with a state-dependent probability
 \begin{equation}
 f(x,b)=x^{q}+b[1-x^{q}-(1-x)^{q}],
\label{probqvoter}
\end{equation}
where $x$ is the fraction of disagreeing (antiparallel) neighbors and
$b$ is a control parameter. Three types of transitions, Ising, DP and
GV can be observed in this model depending on the value of $q$
\citep{qvoter}. Here, we focus on the $q=3$ case, for which a unique GV
transition at $b_c=1/3$ has been reported \citep{qvoter}.

\subsection{The Langevin equation}

\begin{figure} \centering
  \includegraphics[width=0.4\textwidth]{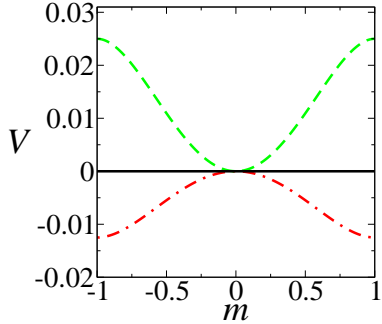}
  \caption{Potential for the GV transition in a mean field approach.
    The dashed, solid and dot-dashed lines correspond to the
    paramagnetic phase, critical point, and the ferromagnetic phase,
    respectively.}
\label{potential-GV}
\end{figure}

In the MF limit (FCN) \footnote{For a fully connected network the number of neighbors has no meaning.
However the MF limit of the model refers to the use of the probability given by Eq.(\ref{probqvoter})}, the fractions of antiparallel
neighbors of the two types of spins $S_i =1$ and $S_i=-1$ are
$x=(1-m)/2$ and $x=(1+m)/2$, respectively. Thus, the transition
probabilities are \begin{equation} \label{w+-}
 \omega_{\pm}(m,b)=\frac{1\mp m}{2}\,f\left(\frac{1\pm m}{2},b\right).
\end{equation}
Following the same steps as in the previous section, we obtain the Langevin equation
\begin{eqnarray}
\label{GVlangevin}
\dot{m}=\frac{1-3b_{0}}{2}m(1-m^{2})+\sqrt{\frac{\left(1-m^{2}\right)(1+6b_{0}+m^{2})}{N}+
  \frac{9 K}{4}\sigma^{2}m^{2}(1-m^{2})^{2}}\,\, \gamma(t). \nonumber \\
\end{eqnarray}
Let us remark that the potential in the nonlinear voter model
(Fig.~\ref{potential-GV}) differs from that for the Ising model. Owing
to the fact that the coefficients of the linear and cubic term in the
deterministic part of Eq.~(\ref{GVlangevin}) coincide (except for
their sign), the system exhibits a discontinuous jump at the
transition point, where the potential minimum changes directly from
$m=0$ in the disordered phase to $m=\pm1$ in the ordered one.
Furthermore, the potential vanishes at the critical point
\citep{alhammal1}.

\subsection{Numerical Results}

\begin{figure}
\centering 
\includegraphics[width=0.7\textwidth]{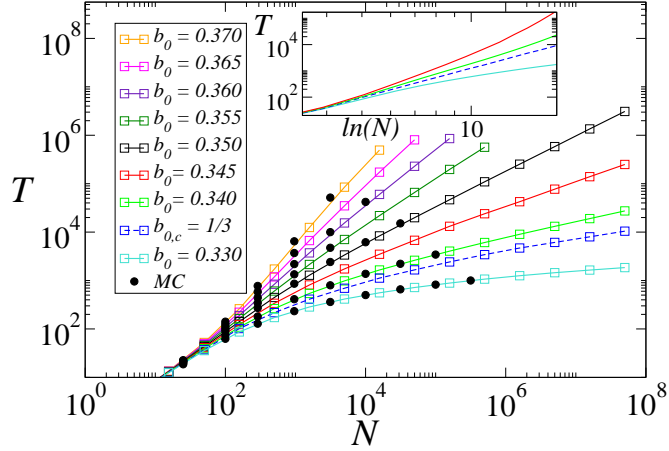}
\caption{Main: Log-log plot of the ordering time as a function of the
  system size $N$ in the MF q-voter model. Monte Carlo simulations on a FCN
  (dots) and numerical integration of the Langevin equation~(\ref{GVlangevin}) $b=0.330$ (bottom) to
  $0.370$ (top), and $\sigma=0.3$ (squares and lines interpolation). In the active phase a
  finite region with power law scaling is observed, $b_{0}\in[0.340,0.370]$. Inset: log-log
  plot of $T$ as a function of $\ln N$. At the critical point (dashed line) is $T
  \sim [\ln N]^{3.68}$.}
\label{timegv}
\end{figure}

The \emph{ordering time}, defined as the averaged time required to
reach a completely ordered configuration (absorbing state) starting
from a disordered configuration, is the equivalent of the crossing
time above. We have measured the mean ordering time $T$ by both,
integrating the Langevin equation~(\ref{GVlangevin}) and running Monte
Carlo simulations of the microscopic dynamics on FCNs and finite
dimensions. In Fig.~\ref{timegv} we show the MF results. We observe
that $T$ has a similar behavior to the one found for the mean crossing
time in the Ising model, and for the mean extinction time for the
contact process \citep{Vazquez-2011}. That is, a critical scaling $T
\sim [\ln N]^{\alpha}$ at the transition point $b_{0,c}=1/3$, with a
critical exponent $\alpha=3.68$ for $\sigma=0.3$, a logarithmic
scaling $T \sim \ln N$ in the absorbing phase $b_0<b_{0,c}$, and a
power law scaling $T \sim N^{\delta}$ with continuously varying
exponent $\delta(b_0)$ in the active phase $b_0>b_{0,c}$.

Monte Carlo simulations on regular lattices of dimensions $d=2$ and
$d=3$ revealed that there is no significant change in the scaling
behavior respect to the pure model (not shown).  The critical point
shifts in $d=2$ and remains very close to its mean-field value in
$d=3$, but results are compatible with the usual critical (pure) voter scaling
$T_{2d}\sim N\ln N$ and $T_{3d}\sim N$.  In the absorbing
phase $T$ grows logarithmically with $N$, while in the active phase
$T$ grows exponentially fast with $N$, as in the pure-model case.
Therefore, in these finite dimensional systems we do not find any
TGP nor other anomalous effects induced by temporal disorder, although
we cannot numerically exclude their existence in $d=3$. Such
effects should be observable only in higher dimensional systems
(closer to the mean-field limit).

\subsection{Analytical results}

The ordering time $T$ can be estimated by assuming that the dynamics
is described by the Langevin equation~(\ref{GVlangevin}), and
calculating the mean first-passage time from $m=0$ to any of the two
barriers located at $|m|=1$.  It turns out useful to consider the
density of up spins rather than the magnetization
\begin{equation} \label{change-rho}
 \rho \equiv \frac{1+m}{2}.
\end{equation}
 $T$ is the mean first-passage time to $\rho=0$
starting from $\rho=1/2$.  The Langevin equation for $\rho$ is
obtained from Eq.~(\ref{GVlangevin}), by neglecting the $1/\sqrt{N}$
term and applying the ordinary transformation of variables (which is done
 employing standard algebra, given that Eq.~(\ref{GVlangevin})
is interpreted in the Stratonovich sense) is
\begin{equation}
 \dot{\rho}=A(\rho)+\sqrt{K}C(\rho)\gamma(t),
\end{equation} 
with
\begin{eqnarray} \label{change-function}
 A(\rho)&=&a_{0}\rho(2\rho-1)(1-\rho), \nonumber \\
 C(\rho)&=&3\sigma\rho(2\rho-1)(1-\rho),
\end{eqnarray}
where $a_{0}=1-3b_{0}$.

Now, we can follow the same steps as in Sec.~\ref{isinganal} for
the Ising model, and find the equation for the mean first-passage time
$T(\rho)$ by means of the Fokker-Planck equation.  The solution is
given by (see Appendix B)
\begin{eqnarray}
\label{timeanagv}
T \sim \left\{ \begin{array}{ll}
{\ln N}/{\left(3b_{0}-1\right)} & \mbox{for $b_{0} < 1/3$} \\
{(\ln N)^{2}}/{3\sigma^{2}} & \mbox{for $b_{0} = 1/3$} \\
N^{\frac{2(b_{0}-1/3)}{\sigma^{2}}} & \mbox{for $b_{0} > 1/3$.}
\end{array}
\right.
\end{eqnarray}
These scalings, which qualitatively agree with the numerical results
of Fig.~\ref{timegv} for the $q$-voter, show that the behavior of $T$
is analogous to the one observed in the Ising transition of section
\ref{secising} and in the DP transition found in \cite{Vazquez-2011}.
Therefore, we conclude that TGPs appear around GV transitions in the
presence of external varying parameters in high dimensional systems.

For the GV universality class the renormalization group fixed point
is a non-perturbative one \citep{Canet}, becoming relevant in a
dimension between one and two. A field theoretical implementation of
temporal disorder in this theory is still missing, hence, theoretical
predictions and sound criteria for disorder relevance are
not available.

\section{Summary and conclusions}
\label{secsummary}

We have investigated the effect of temporal disorder on phase
transitions exhibited by $Z_2$ symmetric systems: the (continuous)
Ising and (discontinuous) GV transitions which appear in many
different scenarios. We have explored whether temporal disorder
induces Temporal Griffiths Phases
as it was previously found in standard (DP) systems
with one absorbing state. By performing mean-field analyses as well as
extensive computer simulations (in both fully connected networks and
 in finite dimensional lattices) we found that TGPs can exist
around equilibrium (Ising) transitions (above $d=2$) and around
discontinuous (GV) non-equilibrium transitions (only in
high-dimensional systems). 

Therefore, we confirm that TGPs may also
appear in systems with two symmetric absorbing states, illustrating
the generality of the underlying mechanism: the appearance of a
region, induced by temporal stochasticity of the control parameter,
where first-passage times scale as power laws of the system size.
The algebraic scaling of the crossing time, compared with the exponential one observed
in pure system, shows that temporal disorder makes the
ordered/active phase less stable. This implies that the system becomes highly
susceptible to perturbations. This appears to be a rather general and
robust phenomenon and an relevant result with applications in ecology.
At a given system size, considering the effect of the
environmental variablity considerably reduces the crossing times, that in some cases can be identified with population lifetimes.
This is a consequence of the change in the growth of this magnitude with the system size, from exponential to algebraic, when external
fluctuations are introduced. From a conservational point of view, the power-law scaling of the crossing times implies that population lifetimes would
decrease less, compared with the exponential behavior of the pure model, when going from larger to smaller systems, which is the main consequence 
of habitat fragmentation.

Additionally, although it is of secondary interest to our focus on ecological systems and it has been not shown in this chapter,
we have also confirmed that the response function of the system (i.e. susceptibility) diverges in a finite region close to the critical point.
This complementary result has been also obtained in \cite{Vazquez-2011}
and confirms the phenomelogical similarities between Temporal Griffiths Phases
and Griffiths Phases in systems with quenched disorder.

It also seems to be a general property that TGPs do not appear in low
dimensional systems, where standard fluctuations dominate over
temporal disorder. In all the cases studied so far, a critical
dimension $d_c$ --at and below which TGPs do not appear-- exist
($d_c=1$ for DP transitions, $d_c=2$ for Ising like systems, and $d_c
\simeq 3$ for GV ones).  Calculating analytically such a critical
dimension and comparing it with the standard critical dimension for
the relevance/irrelevance of temporal disorder at the critical point
(i.e. at the renormalization group non-trivial fixed point of the
corresponding field theory) remains an open and challenging task.

Future research might be oriented to the effect of temporal disorder
 on the formation and dynamics of spatial structures.


        \begin{appendices}
    \chapter{It\^{o}-Stratonovich discussion.} \label{apitost}

The integration of 
stochastic differential equations with multiplicative white noise presents some problems because the integral of the noise is not well defined. 
These problems are solved choosing either the It\^{o}
or the Stratonovich definition of the integral.\footnote{Any definition can be chosen or even made, but these two are the most often used.}
We have chosen one or the other depending on the origin of the noise term.
This Appendix explains how It\^{o} calculus works, and the connection between It\^{o} and Stratonovich schemes. 
We will finish discussing what of them is more suitable in each situation.

\section{Stochastic integration.} \label{secstoint}

Let us start providing a precise definition of the second integral in 
\begin{equation}\label{wiener2}
 M(t)-M(0)=\int_{0}^{t}f[M(s),s]ds+\int_{0}^{t}g[M(s),s]dW(s),
\end{equation}
that is
\begin{equation}\label{stoint}
 \int_{0}^{t}G(s)dW(s).
\end{equation}

The integration interval $[0,t]$ is divided into n subintervals,
\begin{equation}
 0 \leq t_{1} \leq t_{2} \leq t_{3} ... \leq t_{n-1} \leq t_{n}, 
\end{equation}
and intermediate points in each interval $\tau_{i}$ defined
\begin{equation}
 \tau_{i}=t_{i-1}+\alpha(t_{i}-t_{i-1}).
\end{equation}

The stochastic integral in Eq.~(\ref{stoint}) is defined as the limit of the partial sums,
\begin{equation}\label{sumas}
 S_{n}=\sum_{i=1}^{n}G(\tau_{i})(W(t_{i})-W(t_{i-1})),
\end{equation}
where the It\^{o} vs Stratonovich dilemma resides in the fact that the limit of $S_{n}$ depends on the particular set of points
$\tau_{i}$ that are used. It\^{o} stochastic integral is defined taking $\alpha=0$, so Eq.~(\ref{sumas}) becomes 
\begin{equation}
 S_{n}=\sum_{i=1}^{n}G(t_{i-1})(W(t_{i})-W(t_{i-1})),
\end{equation}
that is, the known function $g(x(t))$ is evaluted on the beginning point of the interval while Stratonovich is obtained if $\alpha=1/2$
and
\begin{equation}
 S_{n}=\sum_{i=1}^{n}G\left(\frac{t_{i-1}+t_{i}}{2}\right)(W(t_{i})-W(t_{i-1})).
\end{equation}

\section{It\^{o}'s formula.}\label{itoforsec}

In spite of being much more elegant from a mathematical point of view, Ito's prescription is not always the most suitable choice for
physical interpretation. Calculus we are used to does not work in this scheme, and a different change of variables
must be considered. Let us take an arbitrary function $a[x(t)]$ with $x(t)$ obeying the SDE
\begin{equation}\label{multieq}
\frac{dx(t)}{dt}=f(x,t)+g(x,t)\xi(x,t),
\end{equation}
where $\xi(x,t)$ is a white Gaussian noise. Consider
\begin{eqnarray}
 da[x(t)]&=&a[x(t)+dx(t)]-a[x(t)] \nonumber \\
 &=&a'[x(t)]dx(t)+\frac{1}{2}a''[x(t)]dx^{2}(t)+...\nonumber \\
 &=&a'[x(t)]\left\lbrace f(x,t)+g(x,t)\xi(t)\right\rbrace dt+\frac{1}{2}a''[x(t)]g^{2}(x,t)dW^{2}(t)+..., \nonumber \\
\end{eqnarray}
where higher terms in $dt$ have been neglected. Now, replacing $dW^{2}(t)=dt$ (see \citep{gardiner} for a proof),
\begin{equation}\label{itoformula}
 da[x(t)]=a'[x(t)]\left\lbrace f(x,t)+\frac{1}{2}a''[x(t)]g^{2}(x,t)\right\rbrace dt+a''[x(t)]g(x,t)dW(t),
\end{equation}
which is known as the It\^{o}'s formula and shows that change of variables is not given by ordinary calculus unless $a[x(t)]$ is linear
in $x(t)$.

\section{From Stratonovich to It\^{o}.}

As may be expected, both interpretations of the stochastic integral are somehow related. To show it, consider an stochastic differential equation
\begin{equation}
 \frac{dx}{dt}=\alpha[x(t),t]+\beta[x(t),t]\eta(t),
\end{equation}
where $\eta(t)$ is a white, zero mean, Gaussian noise. Integrating, it is,
\begin{equation}
 x(t)=x(0)+\int_{0}^{t}\alpha[x(s),s]ds+S\int_{0}^{t}\beta[x(s),s]dW(s),
\end{equation}
where $S$ denotes that a Stratonovich integration is used. We will derive the equivalent It\^{o} stochastic differential equation.

Assuming that $x(t)$ is a solution of
\begin{equation}\label{ito2}
 dx(t)=a[x(t),t]dt+b[x(t),t]dW(t),
\end{equation}
the corresponding $\alpha[x(t),t]$ and $\beta[x(t),t]$ will be deduced. The first step is to compute the connection between
 $S\int_{0}^{t}\beta[x(s),s]dW(s)$ and $\int_{0}^{t}\beta[x(s),s]dW(s)$, where the lack of notation in the second integral means an 
It\^{o} interpretation. Then,
\begin{equation} \label{stratoint}
 S\int_{0}^{t}\beta[x(s),s]dW(s) \approx \sum_{i}\beta\left[\frac{x(t_{i})+x(t_{i-1})}{2},t_{i-1}\right]\left[W(t_{i})-W(t_{i-1})\right].
\end{equation}

Taking into an account
\begin{equation}
 x(t_{i})=x(t_{i-1})+dx(t_{i-1}),
\end{equation}
in the Stratonovich integral, then 
\begin{equation}\label{stratointegral}
 \beta\left[\frac{x(t_{i})+x(t_{i-1})}{2},t_{i-1}\right]=\beta\left[x(t_{i-1})+\frac{1}{2}dx(t_{i-1}),t_{i-1}\right].
\end{equation}

Now, the It\^{o} SDE (\ref{ito2}) is used in order to write
\begin{equation}
 dx(t_{i})=a[x(t_{i-1}),t_{i-1}](t_{i}-t_{i-1})+b[x(t_{i-1}),t_{i-1}][W(t_{i})-W(t_{i-1})].
\end{equation}

Using It\^{o}'s formula given by Eq.~(\ref{itoformula}) as well as simplifying the notation writing $\beta(t_{i-1})$ instead of
$\beta[x(t_{i-1}),t_{i-1}]$, Eq.~(\ref{stratointegral}) becomes,
\begin{eqnarray}
\beta\left[\frac{x(t_{i})+x(t_{i-1})}{2},t_{i-1}\right]&=& \beta(t_{i-1})+\left[a(t_{i-1})\partial_{x}\beta(t_{i-1})+\frac{1}{4}b^{2}(t_{i-1})\right]\left[\frac{1}{2}(t_{i}-t_{i-1})\right]+ \nonumber \\
&+&\frac{1}{2}b(t_{i-1})\partial_{x}\beta(t_{i-1})[W(t_{i})-W(t_{i-1})].
\end{eqnarray}

Finally, substituing into Eq.~(\ref{stratoint}), neglecting terms in $dt^{2}$ and $dWdt$ and setting $dW^{2}=dt$,
\begin{equation}
 S\int=\sum_{i}\beta(t_{i-1})(W(t_{i})-W(t_{i-1}))+\frac{1}{2}\sum_{i}b(t_{i-1})\partial_{x}\beta(t_{i-1})(t_{i}-t_{i-1}),
\end{equation}
or going back to integrals,
\begin{equation}
S\int_{0}^{t}\beta[x(s),s]dW(s)=\int_{0}^{t}\beta[x(s),s]dW(s)+\frac{1}{2}\int_{0}^{t}b[x(s),s]\partial_{x}\beta[x(s),s]ds,
\end{equation}
which means that the stochastic integral in Stratonovich representation is equivalent to a stochastic integral in It\^{o}'s and a drift
term. It is also important to remark that this formula gives a connection between both integrals of function $\beta[x(s),s]$, in which
$x(s)$ is the solution of the It\^{o} SDE (\ref{ito2}). It does not give a general connection between the It\^{o} and Stratonovich 
integrals of arbitrary functions.

\begin{eqnarray}
 &\mbox{The It\^{o} SDE} &dx=a(x,t)dt+b(x,t)dW(t) \nonumber \\
 &\mbox{is the Stratonovich SDE} &dx=\left[a(x,t)-\frac{1}{2}b(x,t)\partial_{x}b(x,t)\right]dt+b(x,t)dW(t). \nonumber \\
\end{eqnarray}

Or

\begin{eqnarray} \label{change2}
 &\mbox{The Stratonovich SDE} &dx=\alpha dt+\beta dW(t) \nonumber \\
 &\mbox{is the It\^{o} SDE} &dx=\left[\alpha(x,t)+\frac{1}{2}\beta(x,t)\partial_{x}\beta(x,t)\right]dt+\beta(x,t) dW(t). \nonumber \\
\end{eqnarray}

There are many consequences of this transformation formula, but the more important are
\begin{itemize}
 \item It is always possible to change from the Stratonovich to the It\^{o} interpretation of a SDE by adding 
$\frac{1}{2}\beta(x,t)\partial_{x}\beta(x,t)$ or in the inverse direction subtracting a similar term.
\item In the case of additive noise, i.e., $g(x,t)=const.$ in Eq.~(\ref{multieq}) there is no difference between the It\^{o} and Stratonovich
integral.
\item In the case of multiplicative noise, i.e., $g(x,t)\ne const.$ in Eq.~(\ref{multieq}), where the influence of the random force depends on
the state of the process, the correlation between both the random force and the state of the process is implicit in the Stratonovich integral.
It gives raise to the noise induced drift when moving to It\^{o} appearing in the deterministic part of the equation.
\item The Stratonovich calculus obeys the classical chain rule, It\^{o}'s formula derived in Section \ref{itoforsec} plays a similar role on It\^{o}'s 
calculus.
\end{itemize}

\section{Stratonovich / It\^{o} dilemma.} \label{secstraito}
The long controversy in the physical literature about what is the right definition of the stochastic integral
has created some confusion on this topic. That's why, although a much more mathematically rigorous and longer discussion can be found in the
references, \citep{N.vanKampen2007,Horsthemke1984,jazwinski} some hand waving arguments will be given in this section.

First of all it is important to say that this kind of ambiguity when working with SDE only yields for the particular, but most common, case of differential
equations with multiplicative white noise\footnote{Cases where the rapidly fluctuating external force depends on the state of the system.}.
As a first approach, it is natural to tend to believe that due to invariance of the
equations under ``coordinate transformation'' $y=u(x)$ when working on Stratonovich scheme it is the proper choice. However, it means nothing
but it obeys the classical calculus rules we are familiar with. The only quantities that have to be invariant under a transformation $u=y(x)$,
where $u$ is one to one, are the probabilities,
\begin{equation}
 p(y,t)dy=p(x,t)dx,
\end{equation}
and this is of course guaranteed in both calculi. They lead to a consistent calculus.

It looks sensible, then, to change the question. The matter is not what is the right definition of the stochastic integral,
but how do we model real systems by stochastic processes. That is, in which situation either It\^{o}'s or Stratonovich's choice is the most
suitable.

On the one hand, if the starting point is a phenomenological equation in which some fluctuating parameters represented through colored
noise terms are approximated by Gaussian white noise, then the most appropiate process is the one that is defined by the Stratonovich
interpretation of the equation.

On the other hand, in many systems the appropiate starting point is a discret time equation, as it happens, for instance, in biology when 
working with populations of insects. In these cases the equation reads
\begin{equation} \label{discrete}
 X(t_{i})=X(t_{i-1})+f(X(t_{i-1}))\Delta t+\sigma g(X(t_{i-1}))Q(t_{i-1}),
\end{equation}
where $t_{i}=t_{i-1}+\Delta t$ in every time step and $Q_{i}$ are Gaussian independent random variables with expected values $<Q(t_{i})>=0$
and $<Q^{2}(t_{i})>=\Delta t$.

If times considered are longer compared to $\Delta t$, the continuous time limit can be taken. Then the system is described by
\begin{equation}
 \dot{X}(t)=f[X(t)]+\sigma g[X(t)]\dot{W}(t),
\end{equation}
which is also a SDE where $W(t)$ is the Wiener process. However, due to the asymmetric form of Eq.~(\ref{discrete}) with respect to time it is 
much more appropiate the stochastic process defined according to the It\^{o} interpretation in this case.

To sum up, as a take to home message from this section, two different cases can be considered when working with SDE. When the white Gaussian
noise limit is considered as the limit of a colored noise when the correlation time tends to zero, the Stratonovich interpretation is more
sensible, when It\^{o}'s is more suitable when it represents the continuous limit of a discrete time problem. In any case, there are no 
universally valid theoretical reasons why one or the other interpretation of an SDE should be preferred and the ultimate test must be the
confrontation of the analytical (or numerical) results with the experimental facts.

    \chapter{Analytical calculations on the escape time for the Ising Model} \label{App-A}

We will show here all the analytical calculations done to obtain the result of Eq.~(\ref{timeana}).
To make the integrals analytically solvable we take $g_{1,0}(m)=0$, so the Langevin equation is
\begin{equation}\label{aa}
\dot{m}=f_{0}(m)+\sqrt{K}j_{0}(m)\gamma(t),
\end{equation}
with
\begin{eqnarray}
 f_{0}(m)&=&a_{0}m-c_{0}m^{3},\nonumber \\
 j_{0}(m)&=&\sigma m(1-b_{0}^{2}m^{2}),
\end{eqnarray}
where $a_{0}=b_{0}-1$, $c_{0}=b_{0}^{3}/3$ and $\gamma(t)$ is, again, a white Gaussian noise defined by its autocorrelation function
 $<\gamma(t)\gamma(t')>=\delta(t-t')$, and its mean value $<\gamma(t)>=0$.

The Langevin equation (\ref{aa}) presents one absorbing state in $m=0$ induced by the simplification done when neglecting thermal 
fluctuations.

Working in the Stratonovich scheme \footnote{Because the noise term comes from taking the white noise limit in a colored one},
the associated Fokker-Planck equation is
\begin{eqnarray}
 \frac{\partial P(m,t)}{\partial t}&=&-\frac{\partial}{\partial m}\left[f_{0}(m)+\frac{K}{2}j_{0}(m)j_{0}'(m)\right]P(m,t)+
 \frac{K}{2}\frac{\partial^{2}}{\partial m^{2}}\left[j_{0}^{2}(m)P(m,t)\right], \nonumber \\
\end{eqnarray}
where
\begin{eqnarray}
 f_{0}(m)+\frac{K}{2}j_{0}(m)j_{0}'(m)&=&\left(a_{0}+\frac{\tau}{6}\sigma^{2}\right)m+\frac{\tau\sigma^{2}}{2}b_{0}^{4}m^{5}-\left(c_{0}+\frac{2\tau\sigma^{2}b_{0}^{2}}{3}\right)m^{3}, \nonumber \\
 Dj_{0}^{2}(m)&=&\frac{\tau}{3}\sigma^{2}m^{2}\left(1-b_{0}^{2}m^{2}\right)^{2}. 
\end{eqnarray}

According to \cite{gardiner} and \cite{redner}, the escape time from an starting point $m$ obeys,
\begin{equation}
 \left[f_{0}(m)+\frac{K}{2}j_{0}(m)j_{0}'(m)\right]T'(m)+\frac{1}{2}Kj_{0}^{2}(m)T''(m)=-1.
\end{equation}

As the size of the system does not appear naturally in the problem because of the simplification done when taking $g_{1,0}=0$, the mean escape time will be
defined as that needed to pass through $m=2/N$, which is the length of the jumps of the Brownian particle to whose movement the problem has been mapped.
Then, taking into an account that there is an absorbing barrier in $m=0$ and a reflecting one in $m=1$ and the
initial condition, the solution is \citep{gardiner} 
\begin{equation}\label{esctime}
 T(m_{i}=1)=2\int_{2/N}^{m_{i}=1}\frac{dy}{\psi(y)}\int_{y}^{1}\frac{\psi(z)}{Kj_{0}^{2}(z)}dz,
\end{equation}
with
\begin{equation}\label{psi2}
 \psi(z)={\rm exp}\int_{2/N}^{z}dz'\frac{2f(z')+Kj_{0}(z')j_{0}'(z')}{Kj_{0}^{2}(z')},
\end{equation}
which involves $6^{th}$ and $4^{th}$ order polynomial functions.\\

To make the integral simpler, we expand the functions up to  $3^{rd}$
order, and takre the low integration limit in Eq.~(\ref{psi2}) at $1$ instead of $2/N$. This change
can be done because $\psi(z)$ appears both in the numerator and the denominator of $T(m)$, so the contribution of the lower limit
vanishes, allowing to take it in our interest. The first assumption leads to
\begin{eqnarray}
 f_{0}(m)+\frac{K}{2}j_{0}(m)j_{0}'(m)&\approx&m(r-sm^{2}), \nonumber \\
Kj_{0}^{2}(m)&\approx&\omega m^{2},
\end{eqnarray}
where it has been defined $\omega \equiv \tau\sigma^{2}/3$; $r \equiv a_{0}+\omega/2$; $s\equiv (c_{0}+2\omega b_{0}^{2})$. 
The size of the system will be rescaled too, so the lower limit in the expression of the escape time Eq.~(\ref{esctime}) is $1/N$. This simplifies
the notation and does not affect the qualitative behaviour of the results in the asymptotic limit (only a constant factor appears).\\

Now, it can be written,
\begin{equation}\label{psiy}
 \psi(z)={\rm exp}\int_{1}^{z}\frac{2z'(r-sz'^{2})}{\omega z'^{2}}dz'=z^{\alpha}{\rm e}^{\beta(1-z^{2})},
\end{equation}
where $\alpha\equiv 2r/\omega$ and $\beta\equiv s/\omega$.\\

Lets now define the function
\begin{equation} \label{iy}
 I(y)=\int_{y}^{1}\frac{\psi(z)}{Kj_{0}^{2}(z)}dz=\frac{{\rm e}^{\beta}}{\omega}\int_{y}^{1}z^{\alpha-2}{\rm e}^{-\beta z^{2}}dz,
\end{equation}
which presents a singularity when $\alpha = 1$ as can be seen integrating by parts. With the definition made of the parameters,
it can be shown that it corresponds to $b_{0}=1\equiv b_{0,c}$.\\

Considering the definitions of Eq.~(\ref{psiy}) and Eq.~(\ref{iy}), the mean escape time is given by 
\begin{equation}\label{ty}
 T=2\int_{1/N}^{1}\frac{I(y)}{\psi(y)}dy.
\end{equation}

Each case will be studied separately.

\section{Case \texorpdfstring{$\alpha \neq 1$}. }

Integrating by parts Eq.~(\ref{iy})
\begin{equation}
 I(y)=\frac{{\rm e}^{\beta}}{\omega}\left[\frac{{\rm e}^{-\beta}-{\rm e}^{-\beta y^{2}}y^{\alpha-1}}{\alpha-1}+2\beta\int_{y}^{1}\frac{z^{\alpha}{\rm e}^{-\beta z^{2}}}{\alpha-1}dz\right],
\end{equation}
where the new integral can be solved again integrating by parts. Working recursively this way,
\small
\begin{equation}
 I(y)=\frac{{\rm e}^{\beta}}{\omega}\left[\frac{{\rm e}^{-\beta}-{\rm e}^{-\beta y^{2}}y^{\alpha-1}}{\alpha-1}+2\beta\frac{{\rm e}^{-\beta}-{\rm e}^{-\beta y^{2}}y^{\alpha+1}}{(\alpha-1)(\alpha+1)}+...\right],
\end{equation}
\normalsize
or
\begin{equation}
I(y)=\frac{1}{\omega}\sum_{k=0}^{\infty}(2\beta)^{k}\frac{1-{\rm e}^{-\beta(y^{2}-1)}y^{\alpha-1+2k}}{\prod_{i=0}^{k}(\alpha-1+2i)}.
\end{equation}

The mean escape time is given now by
\begin{equation}
T=\frac{2}{\omega}\sum_{k=0}^{\infty}\frac{(2\beta)^{k}}{\prod_{i=0}^{k}(\alpha-1+2i}\left[I_{1}(N)-I_{2}(k,N)\right], 
\end{equation}
where
\begin{eqnarray}
 I_{1}(N)&\equiv& \int_{1/N}^{1} y^{-\alpha}{\rm e}^{\beta(y^{2}-1)}dy \\
 I_{2}(k,N)&\equiv&\int_{1/N}^{1} y^{2k-1}dy.
\end{eqnarray}

Integrating $I_{1}(N)$ by parts (taking again the exponential part as $u$ and the rest as $dv$) and following the same procedure as in 
Eq.~(\ref{iy}) it is obtained
\begin{equation}
 I_{1}(N)=\sum_{l=0}^{\infty}\frac{(-2\beta)^{l}[1-N^{\alpha-1-2l}{\rm e}^{\beta(1/N^{2}-1)}]}{\prod_{j=0}^{l}(\alpha-1+2j)},
\end{equation}
while $I_{2}(k,N)$ is easily solved
\begin{eqnarray}
I_{2}(k,N) = \left\{ \begin{array}{ll}
-{\rm ln}(N^{-1})={\rm ln}(N) & \mbox{for $k=0$,} \\
\frac{1-N^{-2k}}{2k} & \mbox{for $k \geq 1$.} \\
\end{array}
\right.
\end{eqnarray}

At the end, an expresion for the mean escape time is achieved
\begin{eqnarray}
 T&=&\frac{2}{\omega}\left(\frac{I_{1}(N)-{\rm ln}(N)}{\alpha-1}\right) \nonumber \\
&+&\frac{2}{\omega}\sum_{k=1}^{\infty}\frac{(2\beta)^{k}\left[I_{1}(N)-(1-N^{-2k})/2k\right]}{\prod_{i=0}^{k}(\alpha-1+2i)}. \nonumber \\
\end{eqnarray}

In the asymptotic limit $N \rightarrow \infty$ two different cases must be considered.

\subsection{\texorpdfstring{$\alpha < 1$}. }

Under this prescription, $\alpha-1-2l < 0$ when $l \geq 0$ so in $I_{1}(N)$
\begin{equation}
 1-N^{\alpha-1-2l}{\rm e}^{\beta(1/N^{2}-1)} \sim 1-\frac{{\rm e}^{-\beta}}{N^{\nu}} \sim 1,
\end{equation}
which leads to
\begin{equation}
 I_{1}(N)=\sum_{l=0}^{\infty}\frac{(-2\beta)^{l}}{\prod_{j=0}^{l}(1+2j-\alpha} \equiv C(\alpha,\beta).
\end{equation}

Finally, for the mean escape time,
\small
\begin{equation}
 T \approx \frac{2}{\omega}\left[\frac{C(\alpha,\beta)-{\rm ln}(N)}{\alpha-1}+\sum_{k=1}^{\infty}\frac{(2\beta)^{k}\left(C(\alpha,\beta)-(2k)^{-1}\right)}{\prod_{j=0}^{l}(\alpha-1+2j)}\right], 
\end{equation}
\normalsize
what means,
\begin{equation}
 T \approx \frac{2}{\omega(\alpha-1)}{\rm ln}(N).
\end{equation}

\subsection{\texorpdfstring{$\alpha > 1$}.}

It is taken as a starting point
\begin{equation} \label{I1}
 I_{1}(N)=\sum_{l=0}^{\infty}\frac{(-2\beta)^{l}[1-N^{\alpha-1-2l}{\rm e}^{\beta(1/N^{2}-1)}]}{\prod_{j=0}^{l}(\alpha-1+2j)},
\end{equation}
where considering that $N^{\alpha-1} \gg N^{\alpha-1-2l}, \forall l>0$, only the first term in Eq.~(\ref{I1}) is relevant. It implies
\begin{equation} 
 I_{1}(N)\approx \frac{1-{\rm e}^{-\beta}N^{\alpha-1}}{1-\alpha}\approx \frac{{\rm e}^{-\beta}N^{\alpha-1}}{1-\alpha},
\end{equation}
and in the mean escape time
\begin{eqnarray}
 T \approx K(\alpha,\beta)N^{\alpha-1}-\frac{2{\rm ln}(N)}{\omega(\alpha-1)} \sim N^{\alpha-1} \ \ \ \ (N \gg 1). \nonumber \\
\end{eqnarray}

\section{Case \texorpdfstring{$\alpha=1$}. Critical point.}

It has to be solved now
\begin{equation}
 I(y)=\int_{y}^{1}\frac{\psi(z)}{Kj_{0}^{2}(z)}dz=\frac{{\rm e}^{\beta}}{\omega}\int_{y}^{1}y^{-1}{\rm e}^{-\beta z^{2}}dz,
\end{equation}
using the expansion of the exponential function and integrating it is
\begin{equation}\label{iy2}
  I(y)=\frac{{\rm e}^{\beta}}{\omega}\left[-{\rm ln}(y)+\sum_{k=1}^{\infty}\frac{(-\beta)^{k}(1-2y)^{2k}}{k!2k}\right].
\end{equation}

It makes the mean escape time to obey, taking the form of $I(y)$ Eq.~(\ref{iy2}) into Eq.~(\ref{ty})
\begin{equation}
 T=\frac{2{\rm e}^{\beta}}{\omega}\left[I_{3}(N)+\sum_{k=1}^{\infty}\frac{(-\beta)^{k}}{k!2k}\left(I_{4}(N)+I_{5}(k,N)\right)\right],
\end{equation}
where
\begin{eqnarray}
 I_{3}(N)&=&-\int_{1/N}^{1}{\rm ln}(y)y^{-1}{\rm e}^{\beta(y^{2}-1)}dy, \nonumber \\
I_{4}(N)&=&\int_{1/N}^{1}y^{-1}{\rm e}^{\beta(y^{2}-1)}dy, \nonumber \\
I_{5}(k,N)&=&\int_{1/N}^{1}y^{k-1}{\rm e}^{\beta(y^{2}-1)}dy.
\end{eqnarray}

First of all, lets consider the solution of $I_{3}(N)$ integrating by parts ($u={\rm e}^{\beta y^{2}}$ and $dv={\rm ln}(y)y^{-1}dy$) so,
\begin{equation}
 I_{3}(N)=\frac{({\rm ln}N)^{2}}{2}{\rm e}^{\beta(N^{-2}-1)}+\beta\int_{1/N}^{1}({\rm ln}y)^{2}{\rm e}^{\beta(y^{2}-1)}dy,
\end{equation}
where the new integral is solved again integrating by parts taking 
\begin{eqnarray}
u&=&{\rm e}^{\beta(y^{2}-1)} \rightarrow du=2\beta{\rm e}^{\beta(y^{2}-1)} \nonumber \\
dv&=&({\rm ln}y)^{2}dy \rightarrow v=2y-2y{\rm ln}y+y({\rm ln}y)^{2}. \nonumber \\
\end{eqnarray}
It leads to a solution behaving like
\begin{equation}
 \beta\int_{1/N}^{1}({\rm ln}y)^{2}{\rm e}^{\beta(y^{2}-1)}dy=2\beta-O(N^{-1})+O\left(\frac{{\rm ln}N}{N}\right),
\end{equation}
so finally,
\begin{equation}
 I_{3}(N)=\frac{({\rm ln}N)^{2}}{2}{\rm e}^{\beta(N^{-2}-1)}+2\beta-O(N^{-1})+O\left(\frac{{\rm ln}N}{N}\right),
\end{equation}
which scales in the asymptotic limit as
\begin{equation}
 I_{3}(N)\sim \frac{({\rm ln}N)^{2}}{2}{\rm e}^{-\beta}.
\end{equation}

Secondly, lets focus on $I_{4}(N)$, where, again, an expansion of the exponential function has to be done
\small
\begin{equation}
 I_{4}(N)=\int_{1/N}^{1}y^{-1}{\rm e}^{\beta(y^{2}-1)}dy={\rm e}^{-\beta}\int_{1/N}^{1}y^{-1}\sum_{k=0}^{\infty}\frac{\beta^{k}y^{2k}}{k!}dy
\end{equation}
\normalsize
which can be easily solved
\begin{equation}
 I_{4}(N)={\rm e}^{-\beta}\left[{\rm ln}N+\frac{\beta^{k}}{k!2k}\left(1-N^{-2k}\right)\right].
\end{equation}

The leading behavior when the size of the system is big enough $(N \gg 1)$ is
\begin{equation}
 I_{4}(N)\sim{\rm e}^{-\beta}{\rm ln}N+C_{4}(\beta).
\end{equation}

The last integral to be solved, also using the expansion of the exponential function, is
\begin{eqnarray}
 I_{5}(k,N)={\rm e}^{-\beta}\sum_{l=0}^{\infty}\frac{\beta^{l}}{l!(k+2l)}\left(1-N^{-2l-k}\right) &\sim& cte \nonumber \\
&N&\gg 1. \nonumber \\
\end{eqnarray}

It finally leads to an expression for the mean escape time in the critical point
\small
\begin{equation}
 T \approx \frac{2{\rm e}^{-\beta}}{\omega}\left\lbrace\frac{{\rm e}^{-\beta}({\rm ln}N)^{2}}{2}+\sum_{k=1}^{\infty}\frac{(-\beta)^{k}}{k!2k}\left[{\rm e}^{-\beta}{\rm ln}N+C'_{4}(\beta)\right]\right\rbrace.
\end{equation}
\normalsize

In the limit of very big systems ($N \gg 1$) the mean escape time scales as
\begin{equation}\label{quadraticpol}
 T \sim \frac{({\rm ln}N)^{2}}{\omega}+\frac{1}{\omega}\sum_{k=1}^{\infty}\frac{(-\beta)^{k}}{k!k}{\rm ln}N+K(\beta),
\end{equation}
with asymptotic behaviour
\begin{equation}
 T \sim \frac{({\rm ln}N)^{2}}{\omega}.
\end{equation}

To sum up, it has been obtained analytically the finite size scaling of the mean escape time, defined as the time taken by the system for reaching
$m=0$ from an initial condition $m_{i}=1$. It is

\begin{eqnarray}
T \sim \left\{ \begin{array}{ll}
\frac{2}{\omega(\alpha-1)}{\rm ln}N & \mbox{for $\alpha < 1$,} \\
\frac{({\rm ln}N)^{2}}{\omega} & \mbox{for $\alpha = 1$,} \\
N^{\alpha-1} & \mbox{for $\alpha > 1$.}
\end{array}
\right.
\end{eqnarray}

or in terms of the original parameters
\begin{eqnarray}\label{apA:final}
T \sim \left\{ \begin{array}{ll}
\frac{{\rm ln}N}{b_{0}-1} & \mbox{for $b_{0} < b_{0,c}$,} \\
\frac{3({\rm ln}N)^{2}}{\tau\sigma^{2}} & \mbox{for $b_{0} = b_{0,c}$,} \\
N^{\frac{6(b_{0}-1)}{\tau\sigma^{2}}} & \mbox{for $b_{0} > b_{0,c}$.}
\end{array}
\right.
\end{eqnarray}

     \end{appendices}
    
%
  \part[\textit{\textsc{Conclusions and outlook}}]{\textbf{\textit{\textsc{Conclusions and outlook \\}}}
         }   
 \chapter{Conclusions and outlook}

This thesis has addressed a series of ecological problems from the point of
view of statistical physics, that has provided the theoretical framework to develop different
mathematical models.

The origin of the regular structures of vegetation that are observed
in many regions around the world has been studied. They appear 
in landscapes where there is a limited amount of rainfall during the year,
regardless of the type of soil and vegetation. This scarcity of water is
an important constraint for the establishment of new plants. Traditionally, it has been thought that the emergence of the patterns
comes from the presence of facilitative and competitive interactions among plants acting
simultaneously but at different spatial scales. This phenomenon has been referred to as \textit{scale-dependent feedback} in the literature.
The findings presented in this thesis, using mathematical models that contain only competitive interactions, 
suggest that facilitative interactions could be superfluous if the finite length of
the roots is considered in the equations. As an alternative to 
the \textit{scale-dependent feedback}, we have introduced the concept of \textit{exclusion areas}. They are regions, typically between two maxima
of vegetation density, where the competition is
so strong that it cannot be overcome by new plants. The extension and the location of these areas are
given by the interaction kernel of the different species of plants, which is intimately related to the length of their roots.
This concept allows to know in which regions the vegetation will disappear and in which it will remain given an initial ditribution of
plants. Determining the existence of \textit{exclusion areas} could also have important implications on the design of
farming strategies that minimize the competition in the crops. This would allow an optimal exploitation of the water resources, mainly
in arid regions.

In addition, the proposed models follow previous results that allow the use of these patterns as early warning signals
of desertification in arid regions, allowing the development of conservation strategies by anticipating the death of vegetation.
As the amount of rainfall decreases, the shapes of the distributions show a universal sequence of gaps, stripes and, finally, spots of 
plants when the water is very limited. This sequence is independent on the species in that particular landscape. 
Unveiling the basic mechanisms that drive the formation of these structures becomes essential to change the
natural tendency that would lead arid regions to dessert states. 

The collective searching behavior of some animal species with 
communication skills has been also studied. Although the main focus of the work
is on foraging strategies, our results could be extended to many other situations, such as 
mating or predation. The influence of different classes of random movements on the results is also analysed.
This work constitutes one of the first theoretical approaches tackling the effect that animal interactions have 
on the duration of their daily tasks. The main result is that the effect of communication 
on searching times is maximum when they send information at intermediate length scales. Longer communication ranges,
that suppose interacting with more individuals, overwhelm the searchers
with too much information coming from all the directions. On the other hand, short ranges do not provide 
all the population with information enough to expedite the search.
In both extreme situations the displacements of the foragers lose directionality to the targets. 
This result is robust against changes in the type of movement, either Brownian jumps or Lévy flights.
As a general result, Lévy strategies give faster searches, but Brownian jumps are more influenced
by a communication mechanism.

An application of this model to the foraging behavior of the gazelles inhabiting the Eastern Steppe of Mongolia is also presented. 
The steppe is one of the largest remaining grasslands in the world, where gazelles have to find each other and small areas
of good resources. In addition, because of the orography of this landscape, sound can travel long distances therein.
This, together with the strength of gazelle's vocal tract, allows them to communicate accoustically over long distances.
The model predicts an optimal search for resources when the communication is on a frequency of $1.25$~kHz, a value that lies
in the range measured for gazelles in the wild ($0.4$~kHz to $2.4$~kHz).
This result not only confirms the robustness of the model against changes
in the communication channel, but also gives realistic values for the measured quantities. 
This is our central finding,
and suggests that, during its evolution, the species could have optimized its vocal tract to efficiently communicate in the steppe. This 
work aims to open new research lines in the interrelation between communication, optimal search and mobility patterns. From a theoretical point of view,
we propose a new collective searching strategy that offers a wide range of potential fields of applicability, even far away from an 
ecological context. Similar algorithms, based on collective animal behavior, have been recently implemented in collectivities of robots to tackle 
different problems \citep{penders2011robot,Werfel14022014}. Due to its simplicity, our model could be applied to several searching processes, optimizing
the first hitting times if the individuals are enforced to communicate over intermediate lengths. Furthermore, the comparison between Brownian jumps and L\'evy flights
makes possible to choose the mobility strategy that better works in a given scenario.

Finally, it has been studied the effect of external variability on the diversity, robustness, and evolution
of many interacting particles systems. The behavior of the crossing times  
changes substantially when driven by fluctuating environments. It appears a finite region around the critical point
where this time scales with the system size as a power law with continuously varying exponents. 
These results have clear implications in the mean lifetimes of species in an ecological context (species coexistence, competition...) and also
allow to extend the concept of \textit{Temporal Griffiths Phases}, originally found in epidemic spreading models,
to a larger variety of systems.

There are several open questions to be tackled in the future.
Many of them are related with the influence of patterns and the role that different scales play on its formation.
Most of the future challenges in ecology involve many spatial and temporal scales. In fact, most of the natural
systems do not have a characteristic scale and the observed spatial strucutres most of the times have their origin in phenomena
that take place at smaller scale. The key for understanding and predicting lies in unveiling the mechanisms underlying
these structures \citep{levin1992problem}. 

In this dissertation, vegetation pattern formation has been addressed developing mathematical
models with a single spatial scale. However, trees also present facilitative interactions, related to the size of the tree canopy,
that act at a shorter scale than competition which are mediated by the roots.
Although one of our main results is that positive interactions could be superfluous in the formation
of patterns, they could have further implications on its shape and stability. In addition, nature is full of examples where many interaction scales are involved,
as it is the case of the regular distributions observed in many mussel beds.
Beyond spatial degrees of freedom, ecological systems also show different organizational scales. In the particular case of plants,
they are not isolated in the landscape but in interaction with many other species that influence its evolution.
This is the case of termites or some microbials, that are known as ecosystems engineers. Investigating its influence on the evolution and formation
of the vegetation distributions constitutes a promising challenge.

Establishing relationships between vegetation distributions and animal mobility presents also
many challenging questions. Most of them should focus on merging both research lines,
addressing the influence that grazing could have on the patterns, and how the formation of groups of animals
could modify their shapes or destroy them. While larger groups have clear benefits in terms of group defense and predator swamping, they also lead
to a faster degradation of the vegetation. This is the problem of foraging influencing the vegetation patterns which should be treated in the future.

The study of how an information flow can modify collective searching processes is attracting more attention the last years.
In this work, we have studied how the ranges at which the information is shared modify the duration of foraging.
However, many other questions, such as how informed individuals in a population may adapt its mobility pattern in order to 
increase the success of the group remain still open.

In summary, coming years promise an intense activity trying to answer these and more open questions.
Statistical physics is now much more than a discipline devoted to the study of the macroscopic properties of thermal systems, and theoretical ecology is a well 
established quantitative field. Their development during these years has brought them to a common point, from where natural environment can be better
described and understood.

%

   \addcontentsline{toc}{part}{
%

 \backmatter




%
\end{document}